\documentclass[12pt,letterpaper]{article}

\usepackage[margin=1in]{geometry}
\usepackage{iftex}
\ifPDFTeX
  \usepackage[T1]{fontenc}
  \usepackage{textcomp}
\else
  \usepackage{unicode-math}
\fi
\usepackage{lmodern}
\usepackage{microtype}
\usepackage{parskip}
\usepackage{setspace}
\usepackage{amsmath,amssymb,amsfonts,amsthm,mathtools}
\usepackage{mathrsfs,dsfont}
\usepackage{algorithm,algpseudocode}
\usepackage{booktabs,tabularx,multirow,array}
\usepackage{graphicx,caption,subcaption}
\usepackage[dvipsnames,svgnames,x11names]{xcolor}
\usepackage[hyphens]{url}
\IfFileExists{xurl.sty}{\usepackage{xurl}}{}
\usepackage{natbib}
\usepackage{hyperref}
\usepackage{etoc}

\hypersetup{
  pdftitle={Rectified Fisher-Bingham Model for Compositional Data with Zeros},
  pdfauthor={Eugene Han, Marahi Perez-Tamayo, Hannah D. Holscher, and Ruoqing Zhu},
  pdfkeywords={compositional data, directional statistics, Fisher-Bingham distribution, microbiota data, zero-valued components},
  colorlinks=true,
  linkcolor=blue,
  citecolor=Blue,
  urlcolor=Blue
}
\newtheorem{theorem}{Theorem}

\newcommand{\webappendix}[2]{%
  \section*{Web Appendix #1: #2}%
  \addcontentsline{toc}{section}{Web Appendix #1: #2}%
}
\newcommand{\websubsection}[1]{%
  \subsection*{#1}%
  \addcontentsline{toc}{subsection}{#1}%
}

\DeclareMathOperator{\diag}{diag}

\DeclareMathOperator{\tr}{tr}
\newcommand{\E}{\mathbb{E}}

\begin{document}
\setstretch{1.1}
\raggedbottom

  \title{\bf Rectified Fisher-Bingham Model for Compositional Data with Zeros}
\author{Eugene Han\thanks{Department of Statistics, University of Illinois at Urbana-Champaign, Champaign, IL.}
	\and Marahi Perez-Tamayo\thanks{Division of Nutritional Sciences, University of Illinois at Urbana-Champaign, Champaign, IL.}
	\and Hannah D. Holscher\thanks{Department of Food Science and Human Nutrition, University of Illinois at Urbana-Champaign, Champaign, IL.}\;\footnotemark[2]
	\and Ruoqing Zhu\footnotemark[1]}
\maketitle

\bigskip
\begin{abstract}
	This paper introduces a likelihood-based framework for compositional data with zeros using a rectified and renormalized Fisher-Bingham model, motivated by microbiota studies. The model represents compositions by their componentwise square roots on the nonnegative orthant of the unit sphere. It generates these square-root compositions by taking a latent Fisher-Bingham direction with at least one positive coordinate, setting its nonpositive coordinates to zero, and renormalizing. This Tobit-type construction models zero patterns and positive relative abundances jointly without requiring zero replacement. We establish global identifiability of the model parameters from the observed compositions under a standard Fisher-Bingham normalization. We develop a Monte Carlo expectation-maximization algorithm for likelihood-based estimation and a score test for between-group differences in the directional parameter when the groups share the same quadratic term. Simulations evaluate parameter and distributional recovery, computational performance, and the finite-sample behavior of the score test. Under simulated shifts in the directional parameter with a common quadratic term, the score test matched or exceeded the estimated power of both PERMANOVA benchmarks. We illustrate the framework by comparing gut microbiota compositions between arms at the end of a randomized avocado dietary intervention.
\end{abstract}

\vfill

\newpage
\singlespacing
\setlength{\bibsep}{3pt plus 1pt}


\section{Introduction}
\label{sec:intro}

Microbiome intervention studies often begin by asking whether an intervention changes the microbial community as a whole and how its components change in coordination. Such questions arise in clinical \citep{van2013duodenal}, behavioral \citep{allen2018exercise}, and dietary \citep{thompson2021avocado} interventions. Addressing them using microbial relative abundance data is challenging because these measurements are compositional \citep{gloor2017microbiome}, exhibit dependence induced by the unit-sum constraint, and frequently contain exact zeros. These features motivate methods that can characterize the joint microbial composition and test for structured differences across study conditions. The \textit{Persea americana} for Total Health (PATH) study \citep{thompson2021avocado} provides a concrete illustration. In this randomized dietary intervention, fecal samples were collected from participants assigned to an avocado-containing diet or a matched control diet and characterized using 16S ribosomal RNA gene sequencing. To complement the original analysis, we test for a between-arm difference in gut microbiota composition at the end of the intervention and use an exploratory score diagnostic to guide descriptive comparisons of taxon relative abundances.

Each sample is represented by a $p$-component composition, a vector of nonnegative relative abundances that sum to one. Such compositions lie on the simplex,
\[
\Delta^{p-1} = \left\{(x_1, \ldots, x_p)^\top : x_i \geq 0, \sum_{i=1}^p x_i = 1\right\},
\]
and have been studied extensively in the compositional data analysis literature \citep{aitchison1982statistical}. Many analyses of microbial relative abundances use log-ratio transformations \citep{aitchison1982statistical}, which represent strictly positive compositions in Euclidean coordinates \citep{li2015microbiome,mandal2015analysis}. Applying these transformations to compositions with zeros typically involves replacing zeros with small positive values, a choice that can affect interpretation and downstream inference. Zero replacement warrants particular care in microbiota data because observed zeros may reflect biological absence, limited sequencing depth, or detection thresholds \citep{martin2003dealing, kaul2017analysis}. Regardless of their origin, zeros place observations on lower-dimensional faces of the simplex and motivate methods that can model these observations directly.

A complementary approach represents a composition $x \in \Delta^{p-1}$ by its componentwise square root $y = (\sqrt{x_1}, \ldots, \sqrt{x_p})^\top$, which lies in the nonnegative orthant of the unit sphere,
\[
\mathcal{S}^{p-1}_{\geq 0} := \{y \in \mathbb{R}^p : \|y\|_2 = 1,\ y_i \geq 0 \text{ for all } i\}, 
\]
where directional statistical methods can be applied. A natural starting point is the Fisher-Bingham family, a class of probability distributions on the unit sphere that can capture both directional structure and dependence among components \citep{mardia}. Numerical methods for evaluating the Fisher-Bingham normalizing constant have made it more practical to fit models from the general family \citep{chen2021mle}. The family's full-sphere support poses a separate challenge for modeling square-root transformed compositions. Early directional models for these compositions used simpler Fisher-Bingham subfamilies, often assuming that probability outside the nonnegative orthant was negligible \citep{stephens1982use, scealy2011regression}. Subsequent work used folding \citep{scealy2014fitting} or truncation \citep{schwob2025spatial} to restrict the distribution to that orthant. The resulting continuous distributions assign zero probability to the lower-dimensional faces that correspond to compositions with exact zeros.

To model compositions with exact zeros, we propose the rectified and renormalized Fisher-Bingham (RRFB) model. Conditional on having at least one positive coordinate, a latent Fisher-Bingham direction is rectified by setting its nonpositive coordinates to zero and renormalizing. The latent direction is a modeling device rather than an unobserved biological abundance. This construction resembles a Tobit-type censoring mechanism \citep{tobin1958estimation}, but renormalization couples the observed components and induces a joint distribution on the simplex. The resulting model provides a joint likelihood for zero-containing compositions without requiring zero replacement or separate models for zero and nonzero components. We establish global identifiability of the model parameters from the observed compositions under a standard Fisher-Bingham normalization and develop a Monte Carlo expectation-maximization (MCEM) algorithm \citep{wei1990monte} for likelihood-based estimation. As an application of the RRFB likelihood, we develop a score test for between-group differences in the directional parameter when the groups share the same quadratic term. The test complements distance-based $\beta$-diversity analyses such as PERMANOVA \citep{legendre2013beta, anderson2001new}.

\section{Review of the Fisher-Bingham Distribution}
\label{sec:review}

The Fisher-Bingham distribution can be viewed as the distribution of a multivariate Gaussian random vector conditioned to lie on the unit sphere $\mathcal{S}^{p-1}$. Equivalently, it is a quadratic exponential family on the unit sphere. Let $\sigma$ denote the standard $(p-1)$-dimensional surface-area measure on $\mathcal{S}^{p-1}$. The Fisher-Bingham distribution has density, with respect to $\sigma$,
\begin{equation}
	\label{eq:fb_dens}
	f_{\mathrm{FB}}(y \mid A, \gamma) = \mathcal{C}^{-1}(A, \gamma) \exp(-y^\top A y + \gamma^\top y),
\end{equation}
where $y \in \mathcal{S}^{p-1}$, $\gamma \in \mathbb{R}^p$ governs directional location and concentration, $A = A^\top \in \mathbb{R}^{p \times p}$ allows anisotropic concentration and dependence, and
\[
	\mathcal{C}(A, \gamma) = \int_{\mathcal{S}^{p-1}} \exp(-y^\top A y + \gamma^\top y)\,d\sigma(y)
\]
is the normalizing constant. Since $A$ is symmetric, let $A = Q \Lambda Q^\top$ denote its spectral decomposition where $\Lambda = \diag(\lambda)$ and $\lambda_1 \leq \cdots \leq \lambda_p$. By rotational invariance of the surface measure on $\mathcal{S}^{p-1}$,
\begin{equation}
    \label{eq:nc}
	\mathcal{C}(A, \gamma) = \mathcal{C}(\Lambda, Q^\top \gamma) \eqqcolon \mathscr{C}(\lambda, Q^\top\gamma),
\end{equation}
so the normalizing constant depends on $A$ through its eigenvalues and on $\gamma$ in the eigenbasis of $A$. This representation underlies existing methods for evaluating the normalizing constant \citep{kume2005, kume2018, chen2021mle}. The parameter pairs $(A, \gamma)$ and $(A+cI_p, \gamma)$ define the same Fisher-Bingham distribution for any scalar $c$. The contribution of $cI_p$ to the exponent is constant on the unit sphere and is therefore absorbed into the normalizing constant. We remove this ambiguity by subtracting the smallest eigenvalue from all eigenvalues, which is the unique common shift that gives $\lambda_1 = 0$. Under this normalization, $A$ is singular by construction, and adding any positive multiple of the identity gives a positive-definite representation of the same distribution.

Several well-known distributions for directional data arise as special cases of the Fisher-Bingham distribution. Subfamilies are often described via coordinate-wise restrictions in a fixed eigenbasis \citep[e.g.,][]{kume2018}, but we instead characterize them through invariant structural conditions on $A$ and $\gamma$, thereby separating geometric structures from basis-dependent parameterizations. Since $A$ is identifiable only up to addition of a scalar multiple of the identity, all structural conditions are understood up to this ambiguity.

The von Mises-Fisher distribution \citep{mardia}, Bingham distribution \citep{bingham1974antipodally}, and Kent distribution \citep{kent1982fisher} represent widely used subfamilies of the Fisher-Bingham distribution, each capturing distinct geometric features of directional data. The von Mises-Fisher distribution ($A = 0_{p \times p}$) models unimodal, isotropic concentration around a mean direction and is commonly used in modern machine learning applications involving normalized embeddings \citep{banerjee2005clustering, davidson2018hyperspherical}. The Bingham distribution ($\gamma=0_p$) models axial data, capturing anisotropic dispersion while remaining invariant under antipodal symmetry \citep{bingham1974antipodally}, and is widely used in paleomagnetic data analysis \citep{onstott1980application}. The generalized Kent distribution is characterized by $\gamma$ being an eigendirection of $A$, equivalently $A\gamma = 0_p$ under an appropriate scalar-shift normalization. Because $A$ is symmetric, its nonzero eigendirections are orthogonal to the preferred direction defined by $\gamma$, allowing elliptical dispersion around that direction. The Kent distribution can therefore be viewed as a spherical analogue of a multivariate Gaussian distribution with anisotropic covariance \citep{kent1982fisher}. The widespread use of these subfamilies in fields ranging from geoscience to machine learning highlights the practical importance of modeling directional concentration and anisotropy. Their limitations when considered individually motivate the use of the full Fisher-Bingham family, which provides a unified and more expressive framework for capturing complex directional structure.

\subsection{Numerical Computation}
The Fisher-Bingham family belongs to the exponential family \citep[see, e.g.,][]{mardia}, and likelihood-based inference depends critically on the normalizing constant, which generally must be evaluated or approximated numerically in higher dimensions. \citet{kume2005} developed a saddlepoint approximation that is computationally efficient, numerically stable, and closely approximates the exact value. Subsequently, \citet{kume2018} introduced an adjusted holonomic gradient method that characterizes the normalizing constant through a Pfaffian system evaluated by numerical ODE integration. Although this approach can provide high accuracy, its computational cost increases substantially with dimension. More recently, \citet{chen2021mle} introduced an efficient numerical integration scheme that achieves higher accuracy than the saddlepoint approximation while remaining computationally feasible in high dimensions.

Efficient sampling from the Fisher-Bingham distribution is also an important computational component, particularly for simulation studies and Monte Carlo approximations. Several approaches have been proposed in the literature, including Markov chain Monte Carlo methods based on Gibbs and slice sampling \citep{kume2009}, as well as more recent unified rejection-based frameworks for directional distributions \citep{kent2018new}. 

\section{Rectified and Renormalized Fisher-Bingham Model}
\label{sec:methods}

Let $x \in \Delta^{p-1}$ denote an observed relative abundance composition for a sample, and let $y = \sqrt{x} \in \mathcal{S}^{p-1}_{\geq 0}$ denote its componentwise square-root transformation. Zero components of $x$ place $y$ on the boundary of the nonnegative spherical orthant. To model both interior and boundary observations, we introduce a latent Fisher-Bingham direction $z$ and map it to $y$ through rectification and renormalization. The rectification step has a Tobit-type interpretation: a nonpositive latent coordinate produces an observed zero. Unlike ordinary Tobit censoring, the positive coordinates are subsequently renormalized, coupling all observed components through the unit-norm constraint. Specifically, let $z \sim \mathrm{FB}(A, \gamma)$ on $\mathcal{S}^{p-1}$, and define the valid latent region
\[
    \mathcal{E} = \{z \in \mathcal{S}^{p-1} : \|z^+\|_2 > 0\},
\]
where $z^+ \in \mathbb{R}_{\geq 0}^p$ is the componentwise nonnegative part of $z$, i.e. $z^+_j = \max(z_j, 0)$. Its complement $\mathcal{E}^c$ consists of directions whose coordinates are all nonpositive, for which $z^+ = 0_p$ and the rectified-renormalized transformation is undefined. We therefore condition the latent Fisher-Bingham distribution on the valid region. The resulting density is
\[
	f_{\mathrm{FB}\mid\mathcal E}(z \mid A, \gamma) = P_{A, \gamma}(\mathcal{E})^{-1} f_{\mathrm{FB}}(z \mid A, \gamma), \qquad z \in \mathcal{E}, 
\]
where
\[
	P_{A, \gamma}(\mathcal{E}) = \int_{\mathcal{E}} f_{\mathrm{FB}}(z \mid A, \gamma)\,d\sigma(z)
\]
is the Fisher-Bingham probability assigned to the valid latent region. We then set
\[
    y = T(z) = \frac{z^+}{\|z^+\|_2},
\]
where the transformation $T$ maps valid latent directions to $\mathcal{S}^{p-1}_{\geq 0}$. Because the RRFB distribution assigns positive probability to faces of different dimensions, an ordinary $(p-1)$-dimensional surface measure does not dominate the entire observed distribution. For each nonempty index set $M \subseteq \{1, \ldots, p\}$, define the face
\[
	\mathcal{S}_M = \left\{y \in \mathcal{S}^{p-1}_{\geq 0}: y_j > 0 \text{ for } j\in M, \; y_j=0 \text{ for } j\notin M\right\}.
\]
Let $\sigma_M$ denote the natural $(|M|-1)$-dimensional surface measure on $\mathcal{S}_M$. For $|M| = 1$, $\sigma_M$ is the counting measure on the singleton face. We equip the observed sample space with the stratified dominating measure
\[
	\mu(dy) = \sum_{\emptyset \neq M \subseteq \{1, \ldots, p\}}\sigma_M(dy).
\]
Thus, the component of $\mu$ on each face uses the surface measure appropriate to that face's dimension. For $y\in \mathcal{S}_M$, define its face-specific latent preimage by
\[
	\mathcal{Z}_M(y) = \left\{z \in \mathcal{E}: T(z) = y\right\}.
\]
Let $\nu_{M,y}$ denote the measure on $\mathcal Z_M(y)$ induced by the surface-area measure $\sigma$ under the transformation $T$. Its formal construction is provided in Web Appendix~A. The observed-data density, with respect to $\mu$, is
\begin{equation}
	\label{eq:obs-dens}
	f_{\mathrm{RRFB}}(y \mid A, \gamma) = P_{A, \gamma}(\mathcal{E})^{-1} \int_{\mathcal{Z}_M(y)} f_{\mathrm{FB}}(z \mid A, \gamma)\, d\nu_{M, y}(z), \qquad y \in \mathcal S_M.
\end{equation}

\begin{theorem}[Global identifiability of the RRFB model]
    \label{thm:identifiability}
    Suppose $p \geq 2$. Let $A, A' \in \mathbb{R}^{p \times p}$ be symmetric matrices and let $\gamma, \gamma' \in \mathbb{R}^p$. Then $f_{\mathrm{RRFB}}(y \mid A, \gamma) = f_{\mathrm{RRFB}}(y \mid A', \gamma')$ for $\mu$-almost every $y$ if and only if $\gamma' = \gamma$ and $A' = A + cI_p$ for some $c \in \mathbb{R}$.
\end{theorem}
The key observation is that the strictly positive face has positive probability for every finite $(A,\gamma)$, and $T$ is the identity on this face. Here, the observed density retains the Fisher-Bingham kernel up to a normalizing constant. Web Appendix~B provides the proof of Theorem~\ref{thm:identifiability}, showing that equality of the observed densities on this face forces $\gamma=\gamma'$ and $A'=A+cI_p$, establishing identification under the normalization $\lambda_1=0$. Rectification may nevertheless reduce the information available about $(A,\gamma)$, particularly when most observations lie on boundary faces. Evaluating the likelihood contributions of boundary observations requires integration over their latent preimages, for which we next introduce a block-norm parameterization.

\subsection{Block-Norm Parameterization}
\label{subsec:block}
To evaluate the face-specific integral in \eqref{eq:obs-dens}, we introduce a block-norm parameterization of the preimage $\mathcal{Z}_M(y)$ that separates the positive and nonpositive coordinate blocks. Let $m$ denote the number of zero components of $y$, so that $|M| = p - m$. Suppose first that $m \geq 1$. Conditional on the observed zero pattern, and without loss of generality, assume the first $p-m$ entries of $y$ are positive and the remaining $m$ entries are zero. Any $z \in \mathcal{Z}_M(y)$ can be represented as
\[
    z(\delta, u) = \begin{bmatrix}
        \sqrt{\delta}\, y_{1:p-m} \\
        \sqrt{1 - \delta}\, u
    \end{bmatrix},
\]
where $\delta = \|z^+\|_2^2 \in (0, 1]$ represents the squared norm of the positive latent components and $u \in \mathcal{S}^{m-1}_{\leq 0} = \mathcal{S}^{m-1}\cap \mathbb{R}^m_{\leq 0}$ lies in the nonpositive orthant of the lower-dimensional unit sphere. With the natural surface measures used above, the induced preimage measure is
\[
	d\nu_{M, y}(\delta, u) = \frac{1}{2}\delta^{\frac{p-m}{2}-1}(1-\delta)^{\frac{m}{2}-1}\,d\delta\,d\sigma_{m-1}(u),
\]
where $\sigma_{m-1}$ denotes surface-area measure restricted to $\mathcal{S}^{m-1}_{\leq 0}$. Thus, the spherical geometry contributes a beta-type weight for $\delta$, reducing the preimage integral to integration over a scalar block norm and a lower-dimensional direction. Web Appendix~A derives this measure.

For independent observations $y_1, \ldots, y_n$, the same construction applies. For each sample $y_i$, let $m_i$ denote the number of zero components in $y_i$, and let $M_i = \{j : y_{ij} > 0\}$ denote its positive set. Let $P_i \in \{0, 1\}^{p \times p}$ be the permutation matrix that reorders coordinates so that the positive components of $y_i$ appear first, and define $\widetilde{y}_i = (P_i y_i)_{1:p-m_i}$. For $m_i \geq 1$, the latent preimage in the original coordinate order is parameterized by
\[
    z_i(\delta_i, u_i) = P_i^\top \begin{bmatrix}
        \sqrt{\delta_i}\, \widetilde{y}_i \\
        \sqrt{1-\delta_i}\, u_i
    \end{bmatrix},
\]
where $\delta_i \in (0, 1]$ and $u_i \in \mathcal{S}^{m_i - 1}_{\leq 0}$. For fully positive observations, where $m_i = 0$, rectification is inactive and the preimage reduces to $z_i = y_i$. Since the face-specific integral in \eqref{eq:obs-dens} does not admit a closed-form expression for boundary observations, we estimate $(A, \gamma)$ by the Monte Carlo expectation-maximization (MCEM) algorithm \citep{wei1990monte}.

\subsection{MCEM}
\label{subsec:mcem}

For an observation with $m_i \geq 1$, the posterior density of $(\delta_i, u_i)$ given $y_i$ is proportional to
\[
	p(\delta_i, u_i \mid y_i, A, \gamma) \propto \delta_i^{\frac{p-m_i}{2}-1}(1-\delta_i)^{\frac{m_i}{2}-1} \exp\!\big(-z_i(\delta_i,u_i)^\top A z_i(\delta_i,u_i)+ \gamma^\top z_i(\delta_i,u_i) \big),
\]
with support on $(0,1]\times \mathcal S^{m_i-1}_{\leq 0}$. The factor $P_{A, \gamma}(\mathcal E)^{-1}$ is constant over the preimage of a fixed $y_i$ and therefore cancels from this conditional posterior. Thus, conditioning on $\mathcal E$ does not alter the E-step posterior moments. Ignoring the exponential tilt induced by $(A, \gamma)$, the base measure factorizes into a Beta distribution in $\delta_i$ and a uniform distribution on $\mathcal{S}_{\leq 0}^{m_i-1}$, which provides a convenient proposal distribution for Monte Carlo approximation. Define the posterior moments
\[
M_{1i}^{(t)}=\E[z_i \mid y_i, A^{(t)},\gamma^{(t)}],
\qquad
M_{2i}^{(t)}=\E[z_i z_i^\top \mid y_i, A^{(t)},\gamma^{(t)}],
\]
and their empirical averages
\[
\overline S_1^{(t)}=\frac{1}{n}\sum_{i=1}^n M_{1i}^{(t)}, 
\qquad
\overline S_2^{(t)}=\frac{1}{n}\sum_{i=1}^n M_{2i}^{(t)}.
\]
For a fully positive observation, the latent direction is observed exactly, so $M_{1i}^{(t)}=y_i$ and $M_{2i}^{(t)}=y_i y_i^\top$. For boundary observations, the posterior distribution and its moments depend on the zero pattern of $y_i$ and require pattern-specific numerical evaluation. Although the conditioning probability cancels from the E-step posterior, it depends on the candidate parameters in the M-step and must be retained in the expected complete-data log-likelihood. Define the restricted normalizing constant
\[
\mathcal C_{\mathcal E}(A,\gamma)
=
\mathcal C(A,\gamma)P_{A,\gamma}(\mathcal E).
\]
The MCEM objective is then
\begin{equation}
\mathcal Q(A,\gamma \mid A^{(t)},\gamma^{(t)})
=
-n\left[
\log \mathcal C_{\mathcal E}(A,\gamma)
+\tr\!\left(A \overline S_2^{(t)}\right)
-\gamma^\top \overline S_1^{(t)}
\right].
\label{eq:Q-fun}
\end{equation}
The M-step retains the Fisher-Bingham sufficient-statistic structure with the normalizing constant restricted to the valid latent region. 

\subsection{MCEM Algorithm}
\label{subsec:alg}
Algorithm~\ref{alg:mcem} summarizes the fitting procedure. In the M-step, we write $A=Q\diag(\lambda)Q^\top$ and jointly update $(\lambda, \gamma, Q)$. We enforce $0=\lambda_1\leq\cdots\leq\lambda_p$ by reparameterizing the eigenvalue gaps through an unconstrained vector $v\in\mathbb{R}^{p-1}$. The joint update can therefore be viewed as optimization on the product manifold $\mathbb{R}^{p-1}\times\mathbb{R}^p\times\mathrm{O}(p)$, where the two Euclidean factors represent $v$ and $\gamma$, and $\mathrm{O}(p)=\{Q\in\mathbb{R}^{p\times p}:Q^\top Q=I_p\}$ is the orthogonal group. This formulation accommodates simultaneous updates of all parameter blocks while preserving the identifying and orthogonality constraints. Web Appendix~D.2 specifies the reparameterization and numerical optimization procedure. Web Appendices~D.1, D.3, and D.4 describe posterior-moment evaluation, likelihood evaluation, and initialization, respectively.

Let $\ell^{(t)}$ denote the numerical log-likelihood per observation at iteration $t$, evaluated with the parameter-independent face constants omitted as described in Web Appendix~D.3. At each iteration, we calculate
the scaled changes
\[
\frac{|\ell^{(t)}-\ell^{(t-1)}|}{1+|\ell^{(t-1)}|}, \qquad \frac{\|A^{(t)}-A^{(t-1)}\|_{\mathrm F}}{1+\|A^{(t-1)}\|_{\mathrm F}}, \qquad \frac{\|\gamma^{(t)}-\gamma^{(t-1)}\|_2} {1+\|\gamma^{(t-1)}\|_2}.
\]
Convergence is declared when, for a pre-specified $K$ consecutive iterations, all three quantities fall below their respective prespecified tolerances and the joint M-step gradient norm in scaled product-manifold coordinates falls below its prescribed tolerance.

\begin{algorithm}[htbp]
	\singlespacing
	\caption{MCEM algorithm for RRFB parameter estimation}
	\vspace{-\baselineskip}
	\label{alg:mcem}
	\begin{algorithmic}[1]
		\Require Observed compositions $x_1,\dots,x_n \in \Delta^{p-1}$, with $y_i=\sqrt{x_i}$
		\Ensure Approximate maximum-likelihood estimates $(\widehat A, \widehat\gamma)$
		\State Obtain initial values $(\lambda^{(0)},\gamma^{(0)},Q^{(0)})$ using the procedure in Web Appendix~D.4
		\State Set $A^{(0)} \gets Q^{(0)}\diag(\lambda^{(0)})(Q^{(0)})^\top$ and $t\gets 0$
		\While{the convergence criteria have not been satisfied for $K$ consecutive iterations}
		\State \textbf{E-step:} Compute the conditional sufficient statistics $\overline S_1^{(t)}$ and $\overline S_2^{(t)}$
		\State \textbf{M-step:} Jointly update $(\lambda^{(t+1)}, \gamma^{(t+1)}, Q^{(t+1)})$ by approximately maximizing $\mathcal Q(A, \gamma\mid A^{(t)}, \gamma^{(t)})$ subject to $0 = \lambda_1 \leq \ldots \leq \lambda_p$ and $Q^\top Q = I_p$
		\State Set $A^{(t+1)} = Q^{(t+1)}\diag(\lambda^{(t+1)})(Q^{(t+1)})^\top$
		\State Assess convergence using changes in the observed-data log-likelihood, $A$, and $\gamma$, together with the M-step gradient criterion
		\State $t \gets t+1$
		\EndWhile
		\State \textbf{Return:} $\widehat A = A^{(t)}$ and $\widehat\gamma = \gamma^{(t)}$
	\end{algorithmic}
\end{algorithm}

\section{Score Test for Between-Group Directional Differences}
\label{sec:score}

Theorem~\ref{thm:identifiability} establishes that the RRFB parameters are identifiable under the normalization $\lambda_1=0$. With the observed-data likelihood from \eqref{eq:obs-dens}, we can use standard likelihood-based methods to test hypotheses about these parameters. As an illustration, we use a score test to compare two independent groups with a common quadratic parameter $A$ and group-specific directional parameters $\gamma_0$ and $\gamma_1$. Under this model, the null hypothesis $H_0:\gamma_0=\gamma_1$ is equivalent to equality of the two observed RRFB distributions. The score test requires fitting only the pooled model under $H_0$. A difference in $\gamma$ can affect both the probabilities of zero patterns and the distribution of positive relative abundances within a given zero pattern.

Write $\gamma_j=\overline\gamma+(2j-1)\psi$ for $j=0,1$, where $\overline\gamma=(\gamma_0+\gamma_1)/2$ and $\psi=(\gamma_1-\gamma_0)/2$. Let $\eta$ collect $\overline\gamma$ and the locally identifiable coordinates of $A$ under the normalization $\lambda_1=0$. Under $H_0:\psi=0$, we obtain the restricted estimate $\widehat\theta_0=(0_p,\widehat\eta_0)$ by fitting a single RRFB model to the pooled observations using Algorithm~\ref{alg:mcem}. For an observation $y$ in group $j$, the directional score is
\[
\nabla_{\gamma_j}\log f_{\mathrm{RRFB}}(y\mid A,\gamma_j)
=
\E_{A,\gamma_j}[Z\mid Y=y]
-
\E_{A,\gamma_j}[Z\mid Z\in\mathcal E].
\]
The first term is the posterior first moment used in the MCEM E-step. The second comes from differentiating $\log\mathcal C_{\mathcal E}(A,\gamma_j)$ and accounts for conditioning on $\mathcal E$. Multiplying this directional score by $2j-1$ gives the observation's contribution to the score for $\psi$. Let $\overline s_\psi$ and $\overline s_\eta$ denote the average contrast and nuisance scores at $\widehat\theta_0$, and let $\widehat H$ be the negative Hessian of the two-group log-likelihood at $\widehat\theta_0$ divided by $n=n_0+n_1$. The nuisance-adjusted average score is
\[
\overline s_{\psi\mid\eta}
=
\overline s_\psi-\widehat H_{\psi\eta}\widehat H_{\eta\eta}^{-1}\overline s_\eta.
\]
Let $\widehat V_{\psi\mid\eta}$ be the empirical covariance of the individual score contributions after applying the same adjustment. The test statistic is
\begin{equation}
	T
	=
	n\,\overline s_{\psi\mid\eta}^{\top}
	\widehat V_{\psi\mid\eta}^{-1}
	\overline s_{\psi\mid\eta}.
	\label{eq:score-test}
\end{equation}
For fixed $p$, $T$ converges in distribution to $\chi_p^2$ under $H_0$ and the regularity conditions in Web Appendix~E. The appendix provides the full derivation and details of the numerical implementation.

\section{Simulation Studies}
\label{sec:sims}

Following the ADEMP framework \citep{morris2019using}, we evaluate parameter and distributional recovery, score-test performance, and computational performance and sensitivity to initialization in three simulation studies. We simulate latent Fisher-Bingham directions using the rejection sampler of \citet{kent2018new}, described in Web Appendix~F. Web Appendix~G.1 gives the implementation details and numerical settings used for RRFB fitting and score-test calculations used throughout these studies.

\subsection{Study 1: Estimation and distributional recovery}

We consider a factorial design that crosses three dimensions ($p=5$, $10$, and $20$), four sample sizes ($n=100$, $200$, $500$, and $1000$), and four zero-prevalence configurations. Cases~1, 2, and 4 target per-component zero prevalence of 10--15\%, 20--25\%, and 30--35\%, respectively. Case~3 targets approximately 30\% zeros in 2, 3, and 5 components for $p=5, 10,$ and 20, respectively, and approximately 10\% zeros in the remaining components. Within each dimension, we fixed $A$ and selected four different values of $\gamma$ to induce these approximate marginal zero probabilities. For each configuration, we generated independent latent directions from the Fisher-Bingham distribution conditional on the valid region and applied the rectification and renormalization map to obtain the observed square-root compositions. The parameter settings and resulting componentwise zero proportions are reported in Web Appendix~H.1. The primary estimands are the identifiable generating parameters $(A, \gamma)$, with $A$ represented under the constraint $\lambda_1 = 0$. The secondary estimand is the induced observed-data distribution, \eqref{eq:obs-dens}. Each generated dataset was fitted using the MCEM procedure in Algorithm~\ref{alg:mcem}. Let $a = \operatorname{vech}(A)$ contain the unique entries of $A$, and let $\theta$ denote either $a$ or $\gamma$. We summarize parameter recovery using the relative average absolute bias and relative average empirical standard deviation,
\[
	\frac{\sum_{j=1}^{d} |\overline{\widehat{\theta}}_j - \theta_{0, j}|}{\|\theta_0\|_1}, \qquad
	\frac{\sum_{j=1}^d \widehat{\operatorname{SD}} (\{\widehat{\theta}^{(r)}_j\}_{r=1}^R)}{\|\theta_0\|_1},
\]
respectively, where $d$ is the dimension of $\theta$, $R$ is the number of simulation replicates, and $\theta_0$ is the true parameter. As a complementary distribution-level performance measure, let $f_0$ denote the RRFB density under $(A_0,\gamma_0)$ and let $\widehat f_r$ denote the fitted RRFB density under $(\widehat A^{(r)},\widehat\gamma^{(r)})$. We calculate the average expected log-score difference
\begin{equation}
	\label{expected-log-score}
	\overline{\Delta} = \frac{1}{R} \sum_{r=1}^R \Delta_r, \qquad \Delta_r = \mathbb{E}_{Y \sim f_0} \left[\log f_0(Y) - \log \widehat f_r(Y)\right] = D_{\mathrm{KL}}(f_0 \|\widehat f_r),
\end{equation}
to determine whether parameter error materially affects the overall fitted observed-data distribution.

\begin{table}[htbp]
	\centering
	\footnotesize
	\setlength{\tabcolsep}{3.2pt}
	\caption{Finite-sample recovery of the RRFB parameters based on 200 simulation replicates per configuration. Entries are the relative average absolute bias, with the relative average empirical standard deviation in parentheses. Summaries for $A$ are calculated from its unique entries, $a=\operatorname{vech}(A)$.}
	\label{tab:parameter_recovery}
	\begin{tabular}{cc*{8}{c}}
		\toprule
		& & \multicolumn{4}{c}{Parameter $A$}
		& \multicolumn{4}{c}{Parameter $\gamma$} \\
		\cmidrule(lr){3-6}\cmidrule(lr){7-10}
		$p$ & Case
		& $n=100$ & $n=200$ & $n=500$ & $n=1000$
		& $n=100$ & $n=200$ & $n=500$ & $n=1000$ \\
		\midrule
		5  & 1 & 0.67 (1.18) & 0.26 (0.71) & 0.15 (0.46) & 0.06 (0.32)
		& 0.58 (0.95) & 0.21 (0.53) & 0.14 (0.33) & 0.07 (0.23) \\
		5  & 2 & 0.25 (0.74) & 0.10 (0.46) & 0.03 (0.28) & 0.02 (0.20)
		& 0.28 (0.69) & 0.12 (0.38) & 0.04 (0.23) & 0.02 (0.15) \\
		5  & 3 & 0.44 (1.09) & 0.25 (0.73) & 0.07 (0.40) & 0.04 (0.28)
		& 0.53 (1.10) & 0.34 (0.70) & 0.07 (0.34) & 0.03 (0.23) \\
		5  & 4 & 0.15 (0.59) & 0.07 (0.40) & 0.03 (0.25) & 0.02 (0.17)
		& 0.11 (0.65) & 0.09 (0.44) & 0.02 (0.27) & 0.02 (0.19) \\
		\midrule
		10 & 1 & 0.65 (1.44) & 0.35 (0.92) & 0.19 (0.57) & 0.11 (0.41)
		& 0.61 (1.01) & 0.30 (0.63) & 0.19 (0.38) & 0.11 (0.29) \\
		10 & 2 & 0.37 (1.09) & 0.20 (0.70) & 0.08 (0.41) & 0.04 (0.29)
		& 0.42 (0.85) & 0.24 (0.52) & 0.08 (0.31) & 0.04 (0.20) \\
		10 & 3 & 0.70 (1.47) & 0.39 (0.94) & 0.15 (0.57) & 0.09 (0.39)
		& 0.84 (1.26) & 0.46 (0.81) & 0.16 (0.46) & 0.09 (0.30) \\
		10 & 4 & 0.25 (0.92) & 0.13 (0.59) & 0.04 (0.34) & 0.02 (0.24)
		& 0.24 (0.74) & 0.14 (0.47) & 0.03 (0.26) & 0.02 (0.18) \\
		\midrule
		20 & 1 & 1.41 (3.06) & 0.68 (1.73) & 0.28 (0.98) & 0.16 (0.68)
		& 1.25 (1.55) & 0.56 (0.87) & 0.20 (0.49) & 0.11 (0.35) \\
		20 & 2 & 0.86 (2.49) & 0.38 (1.37) & 0.15 (0.76) & 0.08 (0.52)
		& 0.90 (1.44) & 0.36 (0.75) & 0.12 (0.41) & 0.05 (0.27) \\
		20 & 3 & 1.46 (3.19) & 0.72 (1.81) & 0.32 (1.04) & 0.16 (0.71)
		& 1.53 (1.86) & 0.70 (1.08) & 0.30 (0.63) & 0.12 (0.42) \\
		20 & 4 & 0.65 (2.31) & 0.27 (1.19) & 0.10 (0.66) & 0.05 (0.44)
		& 0.69 (1.40) & 0.26 (0.70) & 0.09 (0.36) & 0.04 (0.24) \\
		\bottomrule
	\end{tabular}
\end{table}

Relative bias and empirical standard deviation decreased with sample size in every configuration and were generally larger at higher dimensions and in Cases~1 and 3 (Table~\ref{tab:parameter_recovery}). Across zero-prevalence cases at a given dimension and sample size, the mean expected log-score difference varied much less than the parameter recovery measures. For example, at $p=20$ and $n=1000$, relative bias for $A$ ranged from 0.05 to 0.16, while the mean expected log-score difference ranged from 0.118 to 0.120. This pattern suggests that some parameter errors have relatively little effect on the fitted observed-data distribution. Web Appendix~H.2 reports the complete log-score results and their Monte Carlo standard errors.

\subsection{Study 2: Calibration, power, and sensitivity of the common-\texorpdfstring{$A$}{A} RRFB score test}
\label{subsec:study2}

To assess null calibration, we used dimensions $p=5$ and $p=10$, with 100, 200, or 500 observations per group and the four zero-prevalence configurations from Study~1. For each configuration, we generated 1,000 replicates of two independent groups from the same $(A, \gamma)$. 

\begin{figure}[htbp]
	\centering
	\includegraphics[width=\linewidth]{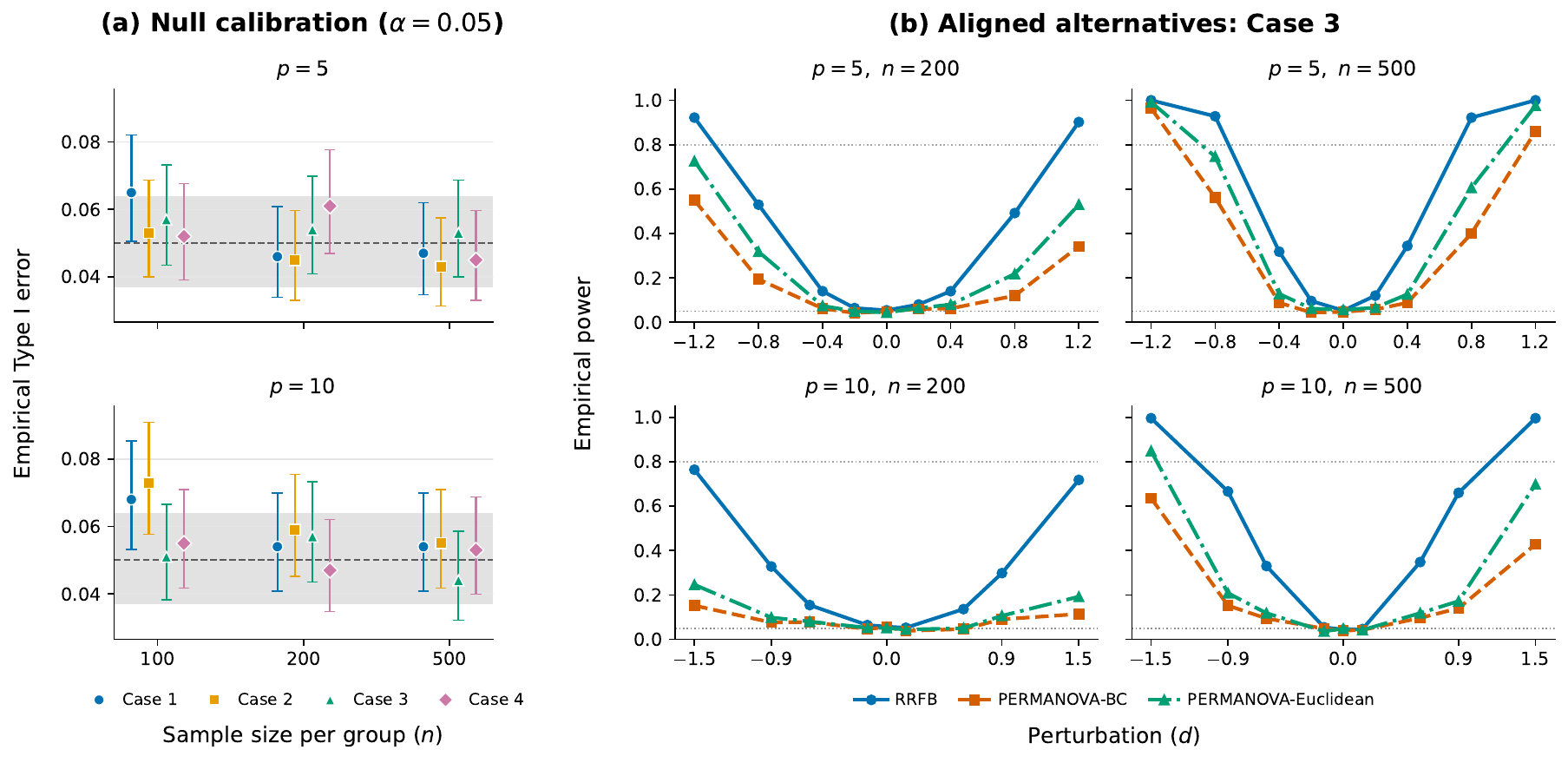}
	\caption{Operating characteristics of the common-$A$ RRFB score test at $\alpha=0.05$. (a) Empirical Type~I error across four zero-prevalence configurations, with $p=5$ or $10$ components and $n=100$, $200$, or $500$ observations per group. Each estimate is based on 1,000 null replicates. (b) Empirical rejection probabilities for the RRFB score test and the two PERMANOVA benchmarks under Case~3, with $p=5$ or $10$ and $n=200$ or $500$ observations per group. The signed perturbation $d$ changes the directional parameter in Group~1 while preserving the common quadratic parameter $A$. Estimates use 500 replicates at each nonzero value of $d$ and the 1,000 matching null replicates at $d=0$. Horizontal reference lines in panel~(b) mark rejection probabilities of 0.05 and 0.80. Web Appendices~I.1 and~I.2 give additional null-calibration results and power curves for all four cases, respectively.}
	\label{fig:study2}
\end{figure}

At the nominal $\alpha=0.05$ level, rejection rates ranged from 0.043 to 0.073, with the largest departures at $p=10$ and $n=100$ per group (Figure~\ref{fig:study2}a). At $n=500$ per group, all rates fell within the pointwise $95\%$ binomial reference band. The score statistic generally followed its $\chi_p^2$ reference distribution, with the clearest discrepancies in the upper tails at smaller sample sizes. Some nuisance sensitivity matrices were poorly conditioned, but all efficient-score covariance matrices were numerically positive definite throughout. Web Appendix~I.1 provides rejection rates at three nominal levels $(0.01, 0.05, 0.10)$, QQ plots, $p$-value uniformity summaries, and score-related matrix diagnostics.

We evaluated power under alternatives aligned with the parameter targeted by the score test. We considered $p \in \{5, 10\}$, group sizes $n \in \{200, 500\}$, and all four zero-prevalence configurations from Study~1. The two groups shared a common $A$, while the second-group directional parameter was $\gamma_1(d) = \gamma_0 - Q\zeta_p(d)$, where $\zeta_p(d)$ was supported on the final two and three eigencoordinates for $p=5$ and $p=10$, respectively. The signed perturbation $d$ varied over a dimension-specific symmetric grid, yielding a sequence of local and moderate departures from the null hypothesis. Web Appendix~I.2 gives the perturbation grid and shows how observed zero prevalence increased with the signed perturbation $d$. Power at level $\alpha=0.05$ was estimated from 500 replicates per configuration. PERMANOVA comparisons used Bray-Curtis distance on relative abundances and Euclidean distance on square-root transformed compositions with 999 permutations.

Estimated RRFB power increased with perturbation magnitude on each side of the null and generally increased with sample size. For these common-$A$ shifts in $\gamma$, estimated RRFB power was at least as high as that of either PERMANOVA procedure at every nonzero grid point. Because these alternatives are aligned with the parameter targeted by the score test, whereas PERMANOVA is an omnibus procedure that spreads power over many types of difference, these comparisons illustrate a power-robustness tradeoff rather than a general power advantage. For example, in Case~3 at $p=10$, $n=500$ per group, and $d=0.9$, estimated power was $0.66$  for RRFB, $0.17$ for Euclidean PERMANOVA, and $0.14$ for Bray-Curtis PERMANOVA (Figure~\ref{fig:study2}b). Web Appendix~I.2 shows the full power curves for all four cases.

We conducted a targeted sensitivity analysis on the common-$A$ assumption using the $p=5$, Case~2 configuration, with $n=200$ observations per group and 500 replicates per setting. The groups had the same ambient-coordinate directional parameter, $\gamma_0=\gamma_1=\gamma$, but different quadratic parameters, $A_0\neq A_1$. We generated departures from the common-$A$ assumption either by scaling the nonzero eigenvalues of $A$ while retaining its eigendirections or by rotating its eigendirections while preserving its eigenvalues and $\gamma$.

For mild departures from common $A$, rejection rates were similar across the three procedures and generally increased as the departures grew. With $\gamma$ held fixed, RRFB rejected in $69.6\%$ of replicates when $A_1 = 0.6 A_0$. Interpreting a rejection as evidence of a difference in $\gamma$ therefore depends on the common-$A$ assumption. Web Appendix~I.3 shows the rejection curves for both eigenvalue scaling and rotation.

\subsection{Study 3: Computational performance and sensitivity to initialization}
\label{subsec:study3}

We characterized the computational burden of RRFB estimation using the factorial simulation design from Study~1. For each combination of dimension, sample size, and zero-prevalence configuration, we recorded the wall-clock time and number of outer MCEM iterations required for model fitting. We separately evaluated the additional cost of two-sample inference using the null-calibration simulations from Study~2, recording the time required to fit the pooled null model and the subsequent time required to calculate the score statistic and its sandwich covariance. We also examined the corresponding timings across the nonnull perturbations used in the power study. Furthermore, to assess sensitivity to initialization, we selected 20 independently generated datasets from each combination of $p\in\{5,10\}$ and the Case~1 and Case~4 zero-prevalence configurations, with $n=200$. Each dataset was fitted from four starting values: the proposed latent-moment initialization, the same initialization with its parameter values scaled by 0.5 or 2, and an observed-moment-based initialization. For each dataset, we calculated the range across the four starts in the final observed log likelihood per observation and in the expected log-score difference. The latter measures the extent to which initialization affected the induced fitted observed-data distribution. Computations were performed at the Illinois Campus Cluster using SLURM. Each simulation replicate was executed as an independent single-core CPU task with 4-16GB of memory allocated per task.

\begin{figure}[htbp]
	\centering
	\includegraphics[width=\linewidth]{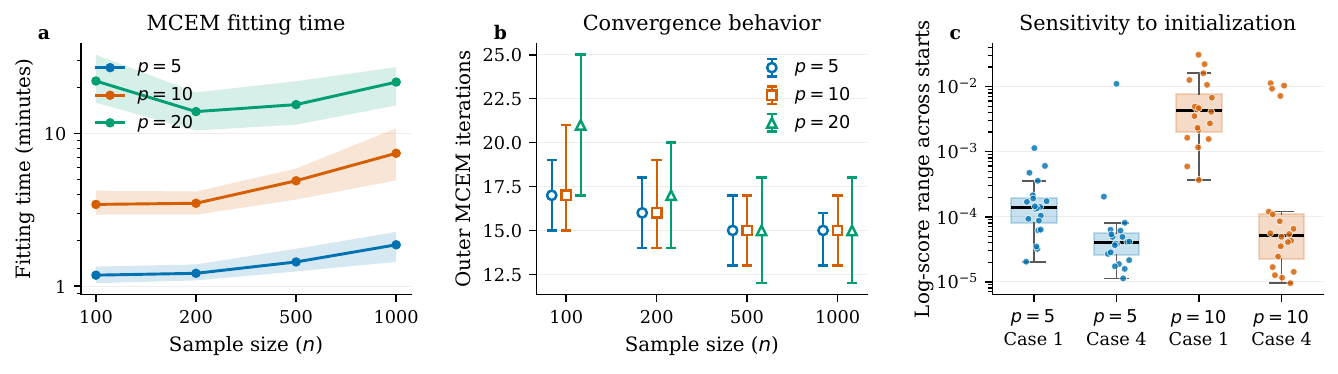}
	\caption{
		Computational performance and sensitivity to initialization.
		(a) Median wall-clock MCEM fitting time by dimension and sample size;
		shaded regions show interquartile ranges across the simulation replicates
		and zero-prevalence configurations.
		(b) Median number of outer MCEM iterations, with interquartile ranges.
		(c) Distribution across 20 independently generated datasets of the range
		across four starting values in expected log-score difference, for $n=200$
		under Cases~1 and 4. Boxes show interquartile ranges, horizontal lines show
		medians, whiskers extend to 1.5 times the interquartile range, and points
		represent individual datasets.
	}
	\label{fig:study3}
\end{figure}

Dimension had the clearest effect on fitting time. At $p=20$,
median fitting times were about 14--22 minutes across sample sizes. Larger samples generally required fewer outer MCEM iterations, and runtime did not increase uniformly with sample size (Figure~\ref{fig:study3}, panels a and b). Web Appendix~J.1 shows fitting times and iteration counts by zero-prevalence case.

Score and covariance calculation added substantial time to the pooled null fit. At $p=10$ and $n=500$ per group, the median was 8.3 minutes for score and covariance calculation and 14.4 minutes for fitting and score calculation combined. Web Appendix~J.1 shows how the additional score calculation time varied with dimension, group size, and zero-prevalence case.

Fitted distributions were typically similar across starts in three of the four settings. The clearest exception was Case~1 at $p=10$, where the median across-start range in expected log-score difference was 0.0043 (Figure~\ref{fig:study3}, panel c). Comparing multiple starts is especially useful when the attained log likelihoods differ materially. Web Appendix~J.2 gives the initialization details and shows how attained likelihoods, log scores, and parameter estimates varied across starts.

\section{Application to an Avocado Dietary Intervention Study}
\label{sec:real data}

We applied the RRFB framework to fecal microbiota data from a parallel, randomized, controlled avocado dietary intervention study (NCT02740439). Study design, intervention procedures, and sample collection have been described previously \citep{thompson2021avocado}. Fecal microbiota data were processed and taxonomically annotated as described in \citet{shinn2021fecal}, and the resulting relative abundances were obtained from the study investigators. Although meals were matched on total energy and macronutrient composition, the avocado arm introduced additional avocado-derived monounsaturated fatty acids and fiber. We analyzed the supplied cohort regardless of reported dietary adherence.

In Study~1, parameter estimates generally had smaller relative bias and variability at lower dimensions. In Study~2, the largest departures from the nominal type~I error rate occurred at the higher dimension and smallest group size. These findings motivated us to reduce the dimension of the avocado data while retaining genera of scientific interest. We retained the ten genera from the HACK-top-17-taxa \citep{goel2025toward} that could be matched to the available taxonomic annotations and summed the relative abundances of all remaining taxa into an ``Other'' category, yielding an eleven-component composition. Web Appendix~K.1 describes the taxonomic mapping and reports the observed nonzero prevalence of each named genus.

We used the common-$A$ RRFB score test to assess equality of the directional parameters $\gamma$ between the control arm ($n=66$) and the avocado arm ($n=67$) at the end of the intervention. The observed score statistic was 21.575, with an asymptotic $p$-value of 0.028 and a permutation $p$-value of 0.040. The permutation reference distribution was based on 999 shuffles of treatment labels across the pooled endpoint sample, with arm sizes held fixed. On the same endpoint compositions, PERMANOVA gave permutation $p$-values of 0.049 using Euclidean distance between square-root-transformed compositions and 0.111 using Bray--Curtis dissimilarity on relative abundances.

\begin{figure}[htbp]
	\centering
	\includegraphics[width=\linewidth]{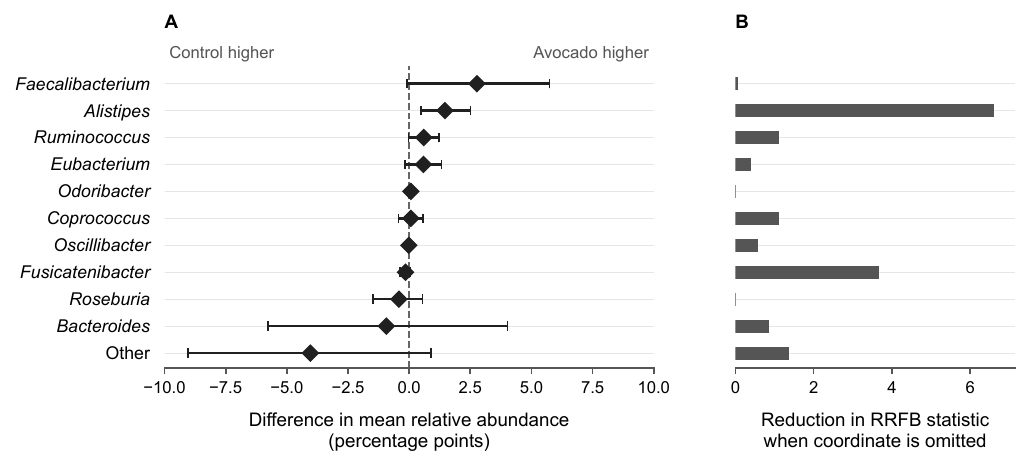}
	\caption{End-of-intervention comparison of the avocado and control arms. (A) Diamonds show avocado-minus-control differences in mean relative abundance in percentage points. Horizontal bars show pointwise 95\% percentile bootstrap confidence intervals from 20,000 resamples of participants within arms. (B) Bar lengths show the reduction in the global RRFB statistic when the corresponding nuisance-adjusted score coordinate is omitted. All reductions use the full eleven-component null fit.}
	\label{fig:avo_figure}
\end{figure}

Figure~\ref{fig:avo_figure} pairs observed mean-abundance contrasts with an exploratory diagnostic of the joint RRFB score. \textit{Alistipes} was detected more often in the avocado arm and had higher mean relative abundance when detected. Omitting its score coordinate produced the largest reduction in the RRFB statistic. The direction of the abundance contrast agrees with the original analysis of this trial \citep{thompson2021avocado}. A subsequent crossover feeding study reported higher abundance of \textit{Alistipes communis} after the avocado diet than after the average American diet \citep{sanabria2026persea}.

The score diagnostic also drew attention to \textit{Fusicatenibacter}, whose marginal mean-abundance contrast was small. Among endpoint samples in which \textit{Fusicatenibacter} was detected, its mean relative abundance was lower in the avocado arm at or below the pooled endpoint median of \textit{Faecalibacterium} abundance and was similar between arms above that median. The diagnostic thus guided an exploratory comparison that revealed a between-arm pattern obscured in the marginal mean contrast. Web Appendix~K.2 gives the score breakdown and descriptive comparisons. In a separate post hoc sensitivity analysis, evidence for a directional difference was stronger when \textit{Alistipes} and \textit{Coprococcus} were retained as separate components at the 85\% prevalence threshold than when both were pooled into ``Other'' at the 90\% threshold (Web Appendix~K.3). For planned randomized comparisons, we recommend specifying the endpoint and compositional representation in the analysis plan and reporting the global RRFB test alongside interpretable between-arm contrasts with confidence intervals.

\section{Discussion}
\label{sec:disc}

In this paper, we introduced the RRFB framework for likelihood-based analysis of compositional data with zeros. By assigning probability to the interior and boundary faces of the simplex, the model describes both which components are zero and how the positive relative abundances vary together. Observed zeros thus enter the likelihood as recorded, without ad hoc replacement by positive values. The two-group score test developed in Section~\ref{sec:score} illustrates how this framework can be used to test a structured hypothesis about differences between compositional distributions, complementary to distance-based comparisons such as PERMANOVA. Furthermore, the application to the avocado dietary intervention study illustrates how an exploratory diagnostic of the joint score can flag taxa for follow-up even when their marginal mean-abundance contrasts are small.

RRFB is most useful when exact zeros are common enough that zero replacement would materially affect the analysis, when the composition has a moderate number of components after aggregation, and when a single model describing how zero patterns and positive relative abundances vary together is of interest. When an omnibus comparison that is robust to many types of difference is the goal, distance-based tests such as PERMANOVA remain appropriate. Log-ratio methods are well suited when zeros are rare, and Dirichlet-multinomial or zero-inflated count models are preferable when read counts and sequencing depth are available and the sampling process should be modeled directly, which RRFB does not do.

Several improvements and extensions are worth exploring. First, the score test studied here targets differences in $\gamma$ under a common quadratic parameter $A$. Performance under other data-generating mechanisms, such as the zero-inflated Dirichlet-multinomial or logistic-normal models with structural zeros, remains to be studied. In our sensitivity analysis, departures from a common $A$ could lead to rejection even when $\gamma$ was equal across groups. Extending the framework to test additional parameters would enable a more complete characterization of compositional changes. For example, one could test for between-group differences in a taxon's mean relative abundance or in its probability of being detected, using contrasts defined under the fitted RRFB model. Second, computational cost increased with dimension in our experiments, with both MCEM fitting and calculation of the score statistic contributing to the runtime. We evaluated model fitting through $p=20$ and score-test calibration and power at $p=5$ and $p=10$. Taxonomic aggregation can reduce the number of components, and the avocado sensitivity analysis illustrates how the choice of taxa retained can affect the results. Faster likelihood calculations and extensions to regression and hierarchical settings are natural directions for future work.

\clearpage
\section*{Supplementary Materials}
\addcontentsline{toc}{part}{Supplementary Materials}
\label{sec:supplementary-materials}

\begingroup
  \etocsetnexttocdepth{subsection}
  \etocsettocstyle{\subsection*{Contents}\small}{\par}
  \localtableofcontents
\endgroup

\setcounter{equation}{0}
\setcounter{figure}{0}
\setcounter{table}{0}
\setcounter{algorithm}{0}
\renewcommand{\theHequation}{web.\arabic{equation}}
\renewcommand{\theHfigure}{web.\arabic{figure}}
\renewcommand{\theHtable}{web.\arabic{table}}
\providecommand{\theHalgorithm}{}
\renewcommand{\theHalgorithm}{web.\arabic{algorithm}}
\captionsetup[figure]{name=Web Figure}
\captionsetup[table]{name=Web Table}

%
%
%
%
%
%
%
%
%
%
%
%

\webappendix{A}{Face-wise Preimages and Posterior Density Representation}
\label{supp:postdens}

We provide the geometric details underlying the face-wise observed-data density in the main text. Recall that
\[
\mathcal{E} = \{z \in \mathcal{S}^{p-1}: \|z^+\|_2 > 0\}
\]
is the valid latent region and that
\[
T(z) = \frac{z^+}{\|z^+\|_2}, \qquad z \in \mathcal{E},
\]
is the rectified-renormalized transformation. For a nonempty index set $M \subseteq \{1, \ldots, p\}$, define the observed face
\[
\mathcal{S}_M = \left\{y \in \mathcal{S}^{p-1}_{\geq 0}: y_j > 0 \text{ for } j\in M, \; y_j=0 \text{ for } j\notin M\right\}.
\]
Thus, $M$ is the positive coordinate set and $m = p-|M|$ is the number of zero components. For $y \in \mathcal{S}_M$, its latent preimage is
\[
\mathcal{Z}_M(y) = \{z \in \mathcal{E}: T(z) = y\}.
\]
Let $\sigma$ denote the surface-area measure on $\mathcal{S}^{p-1}$ and let $\sigma_M$ denote the natural $(|M|-1)$-dimensional surface measure on $\mathcal{S}_M$. For any nonnegative measurable function $h: \mathcal{E} \rightarrow \mathbb{R}$, the induced measures $\nu_{M, y}$ on the preimages $\mathcal{Z}_M(y)$ are characterized by
\begin{equation}
	\label{supp:eq:disintegration}
	\int_{\mathcal{E}} h(z)\,d\sigma(z) = \sum_{\emptyset \neq M \subseteq \{1, \ldots, p\}} \int_{\mathcal{S}_M}\left\{\int_{\mathcal{Z}_M(y)} h(z)\,d\nu_{M,y}(z)\right\} d\sigma_M(y).
\end{equation}
Equation \eqref{supp:eq:disintegration} is the coarea, or disintegration, formula specialized to the many-to-one transformation $T$, which performs rectification followed by renormalization. It decomposes integration over the valid latent region into three steps: selecting an observed face $M$, integrating over possible observed locations $y$ on that face, and integrating over all latent directions $z$ that map to the same $y$. The inner measure $\nu_{M, y}$ accounts for the geometry of this latent preimage. We next derive the exact expression for $\nu_{M, y}$ used in the MCEM calculations.

Let $y \in \mathcal{S}_M$ have $m = p-|M|$ zero components. Suppose first that $m \geq 1$. Conditional on the observed zero pattern and without loss of generality after permuting coordinates, assume that the first $p-m$ components of $y$ are positive and the remaining $m$ components are zero. Then every $z \in \mathcal{Z}_M(y)$ can be written as
\[
z(\delta, u) = \begin{bmatrix}
	\sqrt{\delta}\, y_{1:p-m}\\
	\sqrt{1 - \delta}\, u
\end{bmatrix},
\qquad \delta \in (0, 1], \ u \in \mathcal{S}^{m-1}_{\leq 0},
\]
where $\delta = \|z^+\|_2^2$ and $\mathcal{S}^{m-1}_{\leq 0} = \mathcal{S}^{m-1} \cap \mathbb{R}^m_{\leq 0}$. The reparameterization is one-to-one away from the measure-zero endpoint $\delta=1$; at that endpoint, all values of $u$ represent the same latent direction $z=y$. The endpoint $\delta=0$ is excluded by the valid-region restriction. 

Under the uniform probability measure on $\mathcal S^{p-1}$, the squared norms of the two coordinate blocks have a Dirichlet distribution with parameters $(p-m)/2$ and $m/2$ \citep{fang2018symmetric}. Thus $\delta\sim\operatorname{Beta}((p-m)/2,m/2)$, independently of the two block directions, which are themselves independent and uniform on their respective spheres. Let $s_d=2\pi^{d/2}/\Gamma(d/2)$ be the surface area of $\mathcal S^{d-1}$. Restricting the block directions to the corresponding orthants and converting the probability measures back to surface-area measures gives
\[
d\sigma(z)=\frac{s_p}{s_{p-m}s_m\,\operatorname{B}((p-m)/2,m/2)}
\delta^{\frac{p-m}{2}-1}(1-\delta)^{\frac{m}{2}-1}\,d\delta\,d\sigma_M(y)\,d\sigma_{m-1}(u).
\]
The prefactor equals $1/2$. Hence, by \eqref{supp:eq:disintegration}, the induced preimage measure is
\[
d\nu_{M,y}(\delta,u)=\frac12\delta^{\frac{p-m}{2}-1}(1-\delta)^{\frac{m}{2}-1}\,d\delta\,d\sigma_{m-1}(u).
\]
Let
\[
\mathcal{C}_{\mathcal E}(A,\gamma) = \int_{\mathcal E} \exp\{-z^\top A z+\gamma^\top z\}\,d\sigma(z)
\]
denote the restricted Fisher-Bingham normalizing constant. Combining the conditioned Fisher-Bingham density with the induced preimage measure gives the face-wise observed-data density
\[
f_{\mathrm{RRFB}}(y\mid A,\gamma) = \frac{1}{\mathcal{C}_{\mathcal E}(A,\gamma)} \int_{\mathcal Z_M(y)} \exp\{-z^\top A z+\gamma^\top z\}\,d\nu_{M,y}(z).
\]
When conditioning the latent direction on a fixed observation $y$, the factor $\mathcal{C}_{\mathcal E}(A,\gamma)^{-1}$ is constant over the preimage $\mathcal Z_M(y)$ and cancels upon normalization. Therefore, for $m\geq 1$, the posterior density of $(\delta,u)$ with respect to $d\delta\,d\sigma_{m-1}(u)$ is
\begin{equation}
	\label{supp:eq-posterior}
	p(\delta, u \mid y, A, \gamma) \propto \exp\!\left\{
	-z(\delta, u)^\top A z(\delta, u) + \gamma^\top z(\delta, u)
	\right\}
	\delta^{\frac{p-m}{2}-1}(1-\delta)^{\frac{m}{2}-1}.
\end{equation}
When $m=0$, the observation lies in the strictly positive face, rectification is inactive, and $\mathcal{Z}_M(y) = \{y\}$. Hence, the latent direction is known exactly, with
\[
\E[z \mid y] = y, \qquad \E[zz^\top\mid y]=yy^\top.
\]
Additionally, the induced preimage measure $\nu_{M,y}$ is the unit point mass at $y$.

For $m \geq 1$, conditional on the observed face location $y$, the base preimage measure for $(\delta, u)$ factorizes into a Beta-type measure for the block norm $\delta$ and a uniform surface measure for the direction $u \in \mathcal{S}^{m-1}_{\leq 0}$. The posterior density in \eqref{supp:eq-posterior} can therefore be viewed as an exponential tilt of this product measure by the Fisher-Bingham kernel. This factorization also provides a natural proposal distribution for importance sampling: $\delta$ is sampled from
\[
\mathrm{Beta}\left(\frac{p-m}{2}, \frac{m}{2}\right),
\]
and $u$ is sampled uniformly from $\mathcal{S}^{m-1}_{\leq 0}$ with importance weights proportional to
\[
\exp\!\left\{-z(\delta, u)^\top A z(\delta, u) + \gamma^\top z(\delta, u) \right\}.
\]

\webappendix{B}{Identifiability of the RRFB Parameters}

We establish that the RRFB observed distribution identifies $\gamma$ uniquely and identifies $A$ up to the standard Fisher-Bingham invariance under addition of a scalar multiple of the identity. Recall from Section~3 in the main text that the dominating measure on the observed sample space is the stratified measure
\[
\mu(dy) = \sum_{\emptyset \neq M \subseteq \{1, \ldots, p\}} \sigma_M(dy),
\]
with Web Appendix~A providing details of the geometry of the corresponding faces and their latent preimages.
Also define
\[
\mathcal{C}_{\mathcal{E}}(A, \gamma) = \int_{\mathcal{E}} \exp(-z^\top A z + \gamma^\top z) \,d\sigma(z).
\]

\begin{proof}[Proof of Theorem~1]
	We first prove the forward implication. Let
	\[
	\mathcal{I} = \left\{y \in \mathcal{S}^{p-1}: y_j > 0, j = 1, \ldots, p\right\}
	\]
	denote the relative interior of the positive spherical orthant. This is the face corresponding to $M = \{1, \ldots, p\}$, and its associated component of $\mu$ is the ordinary $(p-1)$-dimensional surface measure. Because the two RRFB densities are equal $\mu$-almost everywhere, their restrictions to $\mathcal{I}$ are equal for surface-almost every $y \in \mathcal{I}$. For any $y \in \mathcal{I}$, the preimage of $y$ under $T$ is the singleton $\{y\}$. It follows that the face-wise RRFB density on $\mathcal{I}$ is
	\[
	f_{\mathrm{RRFB}}(y \mid A, \gamma) = \frac{\exp(-y^\top A y + \gamma^\top y)}{\mathcal{C}_{\mathcal{E}}(A, \gamma)}, \qquad f_{\mathrm{RRFB}}(y \mid A', \gamma') = \frac{\exp(-y^\top A' y + \gamma'^\top y)}{\mathcal{C}_{\mathcal{E}}(A', \gamma')}.
	\]
	Therefore,
	\[
	\frac{\exp(-y^\top A y + \gamma^\top y)}{\mathcal{C}_{\mathcal{E}}(A, \gamma)} = \frac{\exp(-y^\top A' y + \gamma'^\top y)}{\mathcal{C}_{\mathcal{E}}(A', \gamma')}
	\]
	for surface-almost every $y \in \mathcal{I}$. Taking logarithms and rearranging yields
	\[
	-y^\top(A - A')y + (\gamma - \gamma')^\top y = \log \mathcal{C}_{\mathcal{E}}(A, \gamma) - \log \mathcal{C}_{\mathcal{E}}(A', \gamma')
	\]
	for surface-almost every $y \in \mathcal{I}$. Define
	\[
	B = A - A', \quad v = \gamma - \gamma', \quad \kappa = \log \mathcal{C}_{\mathcal{E}}(A, \gamma) - \log \mathcal{C}_{\mathcal{E}}(A', \gamma').
	\]
	Then $-y^\top B y + v^\top y = \kappa$ for surface-almost every $y \in \mathcal{I}$. Note that the function
	\[
	h(y) = -y^\top B y + v^\top y - \kappa
	\]
	is continuous and vanishes surface-almost everywhere on the relatively open set $\mathcal{I}$. Hence, it vanishes everywhere on $\mathcal{I}$. Indeed if $h(y_0) \neq 0$ for some $y_0 \in \mathcal{I}$, continuity would imply that $h$ is nonzero on a relatively open neighborhood of $y_0$. Such a neighborhood has positive surface measure, contradicting the almost-everywhere equality. Moreover, $h$ is real analytic on $\mathcal{S}^{p-1}$, since it is the restriction of a polynomial on $\mathbb{R}^p$. The sphere is connected for $p \geq 2$. The identity theorem for real-analytic functions therefore implies that $h$ vanishes on all of $\mathcal{S}^{p-1}$, i.e. $h(y) = 0, \forall y\in \mathcal{S}^{p-1}$. Hence,
	$-y^\top B y + v^\top y = \kappa$ for every $y \in \mathcal{S}^{p-1}$. Evaluating the equality at $-y$ gives $-y^\top B y - v^\top y = \kappa$. Subtracting the equality at $-y$ from the equality at $y$ gives $2v^\top y = 0$ for every $y \in \mathcal{S}^{p-1}$. Therefore $v = 0$, which implies $\gamma = \gamma'$. Since $B$ is symmetric, let $u_j$ be a unit eigenvector of $B$ with corresponding eigenvalue $\beta_j$. Letting $y = u_j$ gives
	\[
	u_j^\top B u_j = \beta_j = -\kappa.
	\]
	Thus, every eigenvalue of $B$ equals $-\kappa$ and consequently $B = -\kappa I_p$. Therefore, $A' = A + \kappa I_p$. Hence, we have shown the forward implication.
	
	Conversely, suppose that $\gamma' = \gamma$ and $A' = A + cI_p$. For every $z \in \mathcal{S}^{p-1}$,
	\[
	-z^\top A' z + \gamma'^\top z = -z^\top A z + \gamma^\top z - c,
	\]
	which implies
	\[
	\mathcal{C}_{\mathcal{E}}(A', \gamma') = e^{-c} \mathcal{C}_{\mathcal{E}}(A, \gamma).
	\]
	Furthermore, the factor $e^{-c}$ cancels between the kernel and the normalizing constant. Hence
	\[
	f_{\mathrm{FB} \mid \mathcal{E}}(z \mid A', \gamma') = f_{\mathrm{FB} \mid \mathcal{E}}(z \mid A, \gamma).
	\]
	Applying the same deterministic transformation $T$ yields identical observed RRFB distributions, so the corresponding densities are equal $\mu$-almost everywhere. This completes the proof.
\end{proof}

\webappendix{C}{Mathematical Details of the RRFB Estimation}

\websubsection{Web Appendix C.1: MCEM \texorpdfstring{$Q$}{Q}-function}

Using notations and definitions from Section~3.1 and Section~3.2 from the main text, let $z_1, \ldots, z_n$ denote latent variables distributed according to the Fisher-Bingham distribution conditioned on the valid region $\mathcal{E}$, with corresponding observations $y_1, \ldots, y_n$ obtained through the rectified and renormalized transformation $T$. The latent density conditional on $\mathcal E$, with respect to spherical surface measure, is
\[
f_{\mathrm{FB}\mid\mathcal E}(z\mid A,\gamma)
=
P_{A, \gamma}(\mathcal {E})^{-1} \mathcal{C}(A, \gamma)^{-1}\exp(-z^\top A z + \gamma^\top z) \mathbf{1}\{z \in \mathcal E\}.
\]
Define the posterior moments
\begin{equation}
	\label{supp:post_moms}
	M_{1i}^{(t)}=\E[z_i \mid y_i, A^{(t)},\gamma^{(t)}], \qquad M_{2i}^{(t)}=\E[z_i z_i^\top \mid y_i, A^{(t)},\gamma^{(t)}].
\end{equation}
Let $\ell_c$ denote the complete-data log-likelihood. The $Q$-function is therefore
\begin{align*}
	\mathcal Q(A,\gamma \mid A^{(t)},\gamma^{(t)})
	&=
	\E\!\left[\ell_c(A,\gamma; z_1,\dots,z_n)\mid y_1, \ldots, y_n,A^{(t)},\gamma^{(t)}\right] \\
	&=
	-n\left[\log \mathcal C\left(A,\gamma\right) + \log P_{A, \gamma}\left(\mathcal E\right)\right]
	-\sum_{i=1}^n \tr\!\left(A M_{2i}^{(t)}\right)
	+\sum_{i=1}^n \gamma^\top M_{1i}^{(t)} \\
	&=
	-n\left[
	\log \mathcal C_{\mathcal E}\left(A,\gamma\right) + \tr\!\left(A \overline{S}_2^{(t)}\right)
	-\gamma^\top \overline{S}_1^{(t)}
	\right],
\end{align*}
where
\[
\mathcal{C}_{\mathcal E}(A, \gamma) = \mathcal{C}(A, \gamma)P_{A, \gamma}(\mathcal E), \qquad
\overline{S}_1^{(t)}=\frac{1}{n}\sum_{i=1}^n M_{1i}^{(t)},
\qquad
\overline{S}_2^{(t)}=\frac{1}{n}\sum_{i=1}^n M_{2i}^{(t)}.
\]
Thus, the M-step has the same sufficient-statistic structure as the Fisher-Bingham maximum likelihood estimation, but with the full-sphere normalizing constant replaced by the restricted normalizing constant $\mathcal{C}_{\mathcal E}(A, \gamma)$, which we can also write as
\[
\mathcal{C}_{\mathcal E}(A, \gamma) = \mathcal{C}(A, \gamma) - \mathcal{C}_{\mathcal E^c}(A, \gamma),
\]
where 
\[
\mathcal{C}_{\mathcal{E}^c}(A, \gamma) = \int_{\mathcal{E}^c} \exp(-z^\top A z + \gamma^\top z) \, d\sigma(z).
\]
Note that this integral can be evaluated using a Dirichlet representation of the uniform distribution on a spherical orthant. Specifically, let
\[
W = (W_1, \ldots, W_p)^\top \sim \operatorname{Dirichlet}\left(\tfrac12, \ldots, \tfrac12\right), \qquad U = -\sqrt{W},
\]
where the square root is applied componentwise. Then $U$ is uniformly distributed on $\mathcal{E}^c$ with respect to normalized surface measure. Since the surface area of each of the $2^p$ spherical orthants is $2^{-(p-1)}\pi^{p/2}\,\Gamma(p/2)^{-1}$, it follows that
\begin{equation}
	\label{supp:eq:restricted_nc}
	\mathcal{C}_{\mathcal{E}^c}(A, \gamma) = \frac{\pi^{p/2}}{2^{p-1}\Gamma(p/2)} \E\left[\exp\!\left\{-U^\top A U + \gamma^\top U\right\}\right].
\end{equation}

\websubsection{Web Appendix C.2: Posterior Moments Under the Block-Norm Parameterization}

Under the block-norm parameterization defined in Section~3.1 of the main text, the posterior moments $M_{1i}^{(t)}$ and $M_{2i}^{(t)}$ defined in \eqref{supp:post_moms} can be expressed as expectations of functions of $(\delta_i, u_i)$. The form of the posterior moments depends on the number of zero components $m_i$. If $m_i = 0$, then $z_i = y_i$ deterministically and
\[
M_{1i}^{(t)} = y_i, \qquad
M_{2i}^{(t)} = y_i y_i^\top.
\]
For boundary observations ($m_i \geq 1$), we introduce the following auxiliary posterior expectations. Throughout this subsection, expectations are taken with respect to the posterior distribution of $(\delta_i,u_i)$ given $(y_i, A^{(t)}, \gamma^{(t)})$, with the iteration index suppressed in the auxiliary notation:
\begin{gather*}
	\eta_i^{(1)}=\E[\sqrt{\delta_i}], \qquad
	\eta_i^{(2)}=\E[\delta_i], \qquad
	\eta_i^{(3)}=\E[\sqrt{1-\delta_i}], \qquad
	\eta_i^{(4)}=\E[\sqrt{\delta_i(1-\delta_i)}], \\
	\nu_i^{(1)}=\E[\sqrt{1-\delta_i}\,u_i], \qquad
	\nu_i^{(2)}=\E[\sqrt{\delta_i(1-\delta_i)}\,u_i], \qquad
	\nu_i^{(3)}=\E[(1-\delta_i)u_i u_i^\top].
\end{gather*}
Here $\eta_i^{(1)}, \ldots, \eta_i^{(4)}$ are scalars, $\nu_i^{(1)}$ and $\nu_i^{(2)}$ are $m_i$-dimensional vectors, and $\nu_i^{(3)}$ is an $m_i \times m_i$ matrix. If $m_i = 1$, the negative block is one-dimensional and $u_i = -1$, yielding
\[
M_{1i}^{(t)} = P_i^\top
\begin{bmatrix}
	\eta^{(1)}_i \tilde{y}_i \\
	-\eta^{(3)}_i
\end{bmatrix}, \qquad
M_{2i}^{(t)} = P_i^\top
\begin{bmatrix}
	\eta^{(2)}_i \tilde{y}_i \tilde{y}_i^\top & -\eta^{(4)}_i \tilde{y}_i \\
	-\eta^{(4)}_i \tilde{y}_i^\top & 1 - \eta^{(2)}_i
\end{bmatrix}
P_i.
\]
For $m_i \geq 2$,
\[
M_{1i}^{(t)} = P_i^\top
\begin{bmatrix}
	\eta^{(1)}_i \tilde{y}_i \\
	\nu_i^{(1)}
\end{bmatrix},
\qquad
M_{2i}^{(t)} = P_i^\top \begin{bmatrix}
	\eta^{(2)}_i \tilde{y}_i \tilde{y}_i^\top & \tilde{y}_i (\nu^{(2)}_i)^\top \\
	\nu_i^{(2)} \tilde{y}_i^\top & \nu^{(3)}_i
\end{bmatrix} P_i,
\]
where $P_i$ denotes the permutation matrix that reorders the coordinates so that the nonzero components of $y_i$ appear first, and $\tilde{y}_i = (P_iy_i)_{1:p-m_i} \in \mathbb{R}^{p-m_i}$ denotes the subvector of strictly positive components of $y_i$.

\webappendix{D}{Numerical Implementation of the RRFB Estimation}

\websubsection{Web Appendix D.1: E-step}
\label{supp:sec:Estep}

The E-step requires posterior expectations with respect to the conditional distribution of the latent vector $z_i$ given the square-root composition $y_i$. The global conditioning factor $P_{A,\gamma}(\mathcal E)$ is
constant in $z_i$ for fixed $(A,\gamma)$ and therefore cancels from this conditional posterior distribution. Thus, the numerical evaluation of the E-step is based on the face-specific posterior density in \eqref{supp:eq-posterior}. Since the dimension of the preimage depends on the number of zero components $m_i$, the required expectations are evaluated separately for each case.

\paragraph*{Case $m_i=0$}: Rectification is inactive, and the latent vector coincides with the observed vector. Therefore
\[
\E[z_i \mid y_i] = y_i, \qquad \E[z_iz_i^\top \mid y_i] = y_iy_i^\top.
\]

\paragraph*{Case $m_i=1$}: When exactly one component of $y_i$ is zero, the negative block is one-dimensional and  $u_i = -1$ deterministically. The E-step therefore reduces to computing one-dimensional integrals over $\delta_i$ of the form
\[
\mathbb{E}[g(z_i)\mid y_i] = \frac{\int_0^1 g(z_i(\delta))\, w(\delta)\, d\delta}{\int_0^1 w(\delta)\, d\delta},
\]
where
\[
w(\delta) = \delta^{\frac{p-1}{2}-1}(1-\delta)^{-1/2}
\exp\!\left(- z_i(\delta)^\top A z_i(\delta) + \gamma^\top z_i(\delta) \right).
\]
We evaluate these integrals using Gauss-Jacobi quadrature \citep{gautschi2004orthogonal}, which is well suited to this setting because it explicitly accommodates the algebraic endpoint weight $\delta^{\frac{p-1}{2}-1}(1-\delta)^{-1/2}$.

\paragraph*{Case $m_i\geq 2$}: The E-step requires integration over
\[
(\delta_i,u_i) \in (0,1] \times \mathcal{S}_{\leq 0}^{m_i-1}.
\]
Closed-form expressions are not available, so we approximate the required posterior expectations using self-normalized importance sampling based on the face-specific preimage measure described in Web Appendix~A. Ignoring the exponential tilt induced by $(A,\gamma)$, this base measure factorizes into a Beta-type measure for $\delta_i$ and a uniform surface measure on $\mathcal{S}_{\leq 0}^{m_i - 1}$. Accordingly, for each observation $i$, we use $B$ iid proposal pairs $(\delta_{ib}, u_{ib})$, $b = 1,\ldots,B$ with independent components distributed as
\[
\delta_{ib} \sim \mathrm{Beta}\!\left(\frac{p-m_i}{2},\frac{m_i}{2}\right),
\qquad
u_{ib} \sim \mathrm{Uniform}(\mathcal{S}_{\leq 0}^{m_i-1}),
\]
and importance weights are computed using the exponential tilt induced by the Fisher-Bingham kernel,
\[
\omega_{ib} \propto \exp\!\left\{-z_i(\delta_{ib}, u_{ib})^\top A z_i(\delta_{ib}, u_{ib}) + \gamma^\top z_i(\delta_{ib}, u_{ib})\right\}.
\]
Posterior expectations are then approximated by self-normalized importance sampling,
\[
\E\left[g(z_i) \mid y_i\right] \approx \frac{\sum_{b=1}^B \omega_{ib} g(z_i(\delta_{ib}, u_{ib}))}{\sum_{b=1}^B \omega_{ib}}.
\]

\websubsection{Web Appendix D.2: Joint M-step}

At iteration $t$, the E-step provides the posterior sufficient-statistic averages
\[
\overline S_1^{(t)} = \frac{1}{n}\sum_{i=1}^n \E[z_i\mid y_i,A^{(t)},\gamma^{(t)}],
\qquad
\overline S_2^{(t)} = \frac{1}{n}\sum_{i=1}^n \E[z_i z_i^\top\mid y_i,A^{(t)},\gamma^{(t)}],
\]
which are held fixed throughout the subsequent M-step. We write $A=Q\diag(\lambda)Q^\top$ so that existing numerical methods for the full-sphere Fisher-Bingham normalizing constant can be applied in the eigenbasis of $A$. The M-step jointly updates $(\lambda,\gamma,Q)$ by approximately solving
\[
\max_{\lambda,\gamma,Q}
\left\{
-n\left[
\log\widehat{\mathcal C}_{\mathcal E}^{(N_1,N_2,h)}(\lambda,\gamma,Q)
+\tr\!\left(Q\diag(\lambda)Q^\top\overline S_2^{(t)}\right)
-\gamma^\top\overline S_1^{(t)}
\right]
\right\},
\]
subject to $0=\lambda_1\leq\cdots\leq\lambda_p$ and $Q^\top Q=I_p$. By rotational invariance of the full-sphere surface measure, the restricted normalizing constant can be expressed as
\[
\mathcal C_{\mathcal E}\!\left(Q\diag(\lambda)Q^\top,\gamma\right) = \mathscr C(\lambda,Q^\top\gamma) - \mathcal C_{\mathcal E^c}\!\left(Q\diag(\lambda)Q^\top,\gamma\right).
\]
For the full-sphere contribution, we use the numerical integration method of \citet[Equation~(3)]{chen2021mle}, denoting its approximation by
\[
\mathcal C_\omega^{(N_1,h)}(\lambda,Q^\top\gamma) \approx \mathscr C(\lambda,Q^\top\gamma),
\]
where $N_1$ is the truncation parameter of the numerical integration sum and $h$ is its mesh spacing. We approximate the excluded-orthant contribution using the representation in Equation~\eqref{supp:eq:restricted_nc}. Specifically, let
\[
W_i \stackrel{\mathrm{iid}}{\sim} \operatorname{Dirichlet}\!\left(\tfrac12\mathbf 1_p\right),
\qquad
U_i=-\sqrt{W_i},
\qquad i=1,\ldots,N_2,
\]
where the square root is applied componentwise. The resulting Monte Carlo approximation is
\[
\widehat{\mathcal C}_{\mathcal E^c}^{(N_2)}(\lambda,\gamma,Q) = \frac{\pi^{p/2}}{2^{p-1}\Gamma(p/2)}\frac{1}{N_2}\sum_{i=1}^{N_2}\exp\!\left\{-U_i^\top Q\diag(\lambda)Q^\top U_i+\gamma^\top U_i\right\}.
\]
Combining the two approximations gives
\begin{equation}
	\label{supp:eq:restrict_nc_approx}
	\widehat{\mathcal C}_{\mathcal E}^{(N_1,N_2,h)}(\lambda,\gamma,Q) = \mathcal C_\omega^{(N_1,h)}(\lambda,Q^\top\gamma) - \widehat{\mathcal C}_{\mathcal E^c}^{(N_2)}(\lambda,\gamma,Q).
\end{equation}
To enforce the eigenvalue constraints, introduce unconstrained coordinates $v=(v_1,\ldots,v_{p-1})^\top\allowbreak\in\mathbb R^{p-1}$ and set
\[
\lambda_1(v)=0,
\qquad
\lambda_j(v)=\sum_{k=1}^{j-1}\operatorname{softplus}(v_k),
\qquad j=2,\ldots,p,
\]
where $\operatorname{softplus}(a)=\log(1+\exp(a))$. This parameterization fixes the smallest eigenvalue at zero and gives strictly ordered eigenvalues for finite $v$; equal adjacent eigenvalues can be approached as limits. The orthogonality constraint is imposed by treating $Q$ as a point on the square Stiefel manifold
\[
\operatorname{St}(p,p) = \{Q\in\mathbb R^{p\times p}:Q^\top Q=I_p\} = \operatorname O(p).
\]
The joint optimization consequently takes place on the product manifold
\[
\mathcal M = \mathbb R^{p-1}\times\mathbb R^p\times\operatorname{St}(p,p).
\]
Therefore, the M-step approximately maximizes
\begin{equation}
	\label{supp:eq:optimizer}
	\max_{v,\gamma,Q} -\log\widehat{\mathcal C}_{\mathcal E}^{(N_1,N_2,h)}\!\left(\lambda(v),\gamma,Q\right) - \tr\!\left(Q\diag(\lambda(v))Q^\top\overline S_2^{(t)}\right) + \gamma^\top\overline S_1^{(t)}
\end{equation}
jointly over $(v,\gamma,Q)\in\mathcal M$. We use a Riemannian conjugate-gradient method with a line search, implemented in \texttt{pymanopt} \citep{townsend2016pymanopt}. Derivatives of the numerical objective are obtained by automatic differentiation using \texttt{jax} \citep{jax2018github}. Differentiation includes both contributions to the restricted normalizing constant, thereby retaining the parameter-dependent valid-region conditioning factor in the M-step.

\websubsection{Web Appendix D.3: Observed-data Likelihood Evaluation}

The conditioned observed-data log-likelihood is evaluated to assess convergence of the MCEM algorithm and to compute the derivatives required for the score test. For observations $y_1,\ldots,y_n \in \mathcal{S}^{p-1}_{\geq 0}$, let $M_i=\{j:y_{ij}>0\}$ denote the positive coordinate set of observation $i$. The log-likelihood is
\begin{equation}
	\label{supp:eq:observed-loglik}
	\ell(A,\gamma \mid y_1, \ldots, y_n) = -n\log\mathcal C_{\mathcal E}(A,\gamma) + \sum_{i=1}^n \log\left[\int_{\mathcal Z_{M_i}(y_i)} \exp\!\left\{-z^\top A z+\gamma^\top z\right\}\,d\nu_{M_i,y_i}(z)\right],
\end{equation}
where $\mathcal Z_{M_i}(y_i)$ and $\nu_{M_i,y_i}$ are the face-specific preimage and its induced measure described in Web Appendix~A. 

When $m_i=0$, the preimage consists only of $z_i = y_i$, so the integral reduces to $\exp\{-y_i^\top A y_i + \gamma^\top y_i\}$. For observations with $m_i=1$, the preimage integral is evaluated using Gauss-Jacobi quadrature as described in Web Appendix D.1. For observations with $m_i\geq2$ zero components, we approximate the preimage integral by Monte Carlo integration using the block-norm representation from Web Appendix~A. The numerical integration follows the same base-measure construction as the E-step in Web Appendix~D.1, but evaluates the integrated kernel directly rather than a ratio defining a posterior expectation. Specifically, we use $B$ independent proposal pairs $(\delta_{ib},u_{ib})$, $b=1,\ldots,B$, with independent components
\[
\delta_{ib}\sim\operatorname{Beta}\!\left(\frac{p-m_i}{2},\frac{m_i}{2}\right),
\qquad
u_{ib}\sim\operatorname{Uniform}\!\left(\mathcal S_{\leq0}^{m_i-1}\right),
\]
and construct $z_{ib}=z_i(\delta_{ib},u_{ib})$ in the original coordinate order. For $m\geq1$, let $c_{p,m}=\pi^{m/2}\Gamma((p-m)/2)/\{2^m\Gamma(p/2)\}$ denote the total mass of the preimage measure in Web Appendix~A, and set $c_{p,0}=1$. The exact preimage integral is $c_{p,m_i}$ times the expected kernel under the normalized proposal. We approximate this expectation by normalized Gauss-Jacobi quadrature for $m_i=1$ and by the empirical average of the kernel for $m_i\geq2$. The implementation uses the following numerical log-likelihood with the face constants omitted.
\begin{align}
	\label{supp:eq:observed-loglik-numerical}
	\widehat\ell(\lambda,\gamma,Q\mid y_1,\ldots,y_n)
	={}&-n\log\widehat{\mathcal C}_{\mathcal E}^{(N_1,N_2,h)}(\lambda,\gamma,Q) \notag\\
	&+\sum_{i:m_i=0}\left(-y_i^\top A y_i+\gamma^\top y_i\right) \notag\\
	&+\sum_{i:m_i=1}\log\left[
	\sum_{r=1}^{R}a_r
	\exp\!\left\{-z_i(\delta_r,-1)^\top A z_i(\delta_r,-1)+\gamma^\top z_i(\delta_r,-1)\right\}
	\right] \notag\\
	&+\sum_{i:m_i\geq2}\log\left[
	\frac{1}{B}\sum_{b=1}^{B}
	\exp\!\left\{-z_{ib}^\top A z_{ib}+\gamma^\top z_{ib}\right\}
	\right],
\end{align}
where $A=Q\diag(\lambda)Q^\top$, and $\delta_r$ and $a_r$ are the Gauss--Jacobi quadrature nodes and weights for integration with respect to the normalized $\operatorname{Beta}((p-1)/2,1/2)$ distribution. The restricted normalizing constant is approximated as described in Web Appendix~D.2. Adding $\sum_{i=1}^n\log c_{p,m_i}$ to \eqref{supp:eq:observed-loglik-numerical} gives the numerical approximation to the log-likelihood with respect to $\mu$. This offset cancels in likelihood differences evaluated on the same observations and contributes zero to parameter derivatives. The scaled likelihood change in the MCEM stopping rule uses \eqref{supp:eq:observed-loglik-numerical} divided by $n$, because its denominator depends on the level of the log-likelihood.

\websubsection{Web Appendix D.4: Initialization}
We use a neutral-model latent-moment initialization for MCEM estimation. Using the E-step procedures in Web Appendix~D.1, we calculate
\[
\overline S_1 = \frac{1}{n}\sum_{i=1}^n \E[z_i\mid y_i,A=0,\gamma=0],
\qquad
\overline S_2 = \frac{1}{n}\sum_{i=1}^n \E[z_i z_i^\top\mid y_i,A=0,\gamma=0].
\]
We construct preliminary quadratic parameters from $\overline S_2$ using an inverse-moment heuristic. Directions with larger posterior second moments receive smaller quadratic penalties, motivated by the inverse relationship between covariance and precision in a Gaussian model. This construction supplies optimizer starting values rather than an exact RRFB moment estimator. Let
\[
\overline S_2 = V\diag(\xi_1,\ldots,\xi_p)V^\top,
\qquad
\xi_1\geq\cdots\geq\xi_p,
\]
be the eigendecomposition of the posterior second-moment average. To limit excessively large inverse-moment values, we shrink the eigenvalues toward their average and impose a positive lower bound before inversion:
\[
\widetilde\xi_j = \max\left\{(1-\rho)\xi_j+\rho\overline\xi,\,\varepsilon_\xi\right\},
\qquad
\overline\xi = \frac{1}{p}\sum_{j=1}^p\xi_j,
\]
where $\rho\in[0,1]$ is the shrinkage coefficient and $\varepsilon_\xi>0$ is a numerical safeguard. Subtracting the smallest inverse-moment penalty removes the scalar-shift ambiguity in $A$ and enforces the identifying constraint $\lambda_1=0$. We consequently define the preliminary eigenvalues
\[
\lambda_j^{\mathrm{pre}} = \frac12\left(\widetilde\xi_j^{-1}-\widetilde\xi_1^{-1}\right),
\qquad j=1,\ldots,p.
\]
The preliminary eigenvalues are further adjusted to have small positive adjacent gaps, ensuring finite softplus coordinates, and to satisfy a specified upper bound. The numerical settings are given in Web Appendix~G.1. Holding $\overline S_1$ and $\overline S_2$ fixed, we approximately maximize the conditioned $\mathcal Q$-function using the procedure in Web Appendix~D.2, initialized at $Q^{\mathrm{pre}}=V$, $\gamma^{\mathrm{pre}}=0$, and $\lambda^{\mathrm{pre}}$. The resulting parameters define $(A^{(0)},\gamma^{(0)})$ for the subsequent MCEM iterations.

\webappendix{E}{Common-\texorpdfstring{$A$}{A} Score Test Details}

\websubsection{Web Appendix E.1: Observed Likelihood and Scores}
Consider two independent samples of square-root transformed compositional observations $Y^{(j)}=\{y_i^{(j)}\}_{i=1}^{n_j}$, $j=0,1$, arising from RRFB models with a common parameter $A$ and group-specific parameters $\gamma_j$. The two-group log-likelihood is
\[
\ell_2(A, \gamma_0, \gamma_1 \mid Y^{(0)}, Y^{(1)}) = \ell(A, \gamma_0 \mid Y^{(0)}) + \ell(A, \gamma_1 \mid Y^{(1)}) = \sum_{j=0}^1\sum_{i=1}^{n_j}
\log f_{\mathrm{RRFB}}(y_i^{(j)}\mid A,\gamma_j),
\]
where $\ell$ is defined in Equation~\eqref{supp:eq:observed-loglik} and 
\[
\log f_{\mathrm{RRFB}}(y\mid A,\gamma)
=
-\log\mathcal C_{\mathcal E}(A,\gamma)
+
\log\left\{
\int_{\mathcal Z_M(y)}
\exp(-z^\top Az+\gamma^\top z)\,d\nu_{M,y}(z)
\right\}.
\]
As described in Section~4 of the main text, we test
\[
H_0:\gamma_0=\gamma_1 \qquad\text{vs.}\qquad H_1:\gamma_0\neq\gamma_1.
\]
Under $H_0$, the restricted likelihood is the likelihood of a single RRFB model fitted to the pooled observations. The observation-level scores follow by differentiating the conditioned observed log-density. Let $M_{1i}^{(j)}$ and $M_{2i}^{(j)}$ denote the posterior moments defined in Equation~\eqref{supp:post_moms}, evaluated for observation $y_i^{(j)}$ at parameters $(A,\gamma_j)$; here the superscript $j$ indexes groups rather than MCEM iterations. Define the corresponding conditioned latent moments
\[
M_1^{\mathcal E}(A,\gamma_j)=\E_{A,\gamma_j}[Z\mid Z\in\mathcal E],
\qquad
M_2^{\mathcal E}(A,\gamma_j)=\E_{A,\gamma_j}[ZZ^\top\mid Z\in\mathcal E].
\]
Differentiating the preimage integral and restricted normalizing constant gives
\begin{equation}
	\label{supp:eq:gamma-score}
	\nabla_{\gamma_j}\log f_{\mathrm{RRFB}}(y_i^{(j)}\mid A,\gamma_j)
	=
	M_{1i}^{(j)}-M_1^{\mathcal E}(A,\gamma_j).
\end{equation}
With respect to the Frobenius inner product on symmetric matrices, the matrix-valued score contribution for the common parameter $A$ is
\begin{equation}
	\label{supp:eq:A-score}
	\nabla_A\log f_{\mathrm{RRFB}}(y_i^{(j)}\mid A,\gamma_j)
	=
	M_2^{\mathcal E}(A,\gamma_j)-M_{2i}^{(j)}.
\end{equation}
Thus, each score contribution is a difference between a posterior latent moment and its counterpart under the conditioned latent model. The conditioned moments arise from differentiation of $\log\mathcal C_{\mathcal E}$ and therefore retain the parameter-dependent conditioning probability. Let $\xi_A=(\xi_{A,1},\ldots,\xi_{A,q})^\top$ contain locally identifiable coordinates for $A$ after removing its scalar-shift ambiguity. For $B_k=\partial A/\partial\xi_{A,k}$, the chain rule gives
\[
\frac{\partial}{\partial\xi_{A,k}}\log f_{\mathrm{RRFB}}(y_i^{(j)}\mid A,\gamma_j)
=
\operatorname{tr}\left[
B_k\{M_2^{\mathcal E}(A,\gamma_j)-M_{2i}^{(j)}\}
\right].
\]
Since both second moments have trace one, the matrix score is orthogonal to $I_p$, consistent with the invariance $A\mapsto A+cI_p$.

\websubsection{Web Appendix E.2: Reparameterization and Nuisance Adjustment}

To separate the between-group contrast from the common directional parameter under the null, we reparameterize
\[
\psi=\frac{\gamma_1-\gamma_0}{2},
\qquad
\bar\gamma=\frac{\gamma_1+\gamma_0}{2}.
\]
Then $\gamma_j=\bar\gamma+(-1)^{j+1}\psi$, and the null hypothesis becomes $H_0:\psi=0$. Let $\theta$ denote the full parameter vector and $\eta$ collect the nuisance parameters:
\[
\theta=(\psi^\top,\eta^\top)^\top,
\qquad
\eta=(\bar\gamma^\top,\xi_A^\top)^\top,
\]
where $\xi_A$ denotes the locally identifiable coordinates for $A$ introduced in Web Appendix~E.1. Let
\[
g_i^{(j)}
=
M_{1i}^{(j)}-M_1^{\mathcal E}(A,\gamma_j)
\]
denote the directional score in Equation~\eqref{supp:eq:gamma-score}, and let $a_i^{(j)}$ collect the coordinate scores for $A$, with components
\[
a_{ik}^{(j)}
=
\operatorname{tr}\left[
B_k\{M_2^{\mathcal E}(A,\gamma_j)-M_{2i}^{(j)}\}
\right],
\qquad k=1,\ldots,q.
\]
The chain rule gives the observation-level score
\[
s_i^{(j)}(\theta)
=
\nabla_\theta\log f_{\mathrm{RRFB}}(y_i^{(j)}\mid A,\gamma_j)
=
\begin{pmatrix}
	(-1)^{j+1}g_i^{(j)}\\
	g_i^{(j)}\\
	a_i^{(j)}
\end{pmatrix}.
\]
Thus, the directional score enters with opposite signs for the group contrast and with the same sign for the common directional parameter. Let $n=n_0+n_1$ and define the average score
\[
\bar s(\theta)
=
\frac1n\sum_{j=0}^1\sum_{i=1}^{n_j}s_i^{(j)}(\theta)
=
\begin{pmatrix}
	\bar s_\psi(\theta)\\
	\bar s_\eta(\theta)
\end{pmatrix}.
\]
Regard the two-group log-likelihood $\ell_2$ as a function of $\theta$ through the reparameterization above. Write $\widehat\theta_0=(0_p,\widehat\eta_0)$ for the restricted estimator, and define the empirical sensitivity matrix
\[
\widehat H
=
-\frac1n\nabla_\theta^2\ell_2(\theta)\big|_{\theta=\widehat\theta_0}
=
\begin{pmatrix}
	\widehat H_{\psi\psi}&\widehat H_{\psi\eta}\\
	\widehat H_{\eta\psi}&\widehat H_{\eta\eta}
\end{pmatrix}.
\]
To account for estimation of the nuisance parameters, define the nuisance-adjusted average score
\begin{equation}
	\label{supp:eq:eff-score}
	\bar s_{\psi\mid\eta}
	=
	\bar s_\psi-\widehat H_{\psi\eta}\widehat H_{\eta\eta}^{-1}\bar s_\eta
	=
	\widehat{\mathcal A}\bar s,
	\qquad
	\widehat{\mathcal A}
	=
	\begin{pmatrix}
		I_p&-\widehat H_{\psi\eta}\widehat H_{\eta\eta}^{-1}
	\end{pmatrix},
\end{equation}
where all scores are evaluated at $\widehat\theta_0$. At an interior restricted maximum likelihood estimator, $\bar s_\eta=0$, so $\bar s_{\psi\mid\eta}=\bar s_\psi$. The nuisance adjustment nevertheless remains necessary for the covariance calculation. Define the empirical score-variability matrix
\[
\widehat J
=
\frac1n\sum_{j=0}^1\sum_{i=1}^{n_j}
\{s_i^{(j)}(\widehat\theta_0)-\bar s(\widehat\theta_0)\}
\{s_i^{(j)}(\widehat\theta_0)-\bar s(\widehat\theta_0)\}^{\top}.
\]
The sandwich estimator of the limiting covariance of $\sqrt n\,\bar s_{\psi\mid\eta}$ is
\begin{equation}
	\label{supp:eq:score-cov}
	\widehat V_{\psi\mid\eta}
	=
	\widehat{\mathcal A}\widehat J\widehat{\mathcal A}^{\top}.
\end{equation}

\websubsection{Web Appendix E.3: Test Statistic and Calibration}

The score statistic is
\begin{equation}
	\label{supp:eq:robust-score}
	T
	=
	n\,\bar s_{\psi\mid\eta}^{\top}
	\widehat V_{\psi\mid\eta}^{-1}
	\bar s_{\psi\mid\eta}.
\end{equation}
Under $H_0$, for fixed $p$, suppose that $n_j/n\to\rho_j\in(0,1)$, the restricted estimator is consistent for an interior null parameter, and the usual differentiability, moment, and central limit conditions hold. If the limiting nuisance sensitivity is invertible and $V_{\psi\mid\eta}$ is positive definite, then
\[
T\overset{d}{\longrightarrow}\chi_p^2.
\]

Under correct specification, the information-matrix equality gives $J=H$, and
\[
V_{\psi\mid\eta}
=
H_{\psi\psi}
-
H_{\psi\eta}H_{\eta\eta}^{-1}H_{\eta\psi}.
\]
The nuisance-adjusted score is then the usual efficient score, and the sandwich statistic is asymptotically equivalent to the ordinary likelihood score statistic.

The sandwich formulation also permits calibration under a common, possibly misspecified, group distribution when the restricted estimator converges to an interior pseudo-true parameter and the corresponding estimating-equation regularity conditions hold \citep{white1982maximum,newey1994large}. It does not establish calibration under arbitrary group-specific departures from the common-$A$ model.

\websubsection{Web Appendix E.4: Numerical Evaluation}

Numerical evaluation of the score statistic requires local coordinates for $A$, evaluation of the conditioned observed likelihood and its derivatives, and linear systems for the nuisance adjustment and final quadratic form. We retain the spectral representation $A=Q\operatorname{diag}(\lambda)Q^\top$, with $\lambda_1=0$, and use $\lambda_2,\ldots,\lambda_p$ as free coordinates for differentiation. To represent local changes in the fitted eigenvectors $\widehat Q$, we use the matrix-exponential parameterization
\[
Q(\alpha)=\widehat Q\exp\{S(\alpha)\},
\qquad
S(\alpha)\in\mathfrak{so}(p),
\]
where $\mathfrak{so}(p)$ denotes the space of skew-symmetric $p\times p$ matrices and $\alpha\in\mathbb R^{p(p-1)/2}$ collects the independent entries of $S(\alpha)$ \citep{absil2009optimization}. This representation preserves orthogonality and satisfies $Q(0)=\widehat Q$. Together with the free eigenvalues, it gives the local coordinates
\[
\xi_A=(\lambda_2,\ldots,\lambda_p,\alpha^\top)^\top.
\]
These coordinates require distinct fitted eigenvalues. Unlike the softplus coordinates used for M-step optimization in Web Appendix~D.2, the free eigenvalues here describe local perturbations about the restricted estimate. We form the numerical two-group log-likelihood by summing the group-specific contributions from \eqref{supp:eq:observed-loglik-numerical}:
\begin{align*}
	\widehat{\ell}_2(\lambda,\gamma_0,\gamma_1,Q\mid Y^{(0)},Y^{(1)})
	&=\widehat{\ell}(\lambda,\gamma_0,Q\mid Y^{(0)})\\
	&\quad+\widehat{\ell}(\lambda,\gamma_1,Q\mid Y^{(1)}).
\end{align*}
We obtain the observation-level scores by automatic differentiation of the numerical log-density, holding the quadrature nodes and Monte Carlo draws fixed. The derivatives include both the preimage integral and the restricted normalizing constant. Second derivatives of $\widehat{\ell}_2$, scaled as in Web Appendix~E.2, give the sensitivity blocks $\widehat H_{\psi\eta}$ and $\widehat H_{\eta\eta}$. At the null, the directional score for an observation in group $j$ is multiplied by $(-1)^{j+1}$ to obtain its group-contrast score. The empirical variability matrix $\widehat J$ is assembled from the centered observation-level scores as defined in Web Appendix~E.2. The inverse operations in the nuisance adjustment and score statistic are evaluated using direct linear solves, avoiding explicit construction of the inverse matrices. Specifically, we solve
\[
\widehat H_{\eta\eta}^{\top}X
=
\widehat H_{\psi\eta}^{\top},
\qquad
\widehat{\mathcal A}
=
\begin{pmatrix}
	I_p&-X^\top
\end{pmatrix}.
\]
We then calculate $\bar s_{\psi\mid\eta}=\widehat{\mathcal A}\bar s$ and $\widehat V_{\psi\mid\eta}=\widehat{\mathcal A}\widehat J\widehat{\mathcal A}^{\top}$. Finally, we solve
\[
\widehat V_{\psi\mid\eta}b=\bar s_{\psi\mid\eta}
\]
and evaluate $T=n\,\bar s_{\psi\mid\eta}^{\top}b$.

\webappendix{F}{Sampling from the Fisher-Bingham Distribution}

\label{supp:sec:sampling}

Algorithm~\ref{supp:alg:fb-sampling} summarizes the rejection sampling scheme of \citet{kent2018new}, specialized to our Fisher-Bingham parameterization. For $z\in\mathcal S^{p-1}$, let
\[
f^*_{\mathrm{FB}(A,\gamma)}(z)=\exp(-z^\top Az+\gamma^\top z),
\qquad
f^*_{\mathrm{Bing}(A)}(z)=\exp(-z^\top Az)
\]
denote the unnormalized Fisher-Bingham and Bingham density kernels, respectively. Define
\[
A^{(1)}
=
\begin{cases}
	A+\dfrac{\|\gamma\|_2}{2}\left(I_p-\|\gamma\|_2^{-2}\gamma\gamma^\top\right), & \|\gamma\|_2>0,\\[6pt]
	A, & \|\gamma\|_2=0.
\end{cases}
\]
The Fisher-Bingham kernel satisfies the bound
\[
f^*_{\mathrm{FB}(A,\gamma)}(z)
\leq
\exp\{\|\gamma\|_2-z^\top A^{(1)}z\}
\]
given by \citet{kent2018new}. The definition for $\gamma=0$ recovers the Bingham case. Let $\lambda_{\min}$ denote the smallest eigenvalue of $A^{(1)}$, and define
\[
\widetilde A=A^{(1)}-\lambda_{\min}I_p,
\qquad
\Omega(b)=I_p+\frac{2\widetilde A}{b},
\qquad b>0.
\]
Write $0=\widetilde\lambda_1\leq\cdots\leq\widetilde\lambda_p$ for the eigenvalues of $\widetilde A$. Since $z^\top z=1$, combining the Fisher-Bingham bound with Equation~(3.4) of \citet{kent2018new} gives
\begin{align*}
	f^*_{\mathrm{FB}(A,\gamma)}(z)
	&\leq
	\exp(\|\gamma\|_2-\lambda_{\min})
	f^*_{\mathrm{Bing}(\widetilde A)}(z)\\
	&\leq
	\exp\left\{\|\gamma\|_2-\lambda_{\min}-\frac{p-b}{2}\right\}
	\left(\frac{p}{b}\right)^{p/2}
	\{z^\top\Omega(b)z\}^{-p/2},
	\qquad 0<b\leq p.
\end{align*}
The final factor is the kernel of an angular central Gaussian distribution. A draw from this proposal is obtained by sampling $v\sim\mathcal N_p(0,\Omega(b)^{-1})$ and setting $z=v/\|v\|_2$. Following \citet{kent2018new}, we choose $b=b_0$, where $b_0$ is the unique positive solution of
\begin{equation}
	\label{supp:eq:sampling-b0}
	\sum_{i=1}^p\frac{1}{b_0+2\widetilde\lambda_i}=1.
\end{equation}
The left-hand side is strictly decreasing from infinity to zero as $b_0$ increases from zero to infinity. Since $\widetilde\lambda_1=0$ and all shifted eigenvalues are nonnegative,
\[
\frac{1}{b_0}
\leq
\sum_{i=1}^p\frac{1}{b_0+2\widetilde\lambda_i}
=
1
\leq
\frac{p}{b_0},
\]
which gives $1\leq b_0\leq p$. A sharper lower bound follows from the Cauchy-Schwarz inequality:
\[
\frac{k^2}{\sum_{i=1}^k(b_0+2\widetilde\lambda_i)}
\leq
\sum_{i=1}^k\frac{1}{b_0+2\widetilde\lambda_i}
\leq 1,
\qquad k=1,\ldots,p.
\]
Furthermore, multiplying Equation~\eqref{supp:eq:sampling-b0} by $b_0$ and using $b_0\leq p$ gives
\[
b_0
=
\sum_{i=1}^p\frac{b_0}{b_0+2\widetilde\lambda_i}
\leq
\sum_{i=1}^p\frac{p}{p+2\widetilde\lambda_i}.
\]
Combining the bounds yields
\[
\max_{1\leq k\leq p}
\left\{
k-\frac{2}{k}\sum_{i=1}^k\widetilde\lambda_i
\right\}
\leq b_0
\leq
\sum_{i=1}^p\frac{p}{p+2\widetilde\lambda_i}.
\]
To generate an RRFB observation, retain a Fisher-Bingham draw $z$ if $z \in \mathcal E$ and set $y = T(z)$.

\begin{algorithm}[htbp]
	\caption{Rejection sampling from the Fisher-Bingham distribution}
	\label{supp:alg:fb-sampling}
	\begin{algorithmic}[1]
		\Require Symmetric matrix $A\in\mathbb R^{p\times p}$ and vector $\gamma\in\mathbb R^p$
		\Ensure A draw $z$ with density proportional to $\exp(-z^\top Az+\gamma^\top z)$ on $\mathcal S^{p-1}$
		
		\If{$\|\gamma\|_2 > 0$}
		\State Set $A^{(1)}=A+\frac{\|\gamma\|_2}{2}(I_p-\|\gamma\|_2^{-2}\gamma\gamma^\top)$.
		\Else
		\State Set $A^{(1)}=A$.
		\EndIf
		\State Compute $\lambda_{\min}$, the smallest eigenvalue of $A^{(1)}$, and set $\widetilde A=A^{(1)}-\lambda_{\min}I_p$.
		\State Compute the ordered eigenvalues $0=\widetilde\lambda_1\leq\cdots\leq\widetilde\lambda_p$ of $\widetilde A$.
		\State Solve for $b_0>0$ such that $\sum_{i=1}^p(b_0+2\widetilde\lambda_i)^{-1}=1$, using a root-finding method with bounds
		\[
		\max_{1\leq k\leq p}
		\left\{
		k-\frac{2}{k}\sum_{i=1}^k\widetilde\lambda_i
		\right\}
		\leq b_0
		\leq
		\sum_{i=1}^p\frac{p}{p+2\widetilde\lambda_i}.
		\]
		\State Set $\Omega=I_p+2\widetilde A/b_0$.
		\Repeat
		\State Sample $v\sim\mathcal N_p(0,\Omega^{-1})$ and set $z=v/\|v\|_2$.
		\State Accept $z$ with probability
		\[
		\frac{\exp(-z^\top A z + \gamma^\top z)}
		{\exp(\|\gamma\|_2 - \lambda_{\min} - (p - b_0)/2)
			\left(p / b_0 \right)^{p/2}
			\left(z^\top \Omega z \right)^{-p/2}}.
		\]
		\Until{the sample $z$ is accepted.}
		\State \Return $z$
	\end{algorithmic}
\end{algorithm}

\webappendix{G}{Computational Settings and Numerical Accuracy}

\websubsection{Web Appendix G.1: Computational Settings}

Unless otherwise stated, all simulation configurations used the settings in Table~\ref{tab:computational-settings}. Numerical integration followed Web Appendices~D.1--D.3. Monte Carlo draws for integration over the invalid latent region and face-integration proposals were generated once per fit and held fixed across parameter evaluations. For observations with two or more zero components, we generated one set of proposal pairs for each distinct number of zeros and reused that set across observations with the same number of zeros. Posterior moments and the observed-data likelihood used the same integration workspace.

\begin{table}[htbp]
	\centering
	\captionsetup{name=Web Table}
	\caption{Numerical targets and computational settings.}
	\label{tab:computational-settings}
	\small
	\renewcommand{\arraystretch}{1.2}
	\setlength{\tabcolsep}{6pt}
	\begin{tabularx}{\linewidth}{
			@{}
			>{\raggedright\arraybackslash}p{0.27\linewidth}
			>{\raggedright\arraybackslash}X
			>{\raggedright\arraybackslash}p{0.16\linewidth}
			@{}
		}
		\toprule
		Target or calculation
		& Parameters and values
		& Related section \\
		\midrule
		
		Fisher-Bingham normalizing constant,
		$\mathscr C(\lambda,Q^\top\gamma)$
		&
		Method of \citet{chen2021mle};
		$N_1=600$, $h\approx0.448$,
		$t_0=-5$, $d=5$, and
		$(\omega_d,\omega_u)=(0.9,2.2)$.
		&
		Web Appendix~D.2
		\\
		\addlinespace[4pt]
		
		Integral over the invalid latent region,
		$\mathcal C_{\mathcal E^c}$
		&
		$N_2=100{,}000$ Monte Carlo draws.
		&
		Web Appendix~D.2
		\\
		\addlinespace[4pt]
		
		Face integration: one zero component
		&
		50-node Gauss-Jacobi quadrature.
		&
		Web Appendices D.1 and D.3
		\\
		\addlinespace[4pt]
		
		Face integration: two or more zero components
		&
		20{,}000 proposal pairs for each distinct number of zeros, shared across observations with that number of zeros. Posterior moments use self-normalized weights.
		&
		Web Appendices D.1 and D.3
		\\
		\addlinespace[4pt]
		
		Initialization
		&
		Moment-eigenvalue shrinkage $\rho=0.05$ and floor
		$\varepsilon_\xi=10^{-10}$;
		minimum adjacent gap $10^{-3}$ and upper bound $10$
		for the preliminary quadratic eigenvalues.
		These restrictions were used only to initialize the preliminary
		joint M-step and were not imposed during subsequent optimization.
		&
		Web Appendix~D.4
		\\
		\addlinespace[4pt]
		
		Joint M-step
		&
		Requested iteration budget of 3{,}000. Gradient-norm tolerance $10^{-4}$ in scaled
		product-manifold coordinates;
		adaptive scale factors restricted to $[0.2,5]$.
		&
		Web Appendix~D.2
		\\
		\addlinespace[4pt]
		
		Outer MCEM convergence
		&
		Scaled-change tolerances:
		$10^{-6}$ for the conditioned log-likelihood per observation
		and $10^{-5}$ for both $A$ and $\gamma$.
		Monitoring began at iteration 5. The outer iteration cap was 1{,}000. Convergence required all three
		scaled-change criteria and the joint M-step gradient criterion
		to hold for two consecutive iterations. Inexact M-steps were accepted only when they sufficiently decreased the numerical objective and did not count toward the consecutive iterations required for outer convergence.
		&
		Section~3.3
		\\
		
		\bottomrule
	\end{tabularx}
\end{table}

For numerical evaluation of Fisher-Bingham normalizing constant, we shifted the eigenvalues to $\widetilde\lambda_i=\lambda_i+5$ and recovered the original constant using $\mathscr C(\lambda,Q^\top\gamma) = e^5\mathscr C(\widetilde\lambda,Q^\top\gamma)$. In Web Table~\ref{tab:computational-settings}, $t_0$ denotes the contour-shift parameter and $d$ the analytic-strip half-width. Within each joint M-step, the eigenvalue-coordinate and directional-parameter blocks were rescaled using their initial gradient norms relative to the orthogonal-matrix block. Score-test calculations reused the quadrature nodes and Monte Carlo draws from the pooled null fit, held fixed throughout differentiation. Observation-level scores and the second derivatives required for the sensitivity matrix were obtained by automatic differentiation using \texttt{jax}, including derivatives of the restricted normalizing constant. The nuisance adjustment and final score statistic were evaluated using direct linear solves, as described in Web Appendix~E.4.

\websubsection{Web Appendix G.2: Numerical Accuracy Assessment}

We focused the numerical refinement assessment on $p=20$, the largest dimension considered in our simulations, to examine numerical sensitivity at the upper end of the dimensional range studied. For each configuration, we simulated 100 observed compositions and evaluated the restricted normalizing constant and posterior moments at the generating parameters. Quadrature comparisons included all observations with exactly one zero component. For observations with two or more zero components, we selected one observation for each represented number of zeros. Each integration component was refined separately, using the finer calculation as a numerical reference. Web Table~\ref{tab:numerical-refinement} summarizes the maximum differences across the evaluated configurations and observations.
\begin{table}[htbp]
	\centering
	\captionsetup{name=Web Table}
	\caption{Maximum differences under numerical refinement at the four $p=20$ generating configurations. Normalizer quantities summarize the largest differences across configurations; posterior-moment quantities summarize the largest observation-level differences across configurations. First- and second-moment differences use the Euclidean and Frobenius norms, respectively. Values are rounded.}
	\label{tab:numerical-refinement}
	\small
	\renewcommand{\arraystretch}{1.2}
	\setlength{\tabcolsep}{6pt}
	\begin{tabularx}{\linewidth}{@{}>{\raggedright\arraybackslash}p{0.25\linewidth}>{\raggedright\arraybackslash}p{0.25\linewidth}>{\raggedright\arraybackslash}X@{}}
		\toprule
		Calculation & Budget refinement & Maximum differences \\
		\midrule
		
		Full-sphere normalizer truncation
		&
		$N_1$: $600$ to $2{,}000$
		&
		Restricted log normalizer: $1.07\times10^{-14}$;\newline
		scaled gradient: $2.58\times10^{-15}$.
		\\
		\addlinespace[4pt]
		
		Integration over the invalid latent region
		&
		$N_2$: $100{,}000$ to $400{,}000$
		&
		Restricted log normalizer: $4.23\times10^{-12}$;\newline
		scaled gradient: $2.11\times10^{-10}$.
		\\
		\addlinespace[4pt]
		
		Face integration: one zero component
		&
		$50$ to $200$ quadrature nodes
		&
		Posterior first moment: $8.60\times10^{-4}$;\newline
		posterior second moment: $1.21\times10^{-3}$.
		\\
		\addlinespace[4pt]
		
		Face integration: two or more zero components
		&
		$20{,}000$ to $80{,}000$ importance-sampling proposals
		&
		Posterior first moment: $0.0429$;\newline
		posterior second moment: $0.0579$.
		\\
		
		\bottomrule
	\end{tabularx}
\end{table}

For the full-sphere truncation comparison, the invalid-region approximation was fixed at 400{,}000 draws; for the invalid-region comparison, $N_1$ was fixed at $2{,}000$. Log-normalizer differences refer to the restricted normalizing constant $\log\mathcal C_{\mathcal E}$. Scaled gradient differences were calculated as
\[
\frac{\|g-g_{\mathrm{ref}}\|_2}{1+\|g_{\mathrm{ref}}\|_2},
\]
where $g$ concatenates derivatives of the restricted log normalizing constant with respect to its eigenvalue, eigenbasis directional, and orthogonal-matrix arguments. Observation-level posterior first- and second-moment differences were measured using the Euclidean and Frobenius norms, respectively.

Web Table~\ref{tab:numerical-refinement} indicates stability of the restricted normalizer and its first derivatives under the examined refinements. Posterior-moment differences were larger for multiple-zero face integration than for one-zero quadrature, with the largest first- and second-moment differences occurring in the high-zero configuration.

These assessments quantify numerical refinement sensitivity at the evaluated generating parameters. The finer calculations are numerical references rather than exact values, and the comparisons do not establish accuracy throughout the optimization path or stability of score-test second derivatives and final statistics. Convergence with fixed integration proposals concerns the numerical objective and does not itself bound integration error.

\webappendix{H}{Additional Details for Study 1}

\websubsection{H.1 Factorial Design and Parameter Settings}

In Section~5.1 of the main text, we introduced our factorial design that serves as the basis for evaluating our RRFB framework. This design considers three dimensions $p \in \{5, 10, 20\}$, four sample sizes $n \in \{100, 200, 500, 1000\}$, and four targeted zero patterns. For each dimension, we set $A_p = Q_p \diag{\lambda_p} Q^\top_p$, with $Q_p$ and $\lambda_p$ held fixed across the four zero-prevalence configurations. To construct $Q_p$, we start from $I_p$ and successively left-multiply by the Givens rotations for the coordinate pairs in $\mathcal P_p$, in the order listed in Web Table~\ref{tab:simulation_parameters}. Each rotation has angle $\theta=0.35$ radians. For a listed pair $(a,b)$, the rotation has entries $-\sin\theta$ and $\sin\theta$ in positions $(a,b)$ and $(b,a)$, respectively. The vector $\gamma_p^{(c)}$ denotes the directional parameter for Case~$c$, with Cases~1--4 corresponding to diffuse low sparsity, diffuse moderate sparsity, concentrated moderate sparsity, and diffuse high sparsity, respectively. Web Table~\ref{tab:simulation_parameters} provides the complete parameter settings. Furthermore, Table~\ref{supp:tab:sparsity} reports the componentwise zero percentages induced by the generating parameter settings.

\begin{table}[htbp]
	\centering
	\captionsetup{name=Web Table}
	\caption{Generating parameter settings for the factorial simulation design in Study~1. Within each dimension $p$, the eigenvalues $\lambda_p$ and orthogonal matrix $Q_p$ are fixed across Cases~1--4, while the directional parameter $\gamma_p^{(c)}$ varies by case. The coordinate pairs in $\mathcal{P}_p$ are applied in the order listed.}
	\label{tab:simulation_parameters}
	\scriptsize
	\renewcommand{\arraystretch}{1.75}
	\begin{tabular}{@{}c@{\hspace{1.5em}}c@{\hspace{2em}}c@{}}
		\toprule
		Dimension & Parameter & Value \\
		\midrule
		\multirow{6}{*}{$p=5$}
		& $\mathcal{P}_5$
		& $\{(1,2),(2,3),(4,5)\}$ \\
		& $\lambda_5$ 
		& $(0,1,2.5,4.5,7)^\top$ \\
		& $\gamma_5^{(1)}$ & $(2.5,2.5,3.2,3.9,4.2)^\top$ \\
		& $\gamma_5^{(2)}$ & $(1.2,1.2,1.6,2.1,2.3)^\top$ \\
		& $\gamma_5^{(3)}$ & $(2.5,2.5,3.0,1.8,1.8)^\top$ \\
		& $\gamma_5^{(4)}$ & $(0.5,0.5,0.7,1.0,1.2)^\top$ \\
		\midrule
		\multirow{6}{*}{$p=10$}
		& $\mathcal{P}_{10}$
		& $\{(1,2),(2,3),(3,4),(5,6),(7,8),(9,10)\}$ \\
		& $\lambda_{10}$
		& $(0,1.2,2.4,4,6,8.4,11,13.6,16.4,20)^\top$ \\
		
		
		& $\gamma_{10}^{(1)}$
		& $(3.6,3.7,4.0,4.5,5.2,5.7,6.4,6.8,7.4,7.9)^\top$ \\
		& $\gamma_{10}^{(2)}$
		& $(1.9,2.0,2.2,2.7,3.3,3.6,4.2,4.5,4.9,5.2)^\top$ \\
		& $\gamma_{10}^{(3)}$
		& $(3.6,3.6,4.0,4.8,5.8,6.1,7.2,2.9,3.2,3.3)^\top$ \\
		& $\gamma_{10}^{(4)}$
		& $(0.8,0.8,1.0,1.2,1.6,1.8,2.2,2.3,2.6,2.8)^\top$ \\
		
		\midrule

		\multirow{6}{*}{$p=20$}
		& $\mathcal{P}_{20}$
		& $\{(1,2),(2,3),(3,4),(5,6),(7,8),(9,10),(11,12),(13,14),(15,16),(17,18),(19,20)\}$ \\
		
		& $\lambda_{20}$
		& $(0,0.6,1.7,3,4.5,6.2,8,10,12,14,16,18.5,21,23.5,26,28.5,31.5,34,37,40)^\top$ \\
		
		
		& $\gamma_{20}^{(1)}$
		& $(5.6,5.7,5.8,6.0,6.2,6.5,6.8,7.2,7.7,8.1,8.4,8.8,9.3,9.6,9.9,10.2,10.6,11.0,11.3,11.7)^\top$ \\
		
		& $\gamma_{20}^{(2)}$
		& $(3.1,3.1,3.2,3.3,3.5,3.7,4.1,4.4,4.8,5.1,5.3,5.6,6.0,6.3,6.5,6.8,7.0,7.3,7.5,7.7)^\top$ \\
		
		& $\gamma_{20}^{(3)}$
		& $(6.2,6.3,6.4,6.6,6.8,7.1,7.4,7.8,8.3,8.7,9.0,9.4,9.9,10.2,10.5,3.8,4.0,4.2,4.3,4.4)^\top$ \\
		
		& $\gamma_{20}^{(4)}$
		& $(1.3,1.3,1.3,1.4,1.6,1.7,1.8,2.1,2.3,2.5,2.6,2.8,3.1,3.3,3.4,3.5,3.7,3.9,4.0,4.1)^\top$ \\
		\bottomrule
	\end{tabular}
\end{table}

\begin{table}[htbp]
	\captionsetup{name=Web Table}
	\centering
	\caption{Empirical componentwise zero percentages based on 100,000 simulated observed compositions per configuration, after rectification and renormalization. The mean is the overall percentage of zero entries across all observations and components.}
	\label{supp:tab:sparsity}
	\footnotesize
	\setlength{\tabcolsep}{4pt}
	\begin{tabular}{@{}ccc>{\raggedright\arraybackslash}p{0.50\textwidth}@{}}
		\toprule
		$p$ & Case & Mean (\%) & Component-wise zeros (\%) \\
		\midrule
		
		5 & 1 & 12.5 &
		1--5: (11.7, 12.4, 11.6, 13.0, 13.8) \\
		
		& 2 & 23.3 &
		1--5: (21.6, 22.8, 22.6, 24.0, 25.2) \\
		
		& 3 & 18.7 &
		1--5: (9.6, 10.7, 11.8, 29.7, 31.6) \\
		
		& 4 & 34.9 &
		1--5: (34.4, 35.1, 34.9, 35.2, 35.0) \\[0.4em]
		
		10 & 1 & 11.9 &
		1--10: (11.8, 11.8, 11.9, 12.1, 12.2, 11.8, 11.9, 12.0, 11.9, 11.7) \\
		
		& 2 & 19.8 &
		1--10: (19.6, 19.5, 19.8, 19.6, 19.8, 19.9, 19.7, 19.8, 20.2, 20.3) \\
		
		& 3 & 16.5 &
		1--10: (11.3, 11.8, 11.2, 10.3, 9.4, 9.9, 9.8, 29.4, 30.6, 30.9) \\
		
		& 4 & 32.1 &
		1--10: (31.6, 32.6, 31.7, 32.4, 32.2, 32.1, 31.7, 32.2, 32.1, 32.0) \\[0.4em]
		
		20 & 1 & 12.0 &
		1--10: (11.5, 10.9, 11.2, 11.9, 12.8, 12.7, 12.9, 12.5, 12.2, 11.8)
		\newline
		11--20: (12.1, 11.8, 11.8, 11.7, 12.0, 12.0, 12.1, 11.7, 12.0, 11.9) \\
		
		& 2 & 20.2 &
		1--10: (18.6, 18.3, 18.9, 20.3, 21.6, 21.7, 21.0, 20.8, 20.4, 19.9)
		\newline
		11--20: (20.7, 20.3, 20.2, 19.8, 20.2, 20.0, 20.5, 20.2, 20.5, 20.6) \\
		
		& 3 & 16.0 &
		1--10: (9.6, 9.0, 9.3, 9.8, 10.9, 10.8, 11.3, 10.9, 10.8, 10.4)
		\newline
		11--20: (10.8, 10.4, 10.5, 10.4, 11.1, 32.5, 33.1, 32.6, 32.8, 32.8) \\
		
		& 4 & 32.4 &
		1--10: (30.6, 30.6, 31.5, 32.5, 33.2, 33.4, 34.2, 33.0, 33.0, 32.5)
		\newline
		11--20: (32.8, 32.6, 32.1, 31.8, 32.4, 32.3, 32.4, 32.0, 32.4, 32.5) \\
		
		\bottomrule
	\end{tabular}
\end{table}

\websubsection{H.2 Expected Log-Score Evaluation}

Let $f_0$ denote the generating observed-data RRFB density and $\widehat f_r$ the density fitted in simulation replicate $r$. We approximated the expected log-score difference defined in Section~5.1 of the main text using $B=10{,}000$ observations generated independently of the fitting sample:
\[
Y_{rb}\overset{\mathrm{iid}}{\sim}f_0,\qquad
d_{rb}=\log f_0(Y_{rb})-\log\widehat f_r(Y_{rb}),\qquad
\widehat\Delta_r=\frac{1}{B}\sum_{b=1}^{B}d_{rb}.
\]
The generating and fitted log-densities were evaluated using the same numerical integration proposals, and the face constants in Web Appendix~D.3 cancel within each paired difference. We used $B=10{,}000$ evaluation observations per fitted replicate to balance Monte Carlo precision with the computational cost of repeated density evaluation. Standard errors were calculated from the paired log-density differences as $\operatorname{SD}(d_{r1},\ldots,d_{rB})/\sqrt{B}$. Across all replicates and configurations, the maximum standard errors were 0.0104, 0.0148, and 0.0435 for $p=5,10,$ and $20$, respectively. Web Table~\ref{supp:tab:distribution_recovery} reports the mean estimated expected log-score difference across $R=200$ simulation replicates per configuration. We assessed the precision of the reported means using the relative Monte Carlo standard error,
\[
\operatorname{relative\,MCSE}
=
\frac{\operatorname{MCSE}(R^{-1}\sum_{r=1}^R\widehat{\Delta}_r)}{R^{-1}\sum_{r=1}^R\widehat\Delta_r}
=
\frac{\sqrt{R} \cdot \operatorname{SD}(\widehat\Delta_1,\ldots,\widehat\Delta_R)}
{\sum_{r=1}^R\widehat\Delta_r}.
\]
With $R=200$ replicates per configuration, relative Monte Carlo standard errors ranged from approximately 0.7\% to 3.0\%, providing adequate precision for characterizing the overall recovery patterns.

\begin{table}[htbp]
	\centering
	\setlength{\tabcolsep}{5pt}
	\captionsetup{name=Web Table}
	\caption{Recovery of the observed-data RRFB distribution in Study~1.
		Entries are mean estimated expected log-score differences across 200
		simulation replicates, with Monte Carlo standard errors of the means
		in parentheses. Each replicate-specific expectation was approximated
		using 10,000 independent evaluation observations. Smaller values
		indicate better distributional recovery.}
	\label{supp:tab:distribution_recovery}
	\begin{tabular}{cc*{4}{c}}
		\toprule
		$p$ & Case & $n=100$ & $n=200$ & $n=500$ & $n=1000$ \\
		\midrule
		5 & 1 & 0.103 (0.0029) & 0.049 (0.0013) & 0.020 (0.00047) & 0.010 (0.00026) \\
		5 & 2 & 0.113 (0.0034) & 0.052 (0.0012) & 0.019 (0.00048) & 0.009 (0.00023) \\
		5 & 3 & 0.104 (0.0029) & 0.052 (0.0012) & 0.019 (0.00051) & 0.010 (0.00024) \\
		5 & 4 & 0.103 (0.0026) & 0.049 (0.0011) & 0.019 (0.00047) & 0.010 (0.00023) \\
		\midrule
		10 & 1 & 0.383 (0.0065) & 0.174 (0.0024) & 0.069 (0.00086) & 0.034 (0.00049) \\
		10 & 2 & 0.382 (0.0061) & 0.177 (0.0027) & 0.065 (0.00094) & 0.032 (0.00046) \\
		10 & 3 & 0.387 (0.0063) & 0.174 (0.0026) & 0.066 (0.00093) & 0.033 (0.00045) \\
		10 & 4 & 0.376 (0.0056) & 0.178 (0.0024) & 0.067 (0.00090) & 0.033 (0.00046) \\
		\midrule
		20 & 1 & 1.681 (0.017) & 0.688 (0.0058) & 0.247 (0.0020) & 0.118 (0.00083) \\
		20 & 2 & 1.722 (0.017) & 0.698 (0.0057) & 0.245 (0.0018) & 0.120 (0.00085) \\
		20 & 3 & 1.681 (0.015) & 0.678 (0.0052) & 0.248 (0.0018) & 0.120 (0.00089) \\
		20 & 4 & 1.771 (0.018) & 0.699 (0.0059) & 0.249 (0.0019) & 0.119 (0.00090) \\
		\bottomrule
	\end{tabular}
\end{table}

\webappendix{I}{Additional Details for Study 2}

\websubsection{I.1 Null Calibration}

\begin{figure}[htbp]
	\captionsetup{name=Web Figure}
	\centering
	\includegraphics[width=\linewidth]{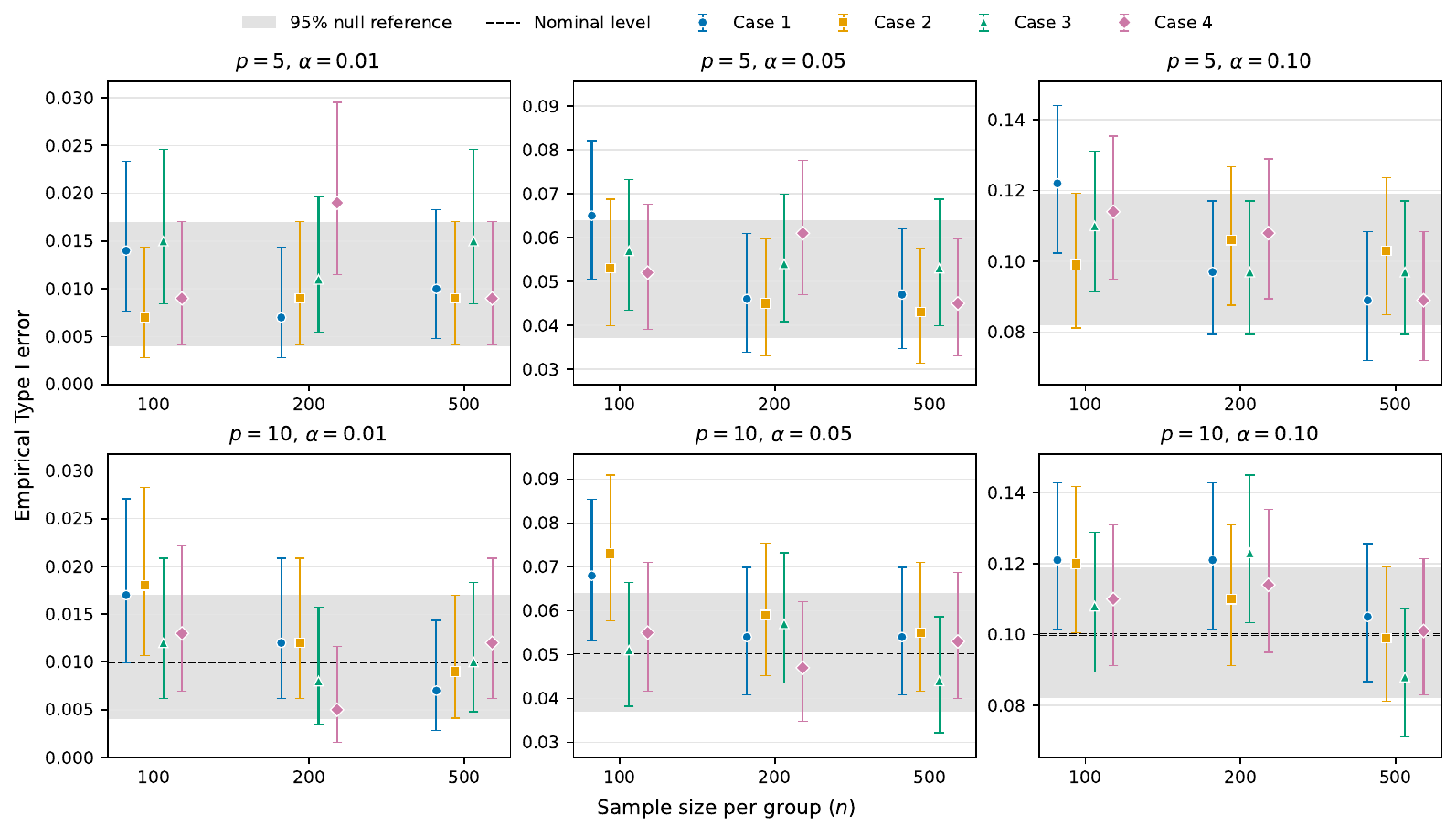}
	\caption{Null rejection proportions for the common-$A$ RRFB score test, based on 1,000 simulation replicates per configuration. Vertical bars are exact 95\% Clopper--Pearson binomial confidence intervals. Dashed lines indicate nominal significance levels, and shaded bands show the central 95\% binomial reference ranges under nominal calibration. Sample sizes are per group.}
	\label{supp:fig:type1}
\end{figure}

\begin{figure}[htbp]
	\captionsetup{name=Web Figure}
	\centering
	\includegraphics[width=\linewidth]{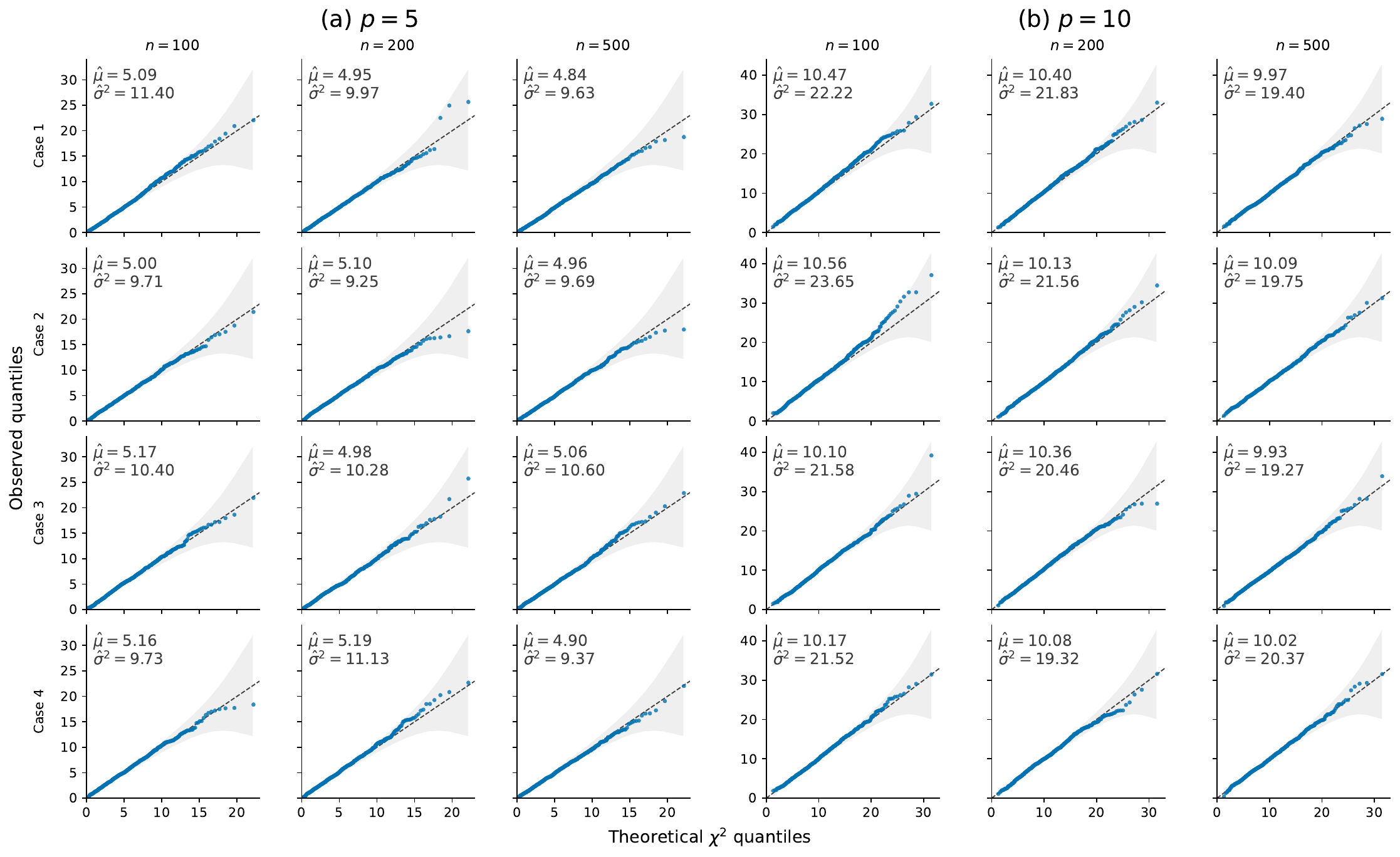}
	\caption{Null score-statistic QQ plots against the $\chi_p^2$ reference distribution, based on 1,000 simulation replicates per configuration. Dashed lines indicate equality. Shaded regions are approximate simultaneous 95\% reference envelopes within each panel, calibrated from 10,000 independent samples of 1,000 $\chi_p^2$ statistics. Panel annotations report the empirical mean and sample variance, whose reference values are $p$ and $2p$, respectively. Sample sizes are per group.}
	\label{supp:fig:qq}
\end{figure}

We assessed null calibration for $p\in\{5,10\}$, sample sizes $n\in\{100,200,500\}$ per group, and the four zero-prevalence configurations described in Study~1. For each configuration, we generated two independent groups from the same RRFB distribution and fitted the pooled null model. We used $R=1{,}000$ simulation replicates per configuration to provide reasonable precision for estimating rejection probabilities. The corresponding binomial Monte Carlo standard error is smaller than 0.01 at all considered nominal rejection probabilities: $\alpha = 0.01, 0.05, 0.10$. Web Figure~\ref{supp:fig:type1} reports empirical rejection proportions at the considered nominal levels, together with binomial confidence intervals and reference bands under nominal calibration. Some smaller-sample configurations exhibited upward departures from the nominal levels, whereas all rejection proportions at $n=500$ fell within the corresponding pointwise 95\% reference bands. Web Figure~\ref{supp:fig:qq} compares the empirical score-statistic quantiles with their $\chi_p^2$ reference quantiles. Agreement was generally close, with some upper-tail departures in smaller-sample configurations. Furthermore, we additionally summarized agreement between the distribution of null $p$-values and the Uniform$(0,1)$ reference distribution using
\[
D_R=\sup_{0\leq u\leq1}\left|\widehat F_R(u)-u\right|,
\]
where $\widehat F_R$ is the empirical distribution function of the $R=1{,}000$ simulated $p$-values. Web Table~\ref{supp:tab:pvalue-discrepancy} reports this discrepancy for each configuration. Values ranged from 0.0159 to 0.0545.

\begin{table}[htbp]
	\captionsetup{name=Web Table}
	\centering
	\small
	\caption{Maximum absolute discrepancies $D_R$ between the empirical distribution function of null $p$-values and the Uniform$(0,1)$ distribution function, based on 1,000 simulation replicates per configuration. Sample sizes are per group. Smaller values indicate closer agreement with the uniform reference distribution.}
	\label{supp:tab:pvalue-discrepancy}
	\begin{tabular}{ccccc}
		\toprule
		$p$ & Case & $n=100$ & $n=200$ & $n=500$ \\
		\midrule
		5 & 1 & 0.0253 & 0.0159 & 0.0422 \\
		5 & 2 & 0.0162 & 0.0315 & 0.0246 \\
		5 & 3 & 0.0440 & 0.0423 & 0.0171 \\
		5 & 4 & 0.0365 & 0.0311 & 0.0294 \\
		\midrule
		10 & 1 & 0.0451 & 0.0489 & 0.0256 \\
		10 & 2 & 0.0545 & 0.0206 & 0.0275 \\
		10 & 3 & 0.0298 & 0.0442 & 0.0207 \\
		10 & 4 & 0.0243 & 0.0338 & 0.0210 \\
		\bottomrule
	\end{tabular}
\end{table}

We examined the nuisance sensitivity matrix $\widehat H_{\eta\eta}$ and the efficient-score covariance matrix $\widehat V_{\psi\mid\eta}$ to assess numerical invertibility, positive definiteness, and conditioning. These diagnostics describe different aspects of the score calculation: full rank indicates numerical invertibility, positive definiteness requires positive eigenvalues bounded away from zero, and conditioning measures sensitivity to perturbations. For a symmetric matrix $M$, we used the numerical tolerance
\[
\tau(M)=10^{-10}\sigma_{\max}(M),
\]
where $\sigma_{\max}(M)$ is the largest singular value. We classified $M$ as having full numerical rank when its smallest singular value exceeded $\tau(M)$, and as numerically positive definite when its smallest eigenvalue exceeded the same tolerance. We summarized conditioning using $\kappa_2(M)=\sigma_{\max}(M)/\sigma_{\min}(M)$, where $\sigma_{\min}(M)$ is the smallest singular value. Larger condition numbers indicate greater sensitivity to perturbations. Condition numbers were calculated after diagonal standardization to account for differences in coordinate scaling and were reported descriptively, without assigning a pass-or-fail classification. Web Table~\ref{supp:tab:null-matrices} summarizes these diagnostics.

\begin{table}[htbp]
	\captionsetup{name=Web Table}
	\centering
	\footnotesize
	\setlength{\tabcolsep}{4pt}
	\caption{Matrix diagnostics for the null-calibration simulations, based on 1,000 replicates per configuration. Condition numbers refer to diagonally standardized matrices and are reported as medians, with 95th percentiles in parentheses. Singular-value entries give the smallest observed singular value across the raw matrices in each configuration. The final column reports the percentage of raw nuisance sensitivity matrices failing the stated numerical positive-definiteness criterion.}
	\label{supp:tab:null-matrices}
	\begin{tabular}{cccccccc}
		\toprule
		& & &
		\multicolumn{2}{c}{$\widehat H_{\eta\eta}$} &
		\multicolumn{2}{c}{$\widehat V_{\psi\mid\eta}$} & \\
		\cmidrule(lr){4-5}
		\cmidrule(lr){6-7}
		$p$ & Case & $n$
		& \shortstack{Standardized\\condition no.}
		& \shortstack{Minimum\\$\sigma_{\min}$}
		& \shortstack{Standardized\\condition no.}
		& \shortstack{Minimum\\$\sigma_{\min}$}
		& \shortstack{Nuisance\\not PD (\%)} \\
		\midrule
		5 & 1 & 100 & 644 (2214) & $8.55\times10^{-8}$ & 4.77 (6.98) & $1.25\times10^{-2}$ & 7.9 \\
		5 & 1 & 200 & 512 (1357) & $1.01\times10^{-6}$ & 4.54 (5.90) & $1.28\times10^{-2}$ & 5.1 \\
		5 & 1 & 500 & 460 (953) & $1.34\times10^{-5}$ & 4.41 (5.30) & $1.75\times10^{-2}$ & 0.7 \\
		5 & 2 & 100 & 56.9 (159) & $1.43\times10^{-5}$ & 2.35 (3.50) & $2.69\times10^{-2}$ & 1.3 \\
		5 & 2 & 200 & 50.9 (102) & $2.35\times10^{-5}$ & 2.23 (2.95) & $3.81\times10^{-2}$ & 0.2 \\
		5 & 2 & 500 & 48.2 (70.9) & $2.17\times10^{-3}$ & 2.14 (2.55) & $4.29\times10^{-2}$ & 0.0 \\
		5 & 3 & 100 & 353 (1512) & $2.33\times10^{-6}$ & 4.45 (6.69) & $1.46\times10^{-2}$ & 6.4 \\
		5 & 3 & 200 & 267 (830) & $1.19\times10^{-5}$ & 4.31 (5.82) & $1.59\times10^{-2}$ & 3.1 \\
		5 & 3 & 500 & 242 (528) & $1.12\times10^{-4}$ & 4.27 (5.25) & $1.93\times10^{-2}$ & 0.7 \\
		5 & 4 & 100 & 13.0 (22.6) & $1.84\times10^{-4}$ & 1.90 (2.45) & $3.10\times10^{-2}$ & 0.0 \\
		5 & 4 & 200 & 11.9 (16.6) & $1.44\times10^{-3}$ & 1.70 (2.05) & $3.93\times10^{-2}$ & 0.0 \\
		5 & 4 & 500 & 11.4 (14.0) & $1.32\times10^{-3}$ & 1.59 (1.79) & $4.47\times10^{-2}$ & 0.1 \\
		\midrule
		10 & 1 & 100 & 6548 (38175) & $1.31\times10^{-8}$ & 5.78 (7.94) & $5.13\times10^{-3}$ & 78.9 \\
		10 & 1 & 200 & 5308 (33070) & $7.39\times10^{-9}$ & 5.39 (6.71) & $6.72\times10^{-3}$ & 78.0 \\
		10 & 1 & 500 & 5174 (26690) & $7.23\times10^{-9}$ & 5.00 (5.87) & $8.53\times10^{-3}$ & 75.1 \\
		10 & 2 & 100 & 594 (3548) & $9.59\times10^{-7}$ & 3.76 (5.20) & $9.99\times10^{-3}$ & 30.4 \\
		10 & 2 & 200 & 438 (1016) & $3.56\times10^{-7}$ & 3.33 (4.25) & $1.27\times10^{-2}$ & 15.5 \\
		10 & 2 & 500 & 372 (691) & $7.31\times10^{-7}$ & 3.04 (3.60) & $1.50\times10^{-2}$ & 8.7 \\
		10 & 3 & 100 & 7376 (66321) & $1.54\times10^{-7}$ & 6.14 (8.25) & $5.75\times10^{-3}$ & 81.7 \\
		10 & 3 & 200 & 3157 (34364) & $1.93\times10^{-8}$ & 5.44 (6.81) & $7.04\times10^{-3}$ & 56.8 \\
		10 & 3 & 500 & 2219 (9528) & $3.15\times10^{-7}$ & 5.07 (5.91) & $8.64\times10^{-3}$ & 33.7 \\
		10 & 4 & 100 & 53.8 (129) & $7.40\times10^{-6}$ & 2.59 (3.35) & $1.19\times10^{-2}$ & 7.2 \\
		10 & 4 & 200 & 42.7 (73.6) & $1.20\times10^{-5}$ & 2.08 (2.50) & $1.48\times10^{-2}$ & 5.3 \\
		10 & 4 & 500 & 37.3 (50.8) & $3.31\times10^{-5}$ & 1.77 (2.01) & $1.69\times10^{-2}$ & 2.9 \\
		\bottomrule
	\end{tabular}
\end{table}

Across all configurations, both the nuisance sensitivity and efficient-score covariance matrices had full numerical rank under the stated criteria, and the efficient-score covariance matrices were numerically positive definite in every replicate. The largest nuisance sensitivity condition numbers occurred for $p=10$ in Cases~1 and 3, with median standardized values reaching 7376. In comparison, median standardized condition numbers for the efficient-score covariance matrices ranged from 1.59 to 6.14, indicating substantially greater sensitivity to perturbations in the nuisance component of the calculation. Increasing sample size reduced the median standardized condition numbers in every configuration, although nuisance conditioning remained comparatively poor in these settings.

Failures of numerical positive definiteness in the nuisance sensitivity matrix were uncommon overall for $p=5$ (2.1\%) but substantially more frequent for $p=10$ (39.5\%). These failures persisted in some larger-sample configurations: for $p=10$, Case~1, the percentage remained 75.1\% at $n=500$. Thus, increasing sample size improved conditioning but did not consistently resolve the positive-definiteness concerns in the nuisance sensitivity matrix, despite its retaining full numerical rank.

\websubsection{I.2 Aligned Alternatives}
We evaluated power against structured changes in $\gamma$ under the common-$A$ RRFB model, using the baseline parameters and four zero-prevalence configurations from Study~1, with $p\in\{5,10\}$ and $n\in\{200,500\}$ observations per group. The construction changes selected linear coefficients in the eigenbasis of $A$, while preserving its eigenvalues and eigendirections. Specifically, let $A=Q\diag(\lambda)Q^\top$, and define the selected-coordinate sets
\[
S_5=\{4,5\}, \qquad S_{10}=\{8,9,10\}.
\]
As described in Section~5.2 (Study~2) of the main text, Group~0 retains the baseline directional parameter $\gamma_0$, while the directional parameter $\gamma_1(d)$ of Group~1 is obtained by subtracting a scalar $d$ from the selected linear coefficients in the eigenbasis of $A$. Specifically,
\[
\gamma_1(d)=\gamma_0-Q\zeta_p(d),
\qquad
\zeta_{p,j}(d)=
\begin{cases}
	d, & j\in S_p,\\
	0, & j\notin S_p.
\end{cases}
\]
Thus, if $q_j$ denotes the $j$th column of $Q$, the directional contrast is $\gamma_1(d)-\gamma_0=-d\sum_{j\in S_p}q_j$. The value $d=0$ gives the null model, and positive and negative values of $d$ produce departures in opposite directions along this fixed direction. The nonzero perturbation grids were
\[
	\mathcal D_5=\{\pm0.2,\pm0.4,\pm0.8,\pm1.2\}, \qquad \mathcal D_{10}=\{\pm0.15,\pm0.6,\pm0.9,\pm1.5\},
\]
Each nonzero setting used 500 independent simulation replicates. The points at $d=0$ are taken from the matching 1,000-replicate null-calibration study in Web Appendix~I.1. All comparisons used $\alpha=0.05$. PERMANOVA used 999 permutations with Bray-Curtis distance on relative abundances or Euclidean distance on square-root compositions.

To explain the effect of the perturbation, let $Z$ denote a latent Fisher-Bingham direction in Group~1, conditional on the valid region $\mathcal E=\{z\in\mathcal S^{p-1}: \|z^+\|_2 > 0\}$, and write
\[
W=Q^\top Z,
\qquad
\eta(d)=Q^\top\gamma_1(d)=Q^\top\gamma_0-\zeta_p(d).
\]
The vector $W$ expresses the same latent direction in the eigenbasis of $A$. In these coordinates, the latent density is proportional to
\[
\exp\left\{-\sum_{k=1}^p\lambda_k W_k^2+\sum_{k=1}^p\eta_k(d)W_k\right\},
\qquad
\|W\|_2=1,\quad QW\in\mathcal E.
\]
The perturbation therefore subtracts $d$ from each selected linear coefficient $\eta_j(d)$, leaving the remaining linear coefficients and the entire quadratic term unchanged.

For a selected coordinate $j\in S_p$, condition on the remaining eigencoordinates $W_{-j}$ and set $r=(1-\|W_{-j}\|_2^2)^{1/2}$. For $0<r<1$, the unit-sphere constraint permits only $W_j=r$ and $W_j=-r$. Provided that both candidate latent directions belong to $\mathcal E$, their conditional odds satisfy
\[
\frac{\Pr_d(W_j=r\mid W_{-j})}
{\Pr_d(W_j=-r\mid W_{-j})}
=
\exp\{2r\eta_j(d)\}
=
\exp\left\{2r\left[(Q^\top\gamma_0)_j-d\right]\right\},
\]
where $\Pr_d$ denotes probability under the perturbed, valid-region-conditioned latent distribution. The quadratic contribution is identical for the two candidates because $r^2=(-r)^2$, and all remaining-coordinate contributions are also unchanged. The normalizing factors cancel in the ratio. Consequently,
\[
\frac{\partial}{\partial d}
\log
\frac{\Pr_d(W_j=r\mid W_{-j})}
{\Pr_d(W_j=-r\mid W_{-j})}
=-2r<0.
\]
The spectral decomposition $A=Q\diag(\lambda)Q^\top$ is central to this calculation: in the eigenbasis of $A$, changing the sign of $W_j$ leaves the quadratic contribution unchanged. Consequently, holding the remaining eigencoordinates fixed and provided that both candidate directions are valid, increasing $d$ decreases the conditional odds of a positive selected eigencoordinate, whereas decreasing $d$ increases those odds. This establishes the direction of the conditional sign change in the eigenbasis of $A$.

Thus, holding the remaining eigencoordinates fixed and provided that both candidate directions are valid, increasing $d$ favors the negative sign of each selected eigencoordinate, whereas decreasing $d$ favors the positive sign. This perturbation was chosen to vary the degree of rectification, and hence observed zero prevalence, while preserving the common quadratic parameter $A$. The localized Givens rotations used to construct $Q$ connect the selected eigendirections to small blocks of ambient coordinates, providing a mechanism through which the directional shift changes observed zero patterns.

\begin{figure}[htbp]
	\captionsetup{name=Web Figure}
	\centering
	\includegraphics[width=0.9\linewidth]{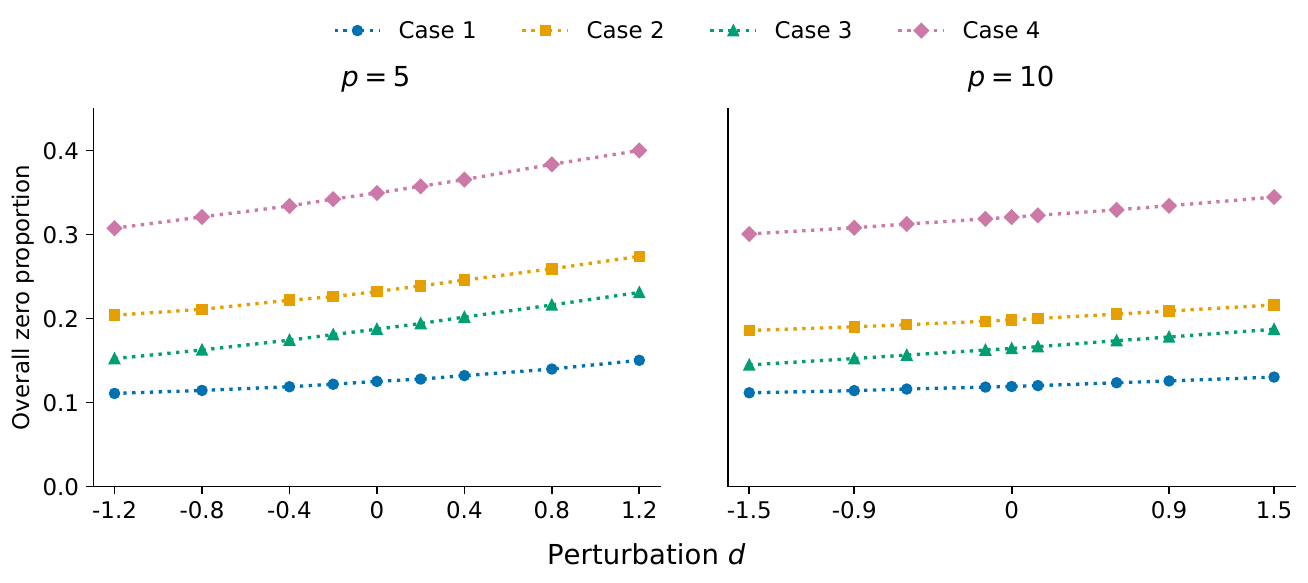}
	\caption{Fraction of observed zero entries in Group~1 as a function of the signed perturbation $d$, with $n=500$ observations per group. For each simulated dataset, the fraction is the number of zero entries among the $np$ observed component values in Group~1 divided by $np$. Points show the mean fraction across 500 replicates at each nonzero value of $d$ and across the 1,000 matching null replicates at $d=0$. Panels correspond to $p=5$ and $p=10$. Colors and marker shapes distinguish the four baseline zero-prevalence configurations.}
	\label{supp:fig:power_zeros}
\end{figure}

Rectification depends on the ambient-coordinate signs, with the $k$th observed component equal to zero precisely when $Z_k=(QW)_k\leq0$. Across the evaluated perturbations, the overall fraction of observed zero entries in Group~1 increased with $d$ in each of the four baseline configurations at both dimensions (Web Figure~\ref{supp:fig:power_zeros}).

\begin{figure}[htbp]
	\captionsetup{name=Web Figure}
	\centering
	\includegraphics[width=0.9\linewidth]{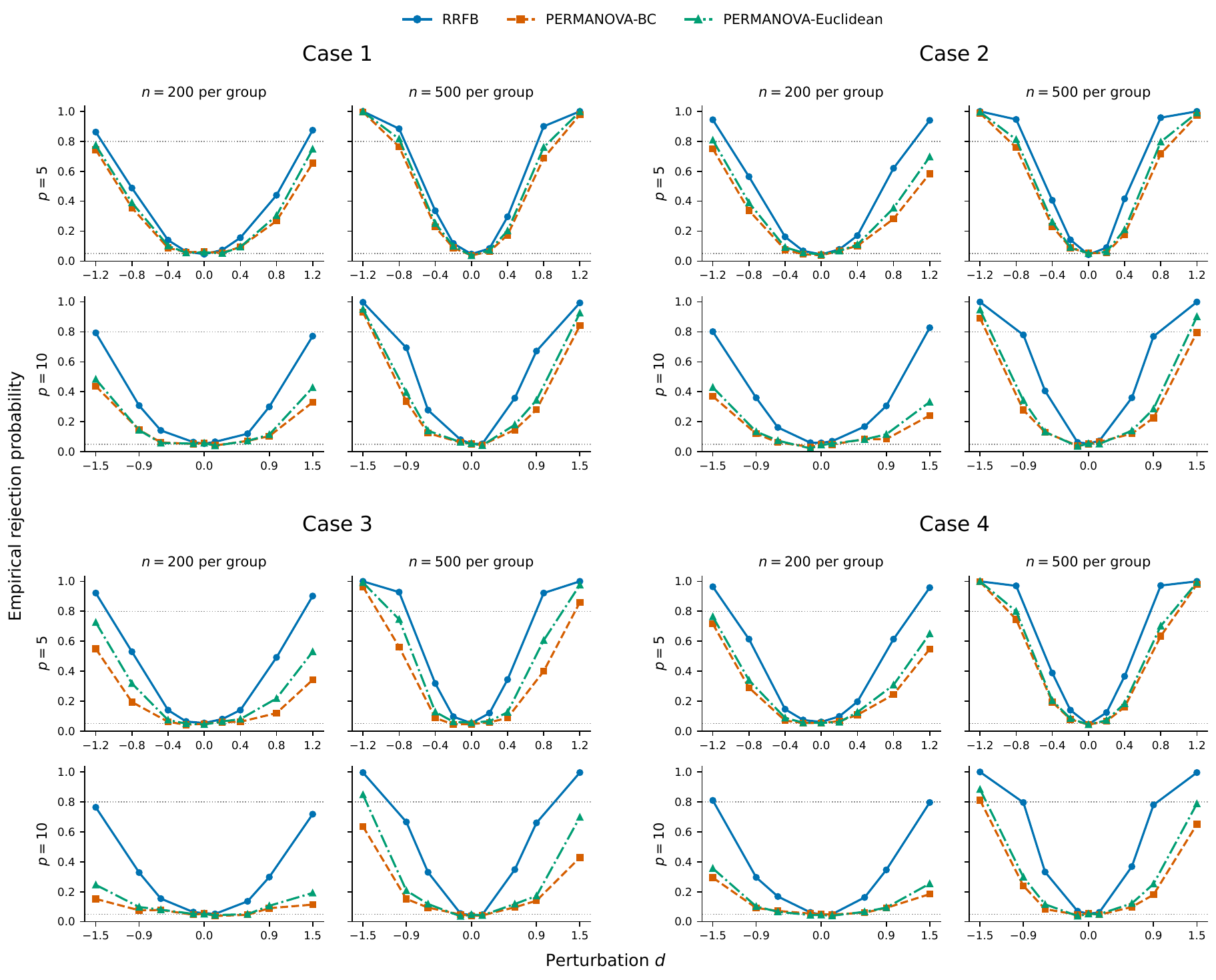}
	\caption{Empirical rejection probabilities at $\alpha=0.05$ under structured perturbations of $\gamma$ with common $A$. Within each case, rows correspond to $p=5,10$ and columns to $n=200,500$ observations per group. PERMANOVA uses Bray--Curtis distance on relative abundances or Euclidean distance on square-root compositions. Nonzero perturbations use 500 replicates per setting; null points use 1,000 matching null-calibration replicates. Lines connect evaluated points; dotted horizontal lines mark 0.05 and 0.80.}
	\label{supp:fig:aligned-power}
\end{figure}

Web Figure~\ref{supp:fig:aligned-power} presents the complete rejection-probability curves. Estimated power increased with perturbation magnitude on each side of the null and with sample size across all four configurations. At every evaluated nonzero grid point, the estimated RRFB rejection probability was at least as large as that of either PERMANOVA procedure. These comparisons concern structured changes in $\gamma$ under the RRFB model with common $A$ and do not establish a general power ordering across other distributional alternatives.

\websubsection{I.3 Violations of the Common-\texorpdfstring{$A$}{A} Assumption}

We examined two departures from the common-$A$ assumption using the $p=5$, Case~2 (moderate zero prevalence) configuration, with $n=200$ observations per group. Group~0 retained the baseline parameters $(A_0,\gamma)$, where $A_0=Q\diag(\lambda)Q^\top$. Group~1 shared the same ambient-coordinate directional parameter, $\gamma_1=\gamma_0=\gamma$, but had a different quadratic parameter. For eigenvalue scaling, we set
\[
A_1(c)=Q\diag(c\lambda)Q^\top=cA_0,
\qquad c\in\{0.60,0.75,0.90,1.10,1.25,1.40\}.
\]
This construction changes the nonzero eigenvalues while retaining the eigendirections and the identifying constraint $\lambda_1=0$. For rotation, we set
\[
A_1(\theta)=R_\theta A_0R_\theta^\top,
\qquad
\theta\in\{\pm5^\circ,\pm10^\circ,\pm15^\circ,\pm20^\circ,\pm25^\circ,\pm30^\circ,\pm50^\circ\}.
\]
Here $R_\theta$ rotates a fixed two-dimensional plane orthogonal to $\gamma$ and acts as the identity on its orthogonal complement, so that $R_\theta\gamma=\gamma$. The plane was constructed by projecting the standard coordinate vectors onto $\gamma^\perp$ in coordinate order and orthonormalizing the first two linearly independent projections. Thus, this construction changes the eigendirections of $A$ while preserving its eigenvalues and leaving $\gamma$ unchanged.

\begin{figure}[htbp]
	\captionsetup{name=Web Figure}
	\centering
	\includegraphics[width=0.9\linewidth]{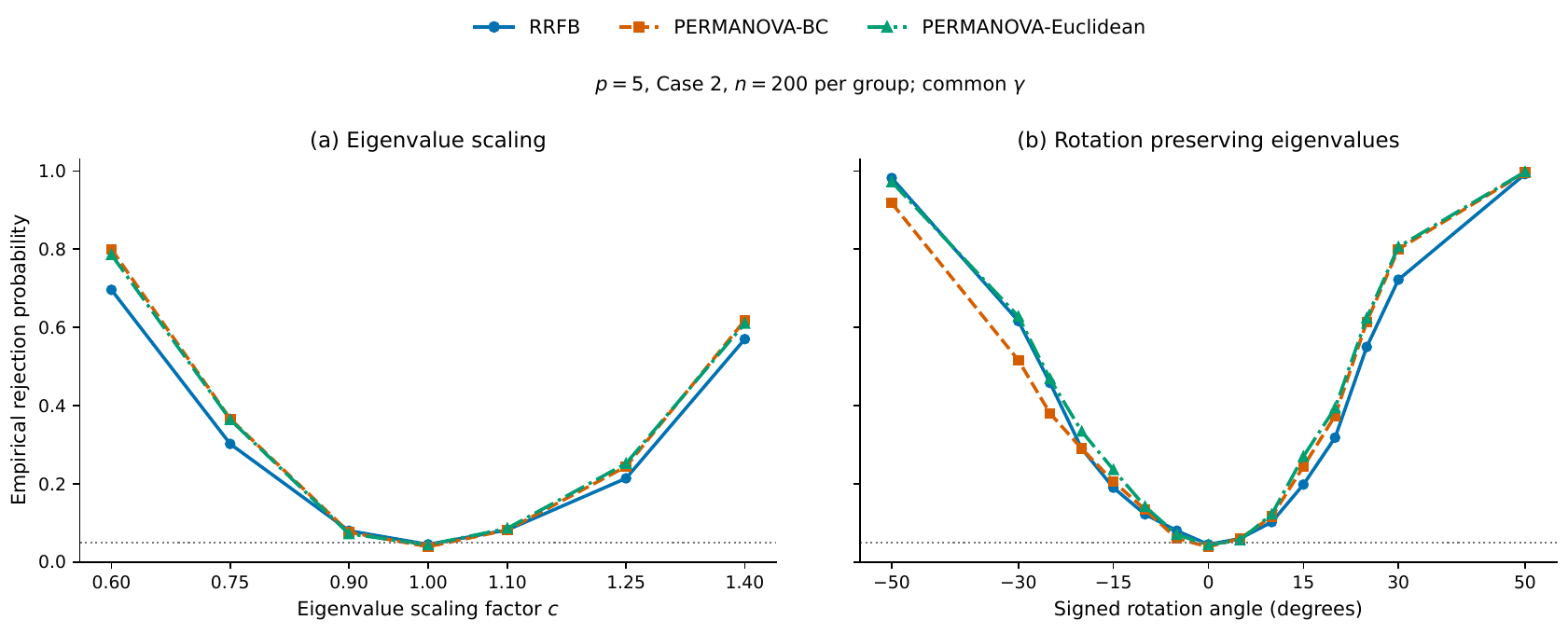}
	\caption{Empirical rejection probabilities at $\alpha=0.05$ under selected common-$A$ violations, for $p=5$, Case~2, and $n=200$ observations per group, with $\gamma$ fixed across groups. (a) Eigenvalue scaling by $c$. (b) Rotation preserving the eigenvalues and $\gamma$. Each departure uses 500 replicates. Unperturbed points use 1,000 matching null-calibration replicates; the same method-specific null estimate is reused in both panels. Lines connect evaluated points, and the dotted horizontal line marks 0.05.}
	\label{supp:fig:common-A}
\end{figure}

Each departure used 500 replicates and the three testing procedures described in Section~5.2 (Study~2). Web Figure~\ref{supp:fig:common-A} shows the rejection probabilities, together with the matching null-calibration estimates at $c=1$ and $\theta=0$. For the mild departures $c\in\{0.90,1.10\}$ and $\theta=\pm5^\circ$, rejection probabilities ranged from 0.060 to 0.082 for RRFB and from 0.056 to 0.086 for the PERMANOVA procedures. Rejection probabilities increased as departures became more pronounced. For example, at $c=0.60$, they were 0.696 for RRFB and 0.786--0.800 for PERMANOVA; at $\theta=\pm50^\circ$, they ranged from 0.918 to 0.998 across the three procedures. These results showcase that RRFB exhibits sensitivity broadly comparable to the two PERMANOVA procedures under the selected violations. Although $\gamma$ is equal between groups, changing $A$ can alter observed means, zero probabilities, and dependence. The RRFB rejection probabilities therefore quantify sensitivity to a violated common-$A$ assumption.

\webappendix{J}{Additional Details for Study 3}

\websubsection{J.1 Computational Performance}

We summarize computational performance using the estimation experiments in Web Appendix~H.1 and the null-calibration experiments in Web Appendix~I.1, under the numerical settings in Web Appendix~G.1. For estimation, we recorded wall-clock fitting time and the number of outer MCEM iterations. For inference, we recorded the additional time required to calculate the score statistic and its sandwich covariance after fitting the pooled null model.

\begin{figure}[htbp]
	\captionsetup{name=Web Figure}
	\centering
	\includegraphics[width=\linewidth]{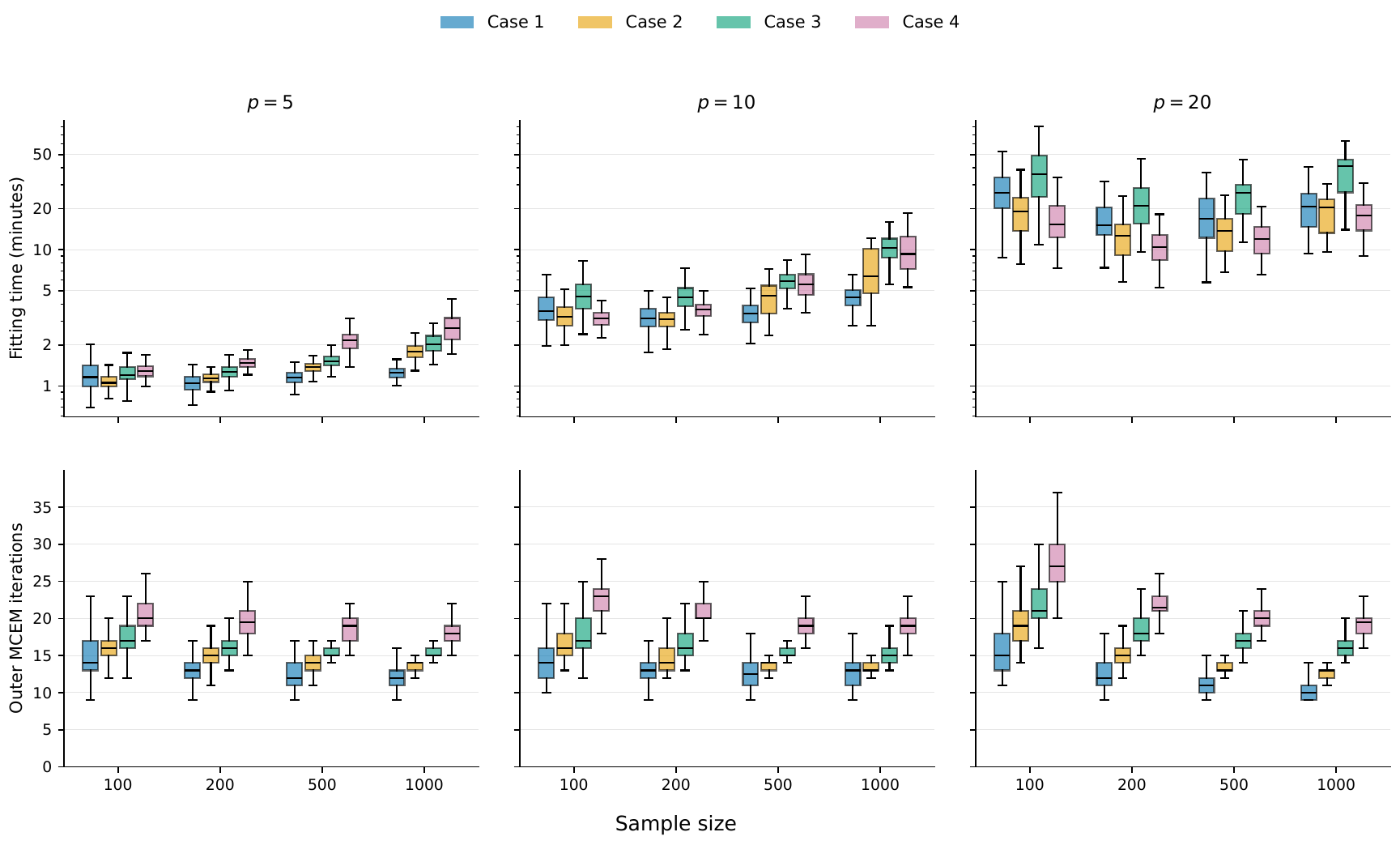}
	\caption{Computational performance of MCEM fitting under the factorial design in Web Appendix~H.1, based on 200 simulation replicates per configuration. Columns identify dimension, and colors identify Cases~1--4. The upper row shows wall-clock fitting time in minutes on a logarithmic scale; the lower row shows the number of outer MCEM iterations on a linear scale. Sample size is the total number of observations in each fitted dataset. Boxes show interquartile ranges, horizontal lines show medians, and whiskers extend to the most extreme observations within 1.5 interquartile ranges of the box. Observations beyond the whiskers are not displayed.}
	\label{supp:fig:j1-fitting}
\end{figure}

Web Figure~\ref{supp:fig:j1-fitting} shows the distributions of fitting times and outer MCEM iteration counts by configuration. Fitting time increased markedly with dimension but was not strictly monotone in sample size. At $p=20$, median fitting times decreased between $n=100$ and $n=200$ in all four cases, alongside reductions in outer-iteration counts. Case~4 generally required the most outer iterations, yet had the shortest median fitting times at $p=20$, whereas Case~3 had the longest. Thus, outer-iteration counts alone did not explain runtime differences across configurations.

\begin{figure}[htbp]
	\captionsetup{name=Web Figure}
	\centering
	\includegraphics[width=\linewidth]{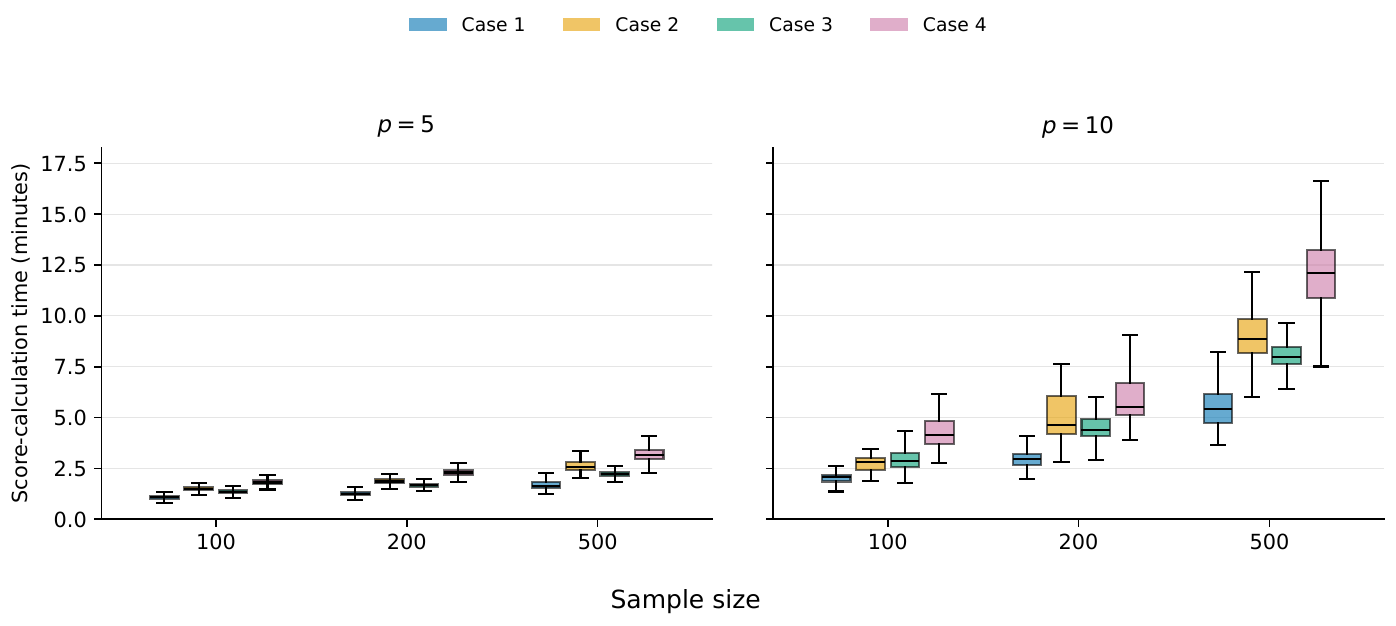}
	\caption{Additional wall-clock time for calculating the common-$A$ RRFB $\gamma$-score statistic and its sandwich covariance after fitting the pooled null model, based on 1,000 null-calibration replicates per configuration. Panels identify dimension, colors identify Cases~1--4, and sample sizes are per group. Pooled null-model fitting time is excluded. Boxes show interquartile ranges, horizontal lines show medians, and whiskers extend to the most extreme observations within 1.5 interquartile ranges of the box. Observations beyond the whiskers are not displayed. Both panels use the same linear runtime scale.}
	\label{supp:fig:j1-score}
\end{figure}

Web Figure~\ref{supp:fig:j1-score} summarizes the additional computational cost of score-test calculation under the null. This stage includes evaluation of the likelihood derivatives, nuisance adjustment, and construction of the sandwich covariance, as described in Web Appendix~E.4. Median score-calculation time increased with sample size in every configuration and was longer for $p=10$ than for $p=5$. Case~4 had the longest median score-calculation time at each dimension and sample size.

\websubsection{J.2 Sensitivity to Initialization}

For the initialization experiment described in Section~5.3 (Study 3), we used the default initialization method described in Web Appendix~D.4 and the following observed-moment based alternative initialization. Let $\overline y=n^{-1}\sum_{i=1}^n y_i$, $R=\|\overline y\|_2$, and $\mu=\overline y/R$. The von Mises--Fisher parameterization $\gamma=\kappa\mu$ motivates aligning the initial directional parameter with $\mu$ and using the resultant length $R$ to determine its magnitude. Using the concentration approximation of \citet{banerjee2005clustering}, we set
\[
\kappa \approx \frac{R(p-R^2)}{1-R^2},
\qquad
\gamma^{(0)}=\min\{0.5\kappa,100\}\mu.
\]
The estimated concentration is halved to moderate the initial directional magnitude, with the resulting norm capped at 100. To initialize $A$, we use observed variation orthogonal to $\mu$. Define
\[
r_i=y_i-(y_i^\top\mu)\mu,
\qquad
C=\frac{1}{n}\sum_{i=1}^n r_i r_i^\top.
\]
An orthonormal tangent basis $b_1,\ldots,b_{p-1}$ is constructed from the leading eigendirections of $C$, giving $Q^{(0)}=[\mu,b_1,\ldots,b_{p-1}]$. This places the empirical mean direction along the unpenalized axis and uses tangent variation to orient the remaining quadratic penalties. Let $\tau_1\geq\cdots\geq\tau_p$ denote the eigenvalues of $C$. We set
\[
\lambda_j^{(0)}
=
10\frac{\tau_1-\tau_j}{\tau_1-\tau_p},
\qquad j=1,\ldots,p,
\qquad
A^{(0)}=Q^{(0)}\diag(\lambda^{(0)})(Q^{(0)})^\top.
\]
This construction gives ordered initial eigenvalues ranging from 0 to 10, with all eigenvalues set to zero when the covariance eigenvalues have no spread. Because conditioning, rectification, and renormalization alter the relationship between observed moments and latent parameters, this construction serves as an initialization heuristic.

We compared two initialization methods: the proposed latent-moment initialization and an alternative based directly on observed moments described above. Each dataset was fitted from four sets of starting values: the proposed initialization, versions scaled by 0.5 and 2, and the observed-moment initialization. We assessed agreement across the four fitted solutions using the attained observed log likelihood, estimated expected log-score difference, and parameter estimates. For each dataset, we calculated the across-start ranges in observed log likelihood per observation and estimated expected log-score difference. The latter summarizes variation in estimated predictive performance relative to the generating distribution. Predictive observations and numerical integration workspaces were shared across starts to support comparisons within each dataset.

We additionally summarized parameter disagreement using the relative diameters
\[
D_A =
\frac{\max_{s<t}\|\widehat A_s-\widehat A_t\|_{\mathrm F}}
{\max\{1,\|\widehat A_{s^*}\|_{\mathrm F}\}},
\qquad
D_\gamma =
\frac{\max_{s<t}\|\widehat\gamma_s-\widehat\gamma_t\|_2}
{\max\{1,\|\widehat\gamma_{s^*}\|_2\}},
\]
where $s$ indexes the four starts and $s^*$ denotes the start with the smallest estimated expected log-score difference. These quantities measure disagreement among fitted parameters rather than estimation error relative to the generating parameters. Comparing $A$ directly also avoids treating eigenvector sign changes as differences in the fitted model.

\begin{figure}[htbp]
	\captionsetup{name=Web Figure}
	\centering
	\includegraphics[width=\linewidth]{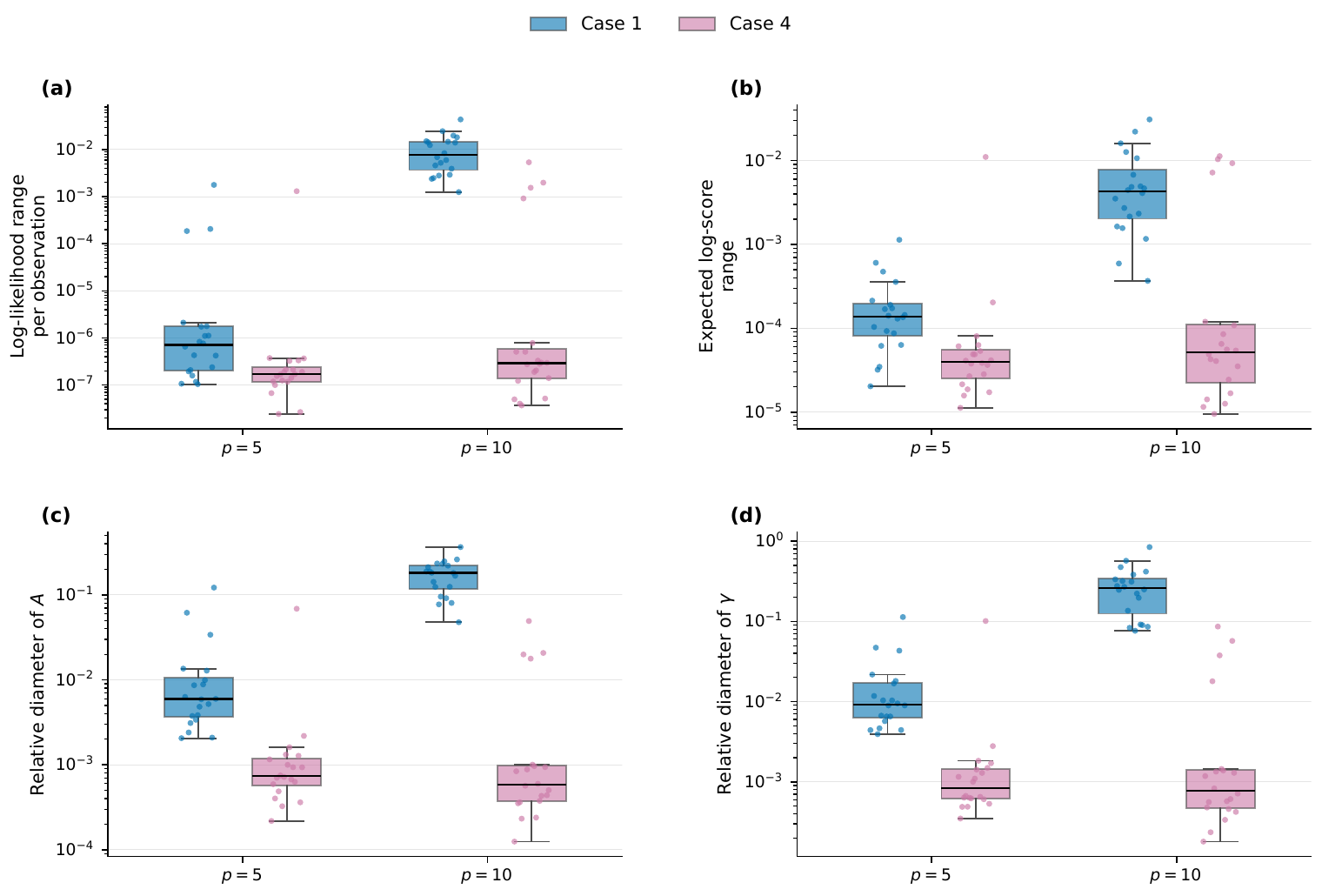}
	\caption{Sensitivity to initialization for 20 independently generated datasets per configuration, with $n=200$, $p\in\{5,10\}$, and Cases~1 and 4. Each dataset was fitted using the four initializations described in the main text. Panels show the across-start range in (a) attained observed log likelihood per observation and (b) estimated expected log-score difference, and the relative diameters of (c) $A$ and (d) $\gamma$, as defined in the text. All vertical axes are logarithmic. Boxes show interquartile ranges, horizontal lines show medians, and whiskers extend to the most extreme observations within 1.5 interquartile ranges of the box. Points represent individual datasets, including those beyond the whiskers.}
	\label{supp:fig:j2-init}
\end{figure}

Web Figure~\ref{supp:fig:j2-init} shows that the different initializations generally yielded similar estimated predictive performance, with small across-start log-score ranges for most datasets in three of the four configurations. Larger differences were observed for $p=10$ under Case~1, and occasional datasets in the other configurations also showed greater sensitivity. Parameter disagreement followed a similar pattern, with greater typical differences under Case~1 than under Case~4, particularly at $p=10$. Overall, predictive performance was typically similar across initialization choices, although the degree of agreement varied by configuration.

\webappendix{K}{Additional Details for the Avocado Intervention Study}

\websubsection{K.1 Feature Selection and Data Description}
\label{subsec:genus_supp}

We based our genus-level feature selection on the eleven genera identified in the HACK-top-17-taxa \citep{goel2025toward}. Taxonomic annotation was performed using the SILVA 132 rRNA database \citep{yilmaz2014silva} where Agathobaculum was not available as a distinct genus-level label and was therefore not included as a separate component. This annotation limitation does not establish its biological absence from the samples. Consequently, we retained the ten genera from the HACK-top-17-taxa that are represented in the SILVA-based annotations and aggregated all remaining taxa into an ``Other’’ category. When multiple labels corresponded to subdivisions of the same genus (e.g., bracketed \textit{Eubacterium} groups or numbered variants such as \textit{Coprococcus} 1–3), their relative abundances were summed to form a single genus-level feature. Web Table~\ref{tab:supp_selected_taxa} summarizes the genera included in the analysis along with their nonzero proportions overall, at baseline, and at the end of intervention.

\begin{table}[htbp]
	\captionsetup{name=Web Table}
	\centering
	\small
	\caption{Observed nonzero prevalence of the ten named compositional components, overall and by randomized arm at baseline and end of intervention. Entries are percentages of samples with relative abundance greater than zero. Each visit includes 133 participants: 66 in the control arm and 67 in the avocado arm. Nonzero prevalence refers to the supplied taxonomic annotations and does not establish biological presence or absence independently of the measurement procedure. The residual ``Other'' component had nonzero abundance in every sample and is omitted from this table; it was retained in the eleven-component analysis.}
	\label{tab:supp_selected_taxa}
	\begin{tabular}{lrrrrrr}
		\toprule
		& \multicolumn{2}{c}{All participants}
		& \multicolumn{2}{c}{Control}
		& \multicolumn{2}{c}{Avocado} \\
		\cmidrule(lr){2-3}
		\cmidrule(lr){4-5}
		\cmidrule(lr){6-7}
		Component & Baseline & End & Baseline & End & Baseline & End \\
		\midrule
		\textit{Bacteroides}        &  99.2 &  99.2 &  98.5 &  98.5 & 100.0 & 100.0 \\
		\textit{Eubacterium}        &  97.7 &  97.0 &  95.5 &  95.5 & 100.0 &  98.5 \\
		\textit{Faecalibacterium}   &  96.2 &  96.2 &  95.5 &  95.5 &  97.0 &  97.0 \\
		\textit{Fusicatenibacter}   &  92.5 &  88.7 &  92.4 &  87.9 &  92.5 &  89.6 \\
		\textit{Roseburia}          &  90.2 &  90.2 &  90.9 &  86.4 &  89.6 &  94.0 \\
		\textit{Alistipes}          &  88.7 &  90.2 &  86.4 &  83.3 &  91.0 &  97.0 \\
		\textit{Coprococcus}        &  87.2 &  85.7 &  89.4 &  83.3 &  85.1 &  88.1 \\
		\textit{Ruminococcus}       &  78.2 &  80.5 &  78.8 &  74.2 &  77.6 &  86.6 \\
		\textit{Oscillibacter}      &  68.4 &  68.4 &  66.7 &  66.7 &  70.1 &  70.1 \\
		\textit{Odoribacter}        &  63.9 &  65.4 &  60.6 &  59.1 &  67.2 &  71.6 \\
		\bottomrule
	\end{tabular}
\end{table}

\websubsection{K.2 Effect Estimates, Permutation Inference, and Score Diagnostics}

For each component, we calculated avocado-minus-control differences in mean relative abundance, nonzero prevalence, and mean relative abundance among observations with nonzero values. All contrasts are expressed in percentage points. We obtained pointwise 95\% confidence intervals from the 2.5th and 97.5th percentiles of 20,000 bootstrap replicates, resampling participants with replacement independently within each arm and retaining each participant's full composition. Web Table~\ref{tab:avocado-component-contrasts} reports the contrasts and intervals for all eleven components.

\begin{figure}[htbp]
	\captionsetup{name=Web Figure}
	\centering
	\includegraphics[width=0.75\linewidth]{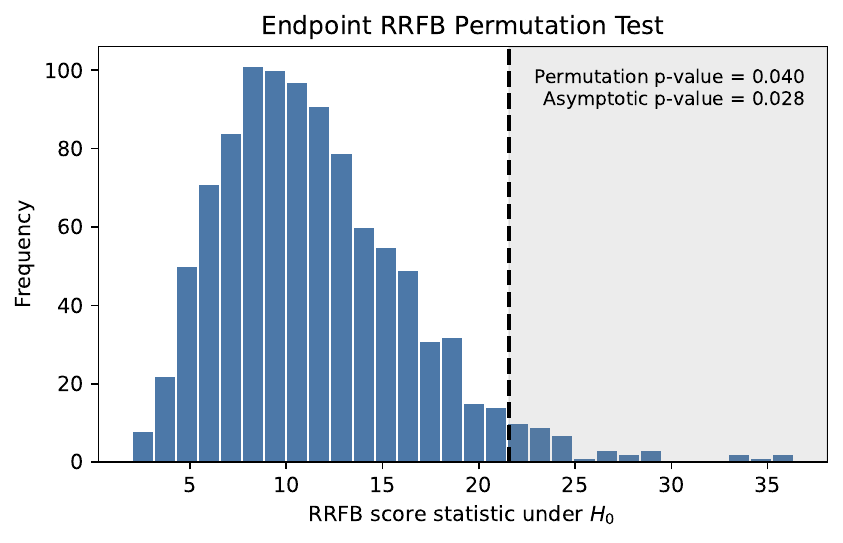}
	\caption{Permutation distribution of the RRFB score statistic for the eleven-component comparison between the avocado and control arms at the end of the intervention. The histogram shows the statistics from 999 treatment-label shuffles. The dashed line marks the observed statistic ($T=21.575$), and the shaded region denotes values at least as large as the observed statistic.}
	\label{supp:fig:avocado-permutation}
\end{figure}

For the permutation analysis, we fitted the null model once to the pooled endpoint sample. The fit met the MCEM stopping criteria after 91 iterations. We shuffled treatment labels across the 133 participants 999 times, preserving the arm sizes of 66 and 67. For each shuffle, we recalculated the statistic using the same pooled fit and numerical integration draws. This pooled permutation reference assumes exchangeability of the analyzed participants under the null. All 999 permutations yielded finite statistics, and 39 were at least as large as the observed statistic of 21.575. The permutation $p$-value was $(1+39)/(1+999)=0.040$. Web Figure~\ref{supp:fig:avocado-permutation} shows the observed statistic and the upper tail of the permutation distribution used to calculate the $p$-value.

Let $u=\bar s_{\psi\mid\eta}$ denote the nuisance-adjusted average score, and let $V=\widehat V_{\psi\mid\eta}$ estimate the covariance of $\sqrt{n}\,u$, as defined in Web Appendix~E. Panel~B of Figure~3 in the main text reports the reduction in the global statistic after omitting score coordinate $j$ and the corresponding row and column of $V$. Writing $u_{-j}$ and $V_{-j,-j}$ for the retained score vector and covariance submatrix, we have
\[
\begin{aligned}
	v_j &= V_{jj}-V_{j,-j}V_{-j,-j}^{-1}V_{-j,j},\\
	\Delta_j &= T-nu_{-j}^{\top}V_{-j,-j}^{-1}u_{-j}
	= n\,\frac{\bigl(u_j-V_{j,-j}V_{-j,-j}^{-1}u_{-j}\bigr)^2}{v_j}.
\end{aligned}
\]
Thus, $\Delta_j$ is the squared standardized residual for the $j$th coordinate of $\sqrt n\, u$ after projection on the remaining coordinates. All reductions use the full eleven-component null fit. Each reduction describes the sensitivity of the global statistic to omitting one coordinate.

\begin{table}[htbp]
	\captionsetup{name=Web Table}
	\centering
	\small
	\setlength{\tabcolsep}{4pt}
	\caption{Observed endpoint contrasts and score reductions for the eleven-component representation. Contrasts are avocado minus control in percentage points, with pointwise 95\% percentile bootstrap confidence intervals in brackets. The final column gives the reduction in the global RRFB statistic upon omission of the corresponding score coordinate.}
	\label{tab:avocado-component-contrasts}
	\begin{tabular}{@{}lcccr@{}}
		\toprule
		Component & \shortstack{Mean relative\\abundance} & \shortstack{Nonzero\\prevalence} & \shortstack{Mean among\\nonzero observations} & $\Delta_j$ \\
		\midrule
		\textit{Faecalibacterium} & $2.77\,[-0.09,\,5.73]$ & $1.56\,[-4.48,\,7.62]$ & $2.66\,[-0.18,\,5.57]$ & 0.073 \\
		\textit{Alistipes} & $1.46\,[0.49,\,2.51]$ & $13.68\,[4.59,\,24.24]$ & $1.16\,[0.14,\,2.24]$ & 6.617 \\
		\textit{Ruminococcus} & $0.60\,[-0.03,\,1.22]$ & $12.32\,[-1.20,\,25.87]$ & $0.40\,[-0.30,\,1.08]$ & 1.104 \\
		\textit{Eubacterium} & $0.58\,[-0.17,\,1.32]$ & $3.05\,[-2.94,\,9.09]$ & $0.50\,[-0.24,\,1.24]$ & 0.407 \\
		\textit{Odoribacter} & $0.07\,[-0.03,\,0.17]$ & $12.55\,[-3.98,\,29.01]$ & $0.05\,[-0.08,\,0.18]$ & 0.005 \\
		\textit{Coprococcus} & $0.07\,[-0.43,\,0.57]$ & $4.73\,[-7.30,\,16.76]$ & $0.00\,[-0.55,\,0.55]$ & 1.114 \\
		\textit{Oscillibacter} & $-0.01\,[-0.07,\,0.03]$ & $3.48\,[-11.62,\,19.92]$ & $-0.03\,[-0.11,\,0.03]$ & 0.586 \\
		\textit{Fusicatenibacter} & $-0.15\,[-0.37,\,0.06]$ & $1.67\,[-8.86,\,12.23]$ & $-0.19\,[-0.42,\,0.04]$ & 3.674 \\
		\textit{Roseburia} & $-0.42\,[-1.49,\,0.54]$ & $7.67\,[-2.83,\,18.20]$ & $-0.70\,[-1.86,\,0.34]$ & 0.020 \\
		\textit{Bacteroides} & $-0.93\,[-5.77,\,4.02]$ & $1.52\,[0.00,\,4.55]$ & $-1.30\,[-6.16,\,3.64]$ & 0.853 \\
		Other & $-4.04\,[-9.03,\,0.90]$ & $0.00\,[0.00,\,0.00]$ & $-4.04\,[-9.03,\,0.90]$ & 1.375 \\
		\bottomrule
	\end{tabular}
\end{table}

To examine the \textit{Fusicatenibacter} result, we decomposed its signed score residual into participant contributions. Let $u_i$ be the nuisance-adjusted score for participant $i$, so that $u=n^{-1}\sum_i u_i$. The signed contribution for coordinate $j$ is
\[
c_{ij}=\frac{u_{ij}-V_{j,-j}V_{-j,-j}^{-1}u_{i,-j}}{\sqrt{n v_j}},
\qquad
\Delta_j=\left(\sum_{i=1}^{n}c_{ij}\right)^2.
\]
For \textit{Fusicatenibacter}, the 118 observations with nonzero abundance contributed $-2.142$ to the signed sum, and the 15 observations with zero recorded abundance contributed $0.225$, giving a total of $-1.917$. Each contribution uses the participant's full composition, including the zero pattern of the other components.

In a descriptive examination of all ten pairwise Spearman correlations involving \textit{Fusicatenibacter}, the largest absolute difference between arms was for its association with \textit{Faecalibacterium}. The correlations were 0.13 in the control arm and 0.58 in the avocado arm. We then restricted attention to samples with nonzero \textit{Fusicatenibacter} abundance and summarized its relative abundance above and below the pooled endpoint median of \textit{Faecalibacterium} abundance (12.62\%). The between-arm contrast was concentrated in the lower stratum (Web Table~\ref{tab:avocado-fusicatenibacter}). The same pattern was observed using the pooled 40th or 60th percentile as the cutoff. These strata were defined post hoc using endpoint measurements, and the comparisons describe the observed joint distribution.

\begin{table}[htbp]
	\captionsetup{name=Web Table}
	\centering
	\small
	\caption{Mean relative abundance of \textit{Fusicatenibacter} among endpoint samples in which it was detected, by arm and observed \textit{Faecalibacterium} abundance. Entries are means in percent, with sample counts in parentheses.}
	\label{tab:avocado-fusicatenibacter}
	\begin{tabular}{@{}lcc@{}}
		\toprule
		\textit{Faecalibacterium} abundance & Control & Avocado \\
		\midrule
		At or below the pooled endpoint median & 0.87 (31) & 0.42 (23) \\
		Above the pooled endpoint median & 0.89 (27) & 0.86 (37) \\
		\bottomrule
	\end{tabular}
\end{table}

The nuisance sensitivity matrix had numerical rank 76 of 76 and was indefinite, with condition numbers of $5.60\times10^9$ on the raw scale and $1.68\times10^6$ after diagonal standardization. Its common-$A$ sub-block had numerical rank 65 of 65, with corresponding condition numbers of $5.71\times10^9$ and $3.69\times10^4$. The efficient-score covariance was positive definite, with numerical rank 11 of 11 and condition numbers of 53.9 and 27.0, respectively.

\websubsection{K.3 Sensitivity to Feature Selection}

We assessed sensitivity to genus-level aggregation using minimum nonzero-prevalence thresholds of 90\% and 85\%. For each named genus, we calculated prevalence as the proportion of the 266 samples with positive relative abundance, pooling both visits and both arms. At each threshold, we summed the relative abundances of genera below the threshold into ``Other,'' preserving the unit sum. We then repeated the between-arm comparison on the 133 endpoint samples using the resulting composition.

The 90\% threshold retained five named genera as separate components, \textit{Bacteroides}, \textit{Eubacterium}, \textit{Faecalibacterium}, \textit{Fusicatenibacter}, and \textit{Roseburia}, together with ``Other.'' The 85\% threshold additionally retained \textit{Alistipes} and \textit{Coprococcus}, which had pooled nonzero prevalences of 89.5\% and 86.5\%, respectively. The primary eleven-component representation also retained \textit{Ruminococcus}, \textit{Oscillibacter}, and \textit{Odoribacter} separately.

For each representation, we fitted a pooled null model and used that fit to calculate the observed and permuted statistics. The six- and eight-component fits met the MCEM stopping criteria after 13 and 21 iterations, respectively. Web Table~\ref{tab:avocado-feature-sensitivity} reports the asymptotic and permutation results. Each permutation calculation used 999 valid label shuffles.

\begin{table}[htbp]
	\captionsetup{name=Web Table}
	\centering
	\small
	\caption{Sensitivity of the endpoint RRFB score test to prevalence-based aggregation. The exceedance count is the number of permuted statistics at least as large as the observed statistic. The permutation $p$-value is the exceedance count plus one, divided by 1,000.}
	\label{tab:avocado-feature-sensitivity}
	\begin{tabular}{@{}lrrrrr@{}}
		\toprule
		Representation & Components & $T$ & Exceedances & Asymptotic $p$ & Permutation $p$ \\
		\midrule
		Primary & 11 & 21.575 & 39 & 0.028 & 0.040 \\
		90\% threshold & 6 & 8.583 & 233 & 0.198 & 0.234 \\
		85\% threshold & 8 & 19.967 & 13 & 0.010 & 0.014 \\
		\bottomrule
	\end{tabular}
\end{table}

To examine the roles of the two genera retained at 85\%, we applied the score diagnostic from Web Appendix~K.2 within the eight-component fit. Omitting the \textit{Alistipes} coordinate reduced the statistic by 8.824, compared with 0.201 for \textit{Coprococcus}. Omitting both coordinates reduced it by 9.103, giving a statistic of 10.863. The larger reduction for \textit{Alistipes} points to its prominence in the eight-component result. The six-component comparison also pools both genera into ``Other'' and fits the null model to that coarser composition, so the threshold comparison reflects both the choice of separate components and the resulting change in the fitted model.

\renewcommand{\refname}{Supplementary References}

\end{document}